\documentclass[12pt]{amsart}

\usepackage{geometry}
\usepackage[utf8]{inputenc}
\usepackage[T1]{fontenc}
\usepackage{amsfonts}

\usepackage{mathtools}

\usepackage{graphicx}
\usepackage{amsmath}
\usepackage{amssymb,color}
\usepackage{mathrsfs}

\def\br{\begin{remark}\rm\small}
\def\er{\end{remark}}
\newcommand{\dd}{\mathrm{d}}
\newcommand{\beq}{\begin{equation}}
\newcommand{\eeq}{\end{equation}}
\newcommand{\bea}{\begin{eqnarray}}
\newcommand{\eea}{\end{eqnarray}}
\newcommand{\Res}{\mathop{\,\rm Res\,}}
\newcommand{\bs}[1]{\ensuremath{\boldsymbol{#1}}}

\newcommand*{\bigcdot}{\raisebox{-0.25ex}{\scalebox{1.2}{$\cdot$}}}

\newtheorem{theorem}{Theorem}
\newtheorem{proposition}[theorem]{Proposition}
\newtheorem{lemma}[theorem]{Lemma}
\newtheorem{corollary}[theorem]{Corollary}

\theoremstyle{definition}
\newtheorem{definition}[theorem]{Definition}
\theoremstyle{remark}
\newtheorem{remark}[theorem]{Remark}
\newtheorem{notation}[theorem]{Notation}

\numberwithin{equation}{section}
\numberwithin{theorem}{section}

\calclayout

\begin{document}

\title[Nesting statistics in the $O(n)$ loop model \ldots ]{Nesting statistics in the $O(n)$ loop model on random maps of arbitrary topologies}
\date{\today}
\author{Ga\"etan Borot}
\address{Max Planck Institut f\"ur Mathematik, Vivatsgasse 7, 53111 Bonn, Germany}
\email{gborot@mpim-bonn.mpg.de}
\author{Elba Garcia-Failde}
\address{Max Planck Institut f\"ur Mathematik, Vivatsgasse 7, 53111 Bonn, Germany}
\email{elba@mpim-bonn.mpg.de}
\thanks{We thank J\'er\'emie Bouttier and Bertrand Duplantier for discussions. We also thank the anonymous referee for the useful comments on this article.}

\begin{abstract}
We pursue the analysis of nesting statistics in the $O(n)$ loop model on random maps, initiated for maps with the topology of disks and cylinders in \cite{BBD}, here for arbitrary topologies. For this purpose we rely on the topological recursion results of \cite{BEOn,BEO13} for the enumeration of maps in the $O(n)$ model. We characterize the generating series of maps of genus $\mathsf{g}$ with $k$ boundaries and $k'$ marked points which realize a fixed nesting graph. These generating series are amenable to explicit computations in the loop model with bending energy on triangulations, and we characterize their behavior at criticality in the dense and in the dilute phase.
\end{abstract}

\maketitle

\section{Introduction}

The enumeration of maps, which are models for discretized surfaces, developed initially from the work of Tutte \cite{Tutte1,Tutte2,TutteQ}. The discovery of matrix model techniques \cite{BIPZ} and the development of bijective techniques based on coding by decorated trees \cite{CoriV,Schae} led in the past 30 years to a wealth of results. An important motivation comes from the conjecture that the geometry of large random maps is universal, \textit{i.e.}, there should exist ensembles of random metric spaces depending on a small set of data (like the central charge and a symmetry group attached to the problem) which describe the continuum limit of random maps. Two-dimensional quantum gravity aims at the description of these random continuum objects and physical processes on them, and the universal theory which should underly is Liouville quantum gravity possibly coupled to a conformal field theory \cite{KPZ,Ginsparg-Moore,Ginsparg}. Understanding rigorously the emergent fractal geometry of such limit objects is nowadays a major problem in mathematical physics. Another important problem is to establish the convergence of random maps towards such limit objects. Solving various problems of map enumeration is often instrumental in this program, as it provides useful probabilistic estimates.

As of now, the geometry of large random planar maps with faces of bounded degrees (\textit{e.g.}, quadrangulations) is fairly well understood. In particular, their scaling limit is the Brownian map \cite{MarckertMokkadem,GallMap,MierMapP,Gall1},  the complete proof of convergence in the Gromov-Hausdorff sense being obtained in \cite{MierMapP,Gall1}. This universality class is often called in physics that of ``pure gravity''. Recent progress generalized part of this understanding to planar maps containing faces whose degrees are drawn from a heavy tail distribution. In particular, the limiting object is the so-called $\alpha$-stable map, which can be coded in terms of stable processes whose parameter $\alpha$ is related to the power law decay of the degree distribution \cite{MierGall}.

The next class of interesting models concerns random maps equipped with a statistical physics model, like percolation \cite{Kazakov}, the Ising model \cite{1986PhLA..119..140K,BouKa}, or the $Q$-Potts model \cite{Daul,BonnetEynard,PZinn}, \ldots To make the distinction explicit, maps without a statistical physics model will be called \emph{usual maps}. It is well-known, at least on fixed lattices \cite{FKcluster,BaKeWu,Truong,PerkWu,NienhuisCG}, that the $Q$-state Potts model can be reformulated as a fully packed loop model with a fugacity $\sqrt{Q}$ per loop, for random maps this equivalence is explained in detail in \cite{BBG12c}. The $O(n)$ model also admits a famous representation in terms of loops \cite{DoMuNiSc81,NienhuisCG} with $n$ the fugacity per loop.
The interesting feature of the $O(n)$ model is that it gives rise to two universality classes which depend continuously on $n$, called \emph{dense} or \emph{dilute} in respect to the behavior of macroscopic loops, as can be detected at the level of critical exponents \cite{1982PhRvL..49.1062N,1984JSP....34..731N,NienhuisCG,1988PhRvL..61.1433D,GaudinKostov,KOn,1990NuPhB.340..491D,KSOn,PagesjaunesOn}. The famous KPZ relations \cite{KPZ} (see also \cite{MR981529,MR1005268}) relate, at least from the physics point of view, the critical exponents of these models on a fixed regular lattice, with their corresponding critical exponent on random planar maps, as was repeatedly checked for a series of models \cite{KPZ,1986PhLA..119..140K,1988PhRvL..61.1433D,1990NuPhB.340..491D,KOn,MR2112128,BBD}. These exponents for random planar maps are summarized in \cite[Figure 4]{BBD}.

 It is widely believed that after a Riemann conformal map to a given planar domain, the proper conformal structure for the continuum limit of random planar maps  weighted by the partition functions of various statistical models is described by the \textit{Liouville theory of quantum gravity} (see, {\it e.g.}, the reviews \cite{Ginsparg-Moore,Ginsparg,MR2073993} and  \cite{DuplantierICM2014,LeGallICM2014}). In the particular case of \emph{pure} random planar maps, the universal metric structure of the Brownian map \cite{Gall1,MierMapP,LeGallICM2014} has very recently been identified with that directly constructed from Liouville quantum gravity \cite{map_making,quantum_spheres,qlebm}. After this Riemann conformal mapping,  the configuration of critical $O(n)$ loops is believed to be described in the continuous limit by the so-called \emph{conformal loop ensemble} \cite{sheffield2009,zbMATH06121652}, denoted by ${\rm CLE}_{\kappa}$  and depending on a continuous index $\kappa\in(8/3,8)$,  with the correspondence $n = 2\cos \pi(1 - \frac{4}{\kappa})$ for $n\in (0,2]$  \cite{MR1964687,kg2004guide_to_sle,MR2112128}. Yet, little is known on the \emph{metric} properties of large random maps weighted by an $O(n)$ model, even from a physical point of view. 
 
Most of the works described above are restricted to planar maps. In \cite{BBD}, one of the authors jointly with Bouttier and Duplantier investigated the nesting properties of loops in maps with the topology of a disk or a cylinder weighted by an $O(n)$ model, and showed them to be in perfect agreement with the known nesting properties of ${\rm CLE}_{\kappa}$ \cite{MWW} after taking into account a suitable version of the KPZ relations \cite{KPZDS}. In this article, we push this analysis forward and investigate rigorously the nesting properties of maps of any topology weighted by an $O(n)$ model. This includes as a special case the description of the critical behavior of maps without loops (i.e.~in the class of pure gravity) having possibly marked points, microscopic and macroscopic boundaries. This generalization is non-trivial as the combinatorics of maps with several boundaries, marked points, and arbitrary genus, is much more involved than in the cases of disks and cylinders. Our approach is based on analytic combinatorics, and relies on two main ingredients: (1) the \textbf{substitution approach} developed in \cite{BBG12a,BBG12b} for planar maps; and (2) the \textbf{topological recursion} of \cite{EOFg,BEO} to reduce by a universal algorithm the enumeration of maps -- possibly carrying an $O(n)$ loop model -- of any topology to the enumeration of disks and cylinders. Obtaining the desired asymptotics for generating series of maps subjected to various constraints is then a matter of careful analysis of singularities.

\subsection{Outline and main results}

\subsubsection{Combinatorics of maps and their nesting} We introduce in detail the $O(n)$ loop model and the notion of nesting graph of a map in Section~\ref{S2}. To present informally our findings, let us say that the primary nesting graph $(\Gamma_0,\star)$ of a map $\mathcal{M}$ has vertices corresponding to the connected components of the complement of the loops, and edges between vertices which correspond to connected components adjacent to the same loop. Each vertex carries a genus (of the connected component it refers to) and may carry marks -- here denoted $\star$ -- remembering to which connected components the marked points or boundaries of $\mathcal{M}$ belong to. The nesting graph $(\Gamma,\star)$ is obtained from $\Gamma_0$ by collapsing all genus $0$ univalent vertices which do not carry a mark, and collapsing any maximal sequence of $P$ consecutive edges with at least one of its endpoints being a genus $0$ unmarked vertex to a single edge remembering $P$ -- which we call an arm of length $P$. These two steps are repeated until one of them leaves the graph unchanged. The collection $\mathbf{P}$ of arm lengths is conventionally not included in the data of $\Gamma$. In other words, every edge in $\Gamma_0$ refer to a loop in $\mathcal{M}$, and an arm of length $P$ in $\Gamma$ represents $P$ consecutive loops disposed along a cylindric part of $\mathcal{M}$ which ``separate'' the marks. Precise definitions are provided in \S~\ref{Markm}.

The main goal of the article is to study maps in the $O(n)$ loop model realizing a fixed nesting graph. Their generating series are typically denoted with script letters $\pmb{\mathscr{F}}$. We will also encounter generating series of maps in the $O(n)$ model which are not keeping track of nestings, and denoted $\mathbf{F}$. We review the substitution approach of \cite{BBG12a} in Section~\ref{SSub}, describing disks with an $O(n)$ loop model as usual maps whose faces can also be disks with an $O(n)$ loop model. Generically, generating series of usual maps whose faces can also be disks with an $O(n)$ loop model will be denoted $\bs{\mathcal{F}}$. We may impose geometric constraints on the maps under consideration, by fixing the genus $\mathsf{g}$, the number $k'$ of marked points, the number of boundaries $k$ and their respective perimeters $(\ell_i)_{i = 1}^k$, the volume (\textit{i.e.} the total number of vertices) $V$, and maybe the arm lengths $(P(\mathsf{e}))_{\mathsf{e}}$. This is conveniently handled at the level of generating series by including extra Boltzmann weights, respectively $u^{V}$, $\prod_{i} x_i^{-(\ell_i + 1)}$, and $\prod_{\mathsf{e}} s(\mathsf{e})^{P(\mathsf{e})}$. The precise definitions of these generating series and the easy combinatorial relations between them are described in Section~\ref{S2}. In particular, the basic formula for the generating series of maps with a fixed nesting graph is Proposition~\ref{magi}.

In Section~\ref{Sanal}, we review the analytic properties of these generating series, \textit{i.e.} in which sense the Boltzmann weights can be considered as nonnegative real-valued parameters instead of formal parameters, and their characterization by functional equations already known in the literature. We state the topological recursion formula for $\mathbf{F}^{(\mathsf{g},k)}$ and $\bs{\mathcal{F}}^{(\mathsf{g},k)}$ from \cite{BEO} which will be our second basic formula, and explain how it can be used in practice. We also explain in Section~\ref{addinm} how the addition of extra marked points can easily be handled at the level of generating series. These results are valid in the general $O(n)$ loop model, where loops are allowed to cross faces of any degree.

We specialize in Section~\ref{S4} them in the $O(n)$ loop model on triangulations with bending energy $\alpha$. It also depends on the parameter $h$ per triangle visited by a loop, and $g$ per empty triangle. This model is the simplest one which is amenable to an explicit solution in terms of theta functions, and still contains the dense and dilute universality classes which are specific to loop models. At this point, it is very useful to introduce the parameter $b \in (0,\tfrac{1}{2})$ such that
$$
n = 2\cos(\pi b).
$$
We review the expression for the generating series of disks and cylinders (Section~\ref{elparam}), which are the non-trivial initial data allowing to reach higher topologies. We also transform (Section~\ref{TRtr}-\ref{diagTR}) the topological recursion formula for $\mathbf{F}^{(\mathsf{g},k)}$ into a more explicit sum over trivalent graphs, which will be suited for later analysis.

\subsubsection{Critical behavior}
\label{Sintrocrit} Following \cite{BBG12b}, we review in Section~\ref{critit} the phase diagram of this bending energy model. The properties of the special functions, and some details necessary to obtain this phase diagram as well as for later use, are collected in Appendix~\ref{AppA}-\ref{AppD} which are mostly taken from \cite{BBD}. For fixed $n \in (0,2)$, $\alpha$ not too large and vertex weight $u = 1$, it features in the $(g,h)$ plane a \emph{non-generic} critical line, beyond which the generating series of pointed disks are divergent. As is well-known, the radius of convergence is actually the same for generating series of maps of any topology. The critical exponents in the interior (resp. at the tip) of the non-generic critical line pertain to the dense (resp. dilute) universality class. Beyond this point, the critical line continues to a \emph{generic} line, \textit{i.e.} corresponding to the universality class of pure gravity. We focus on the non-generic critical line as it is specific to the loop models. If $(g,h)$ is chosen on the non-generic critical line but we keep the vertex weight $u < 1$, the model remains off-critical. The distance to criticality is governed by $(1 - u) \rightarrow 0$. At the level of the explicit solution in terms of theta functions, approaching criticality corresponds to a trigonometric limit with a modulus scaling like
\beq
\label{ququ} q \sim [(1 - u)/q_*]^{c},
\eeq
with an exponent distinguishing between the dense and the dilute phase
$$
c = \left\{\begin{array}{lll} \tfrac{1}{1 - b} & & {\rm dense}, \\ 1 & & {\rm dilute}. \end{array}\right.
$$
It is related to the famous string susceptibility exponent $\gamma_{{\rm str}}$ by $c = -\gamma_{{\rm str}}b$. All other exponents can be expressed in terms of $b$ and $c$, and we will give expressions valid for both universality classes using
$$
\mathfrak{d} = \left\{\begin{array}{lll} 1 & & {\rm dense}, \\ -1 & & {\rm dilute}. \end{array} \right.
$$

The remainder of the text, and the main contribution of this article, is devoted to the analysis of singularities of the generating series under consideration for $(g,h)$ on the non-generic critical line, in the limit $u \rightarrow 1$, here conveniently traded for $q \rightarrow 0$ according to \eqref{ququ}. This is done in several steps in Section~\ref{critit}-\ref{S6} summarized below. We then perform in Section~\ref{FixedV} a saddle point analysis to extract the asymptotics of the desired generating series of maps with fixed volume $V \rightarrow \infty$. The analysis reveals two interesting regimes for boundary perimeters: either we impose the boundary to be
\begin{itemize}
\item[$\bullet$] microscopic (``small''), \textit{i.e.} $\ell_i$ finite,
\item[$\bullet$] or macroscopic (``large''), here corresponding to $\ell_i\,V^{c/2}$ for fixed $\ell_i$.
\end{itemize}
We argue in Section~\ref{crititmarked} that, as far as critical exponents for asymptotics are concerned, marked points behave like small boundaries. So, we can present here our results in a simpler form in absence of marked points.

\begin{notation}
We use $F \stackrel{\bigcdot}{\sim} G$ to indicate that there exists a constant $C > 0$ such that $F \sim CG$ in the asymptotic regime under study.
\end{notation}

Our first main result (Theorem~\ref{LAPA}) concerns generating series of maps with fixed nesting graph.

\begin{center}
\begin{figure}[h!]
\includegraphics[width=\textwidth]{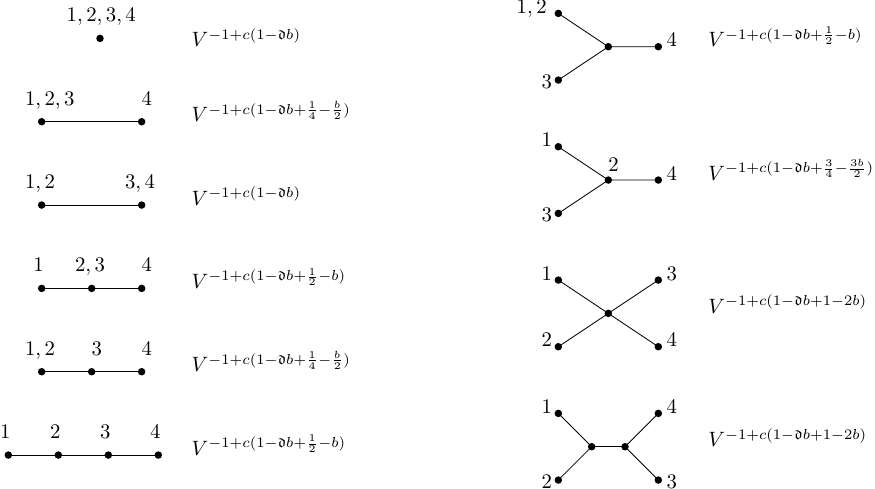}
\caption{\label{04graph} The possible nesting graphs for planar maps with  $4$ microscopic boundaries labeled $1,2,3,4$ (up to permutations of the labels), and the order of magnitude of the number of maps realizing them for large volume $V$. For $n \in (0,2)$, \textit{i.e.} $b \in (0,\tfrac{1}{2})$, the greatest order of magnitude is achieved for the two last graphs in the right column.}
\end{figure}
\end{center}

\begin{theorem}
Assume $2\mathsf{g} - 2 + k > 0$. Let $k_{{\rm L}}$ be the number of macroscopic boundaries, $k_{{\rm S}}$ the number of microscopic boundaries, and $k = k_{{\rm L}} + k_{{\rm S}}$. Let also $k_{{\rm S}}^{(0,2)}$ be the number of microscopic boundaries and marked points that belong to a genus $0$ connected component of the complement of all loops which does not contain any other mark and was adjacent to exactly one loop. The generating series of connected maps of genus $\mathsf{g}$ realizing the nesting graph $(\Gamma,\star)$ behaves as
$$ 
\Big[ u^{V} \prod_{i = 1}^{k_{{\rm L}}} x_i^{-(\ell_iV^{c/2}+1)} \prod_{i = k_{{\rm L}} + 1}^{k_{{\rm L}} + k_{{\rm S}}} x_i^{-(\ell_i + 1)}\Big] \pmb{\mathscr{F}}^{(\mathsf{g},k)}[\Gamma,\star] \mathop{\sim}^{\bigcdot} V^{[-1 + c((2\mathsf{g} - 2 + k)(1 - \mathfrak{d}\frac{b}{2}) - \frac{1}{4}k_{{\rm S}} + (\frac{1}{4} - \frac{b}{2})k_{{\rm S}}^{(0,2)})]}
$$
when $V\rightarrow \infty$.
\end{theorem}
As $b \in (0,\tfrac{1}{2})$, we see that the nesting graphs most likely to occur are those in which each microscopic mark -- either a marked point or a microscopic boundary -- belongs to a genus $0$ univalent vertex which does not carry any other mark. This is exemplified in Figure~\ref{04graph} for maps of genus $0$ with $4$ microscopic marks. The analog statement for cylinders can easily be extracted from \cite{BBD} and is here rederived as Theorem~\ref{LAPB}.

\begin{figure}[h!]
\begin{center}
\includegraphics[width=0.8\textwidth]{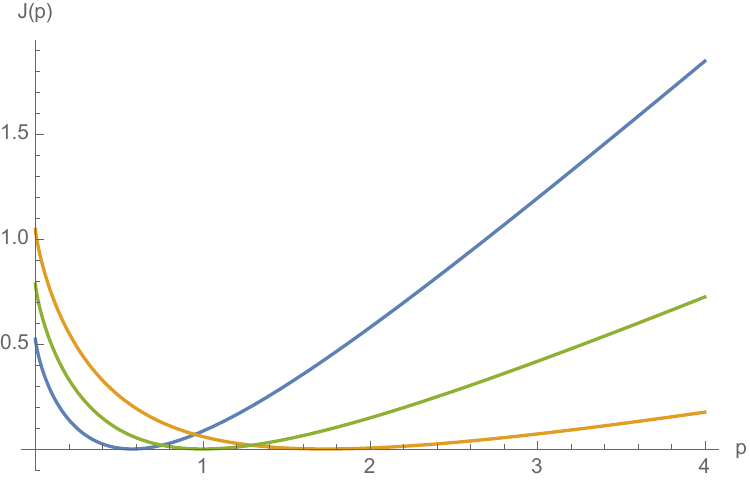}
\caption{\label{Jplotref}The function $J(p)$ for $n = 1$, $n = \sqrt{2}$ (Ising) and $n = \sqrt{3}$ (3-Potts).}
\end{center}
\end{figure}

Our second main result (Theorem~\ref{igfsgb} in Section~\ref{Fixedas}) describes the large deviation function of (large) arm lengths in maps realizing a given nesting graph. It is instructive to first review the result for cylinders obtained in \cite{BBD}, which is expressed in terms of the function
\bea
J(p) & = & \sup_{s \in [0,2/n]} \big\{p\ln(s) + {\arccos}(ns/2) - {\rm arccos}(n/2)\big\} \nonumber \\
\label{Jdef222} & = & p\ln\Big(\frac{2}{n}\,\frac{p}{\sqrt{1 + p^2}}\Big) + {\rm arccot}(p) - {\rm arccos}(n/2).
\eea
plotted in Figure~\ref{Jplotref}. It has the following properties:
\begin{itemize}
\item[$\bullet$] $J(p) \geq 0$ for positive $p$, and achieves its minimum value $0$ at $p_{{\rm opt}} = \frac{n}{\sqrt{4 - n^2}}$ given below.
\item[$\bullet$] $J(p)$ is strictly convex, and $J''(p) = \frac{1}{p(p^2 + 1)}$.
\item[$\bullet$] $J(p)$ has a slope $\ln(2/n)$ when $p \rightarrow \infty$.
\item[$\bullet$] When $p \rightarrow 0$, we have $J(p) = {\rm arcsin}(n/2) + p\ln(2p/n) + O(p)$.
\end{itemize}

\begin{theorem} 
The probability that, in a cylinder with volume $V\rightarrow\infty$, the two boundaries of perimeters $(L_1,L_2)$ are separated by $P$ loops admits the following asymptotics
\bea
\mathbb{P}\Big[P = \frac{c \ln V}{\pi}\,p\,\Big|\,V,L_1 = \ell_1,L_2 = \ell_2\Big] & \stackrel{\bigcdot}{\sim} & (\ln V)^{-\frac{1}{2}}\,V^{-\frac{c}{\pi}\,J(p)}, \nonumber \\
\mathbb{P}\Big[P = \frac{c\ln V}{2\pi}\,p\,\Big|\,V,L_1 = \ell_1,L_2 = \ell_2 V^{\frac{c}{2}}\Big] &\stackrel{\bigcdot}{\sim}  & (\ln V)^{-\frac{1}{2}}\,V^{-\frac{c}{2\pi}\,J(p)}, \nonumber \\
\mathbb{P}\Big[P = \frac{c \ln V}{\pi}\,p\,\Big|\,V,L_1 = \ell_1 V^{\frac{c}{2}},L_2 = \ell_2 V^{\frac{c}{2}}\Big] & \stackrel{\bigcdot}{\sim} & V^{-\frac{c}{\pi}\,p\,\ln(2/n)}. \nonumber  
\eea
\end{theorem} 
We observe that the typical order of magnitude of the number of separating loops between the two boundaries is $\ln V$ when at least one of them is small, and is finite when both of them are large. In the first case, more precisely, $\frac{\pi \jmath P}{c\ln V}$ is almost surely equal to the value $p_{{\rm opt}}$, at which the large deviation reaches its minimum value zero, and the fluctuations of $P$ are Gaussian of order $\sqrt{\ln V}$ due to the quadratic behavior of $J(p)$ near $p = p_{{\rm opt}}$. Here $\jmath$ is a normalization constant depending on the nature (microscopic or macroscopic) of the boundaries.

In this article, for maps of any topology and any nesting graph, we show (Theorem~\ref{igfsgb} in Section~\ref{Fixedasarms}) that individual arms have exactly the same behavior. Arm lengths are asymptotically independent from one another. The analysis of the generating series of maps with a fixed nesting graph will reveal that the arm boundaries that lie in the interior of the map, \textit{i.e.~}~gluing boundaries (instead of boundaries of the original map), are typically  large. Therefore, for maps with $2\mathsf{g} - 2 + k > 0$, we only consider arms incident to a small and a large (gluing) boundary, and to two large (gluing or not) boundaries. For arms incident to a small boundary the length distribution has a large deviation function at rate $\ln V$ universally given by \eqref{Jdef222}. The length of the other arms has an exponential tail with rate $\ln(2/n)$.
\begin{theorem}
\label{Main2}Assume $2\mathsf{g} - 2 + k > 0$, fix a nesting graph $(\Gamma,\star)$, and choose which boundaries are microsopic or macroscopic. Let $E(\Gamma)$ the set of edges of $\Gamma$, $E_{0,2}^{{\rm S}}$, the set of edges incident to a genus $0$ univalent vertex carrying as only mark one microscopic boundary. Consider the regime
\beq
P(\mathsf{e}) = \frac{c\ln V\,p(\mathsf{e})}{\jmath(\mathsf{e})\pi},\qquad \jmath(\mathsf{e}) = \left\{\begin{array}{lll} 2 & & {\rm if}\,\,\mathsf{e} \in E_{0,2}^{{\rm S}}(\Gamma), \\ 1 & & {\rm otherwise}, \end{array}\right.
\eeq
where $p(\mathsf{e})$ may depend on $V$ but remains bounded away from $0$, and negligible in front of $\ln V$, in the sense that $\frac{p(\mathsf{e})}{\ln V}\rightarrow 0$, if $V\rightarrow \infty$. The probability to have arm lengths $\mathbf{P}$ in maps realizing $(\Gamma,\star)$, of volume $V$ with boundary perimeters $L_i = \ell_i$ for the microscopic ones, and $L_i = \ell_i\,V^{\frac{c}{2}}$ for the macroscopic ones, with fixed $\ell_i > 0$, behaves as
$$
\mathbb{P}^{(\mathsf{g},k)}\big[\mathbf{P}|\Gamma,\star,V,\mathbf{L}\big] \stackrel{\bigcdot}{\sim} \prod_{\mathsf{e} \in E_{0,2}^{{\rm S}}(\Gamma)} (\ln V)^{-\frac{1}{2}}\,V^{-\frac{c}{2\pi} J[p(\mathsf{e})]} \cdot \prod_{\mathsf{e} \in E(\Gamma) \setminus E_{0,2}^{{\rm S}}(\Gamma)} V^{-\frac{c}{\pi}\,p(\mathsf{e})\,\ln(\frac{2}{n})}\,.
$$
\end{theorem}

The Gaussian fluctuations of arm lengths at order $\sqrt{\ln V}$ around $\frac{cp_{{\rm opt}}}{\jmath(\mathsf{e})\pi}\,\ln V$ are precisely described in Corollary~\ref{Gaussflu}. If ${\rm CLE}_{\kappa}$ were properly defined on Riemann surfaces of any topology, Theorem~\ref{Main2} could be converted into a prediction of extreme nestings of any topology for ${\rm CLE}_{\kappa}$ thanks to the scheme of KPZ transformations described in \cite[Section 7]{BBD}.

\subsubsection{Steps of the proofs}

The task of Section~\ref{S5} is to derive, for $2\mathsf{g} - 2 + k > 0$, the non-generic critical behavior of:
\begin{itemize}
\item[$\bullet$] the generating series $\bs{\mathcal{F}}^{(\mathsf{g},k)}$ of maps whose faces are disk configurations of the $O(n)$ model, and
\item[$\bullet$] the generating series $\mathbf{F}^{(\mathsf{g},k)}$ of maps in the $O(n)$ model,
\end{itemize}
in presence of an arbitrary fixed number of microscopic and macroscopic boundaries (Theorem~\ref{ouqusf}). Here we work in the canonical ensemble, \textit{i.e.} considering the generating series depending on Boltzmann weights $u$ for vertices and $x_i$ for boundary perimeters. When all boundaries are macroscopic, the result easily follows from the property ``commuting with singular limits'' of the topological recursion, see e.g. \cite[Theorem 5.3.2]{Eynardbook}. The situation is more tricky is presence of microscopic boundaries, and our analysis in this case is new. Our scheme analysis of the topological recursion is in fact more general than the $O(n)$ model, and it may be of use for other problems in enumerative geometry. Concretely, we start from the sum over colored trivalent graphs for $\bs{\mathcal{F}}^{(\mathsf{g},k)}$ and $\mathbf{F}^{(\mathsf{g},k)}$ described in Section~\ref{diagTR}. We analyse the critical behavior of the weights of vertices and of edges in Appendix~\ref{AppBB}, and collect the result in Section~\ref{Sbuildi}. The difference between $\bs{\mathcal{F}}$ and $\mathbf{F}$ only comes from the edge weights, so both cases can be treated in parallel. Then, we determine in Section~\ref{NextS} for fixed genus $\mathsf{g}$, fixed number of boundaries $k$, and \emph{fixed coloring of the $k$ legs}, which graphs give the leading contribution in the critical regime. This is the most technical part, the formula for the critical exponent of this leading contribution in Lemma~\ref{Cbehavior} is quite intrincate, but we should remember that it does not have \textit{a priori} a combinatorial meaning. The quantities which have a meaning are $\mathbf{F}$ and $\bs{\mathcal{F}}$, and they are obtained by summing all these contributions over the colorings of the legs. We find the final result for the critical behavior of $\mathbf{F}$ and $\bs{\mathcal{F}}$ in Theorem~\ref{ouqusf} is much simpler. This result does not concern nesting but is interesting \textit{per se}. It clearly displays the affine dependence of the critical exponents on the Euler characteristic of the maps.

We proceed in Section~\ref{S6} to examine the dominant contribution to the critical behavior of $\pmb{\mathscr{F}}$, the generating series for fixed nesting graph $(\Gamma,\star)$. The starting point is the combinatorial formula of Proposition~\ref{P212}, which is an appropriate glueing along the given nesting graph $\Gamma$ of vertex weight and edge weights. The vertex weights are the $\bs{\mathcal{F}}$'s for which we have already obtained the critical behavior in Theorem~\ref{ouqusf}. The edge weights are the generating series $\mathbf{F}_{s}^{(2)}$ for cylinders remembering the number of separating loops between two boundaries, and some of their variants obtained by attaching a loop around one ($\hat{\mathbf{F}}_{s}^{(2)}$) or both ($\tilde{\mathbf{F}}_{s}^{(2)}$) of their boundaries which are defined in Section~\ref{Srefff}; we determine their critical behavior in Section~\ref{armsS}, thanks to the explicit formula for $\mathbf{F}_{s}^{(2)}$ from Proposition~\ref{p15}. We deduce the critical behavior of $\pmb{\mathscr{F}}$'s by a saddle point analysis in Section~\ref{CoscrFs} and Theorem~\ref{CoscrF}. This is then converted, as explained in \S~\ref{Sintrocrit}, into asymptotics in the microcanonical ensemble, \textit{i.e.} for fixed and large volume, boundary perimeters, and then for arm lengths as well in Section~\ref{FixedV}.

A word of caution relevant in Section~\ref{S6} concerning the canonical ensemble: the dominant contributions depend on the set of variables for which one wishes to study the singularities. If one is only interested later on in fixing the volume and boundary perimeters, one should study singularities with respect to $u$ -- via the variable $q$ -- and $x_i$'s. If one is interested later on in fixing as well the arm lengths, one wants to study singularities with respect to $u$, $x_i$ and the collection of Boltzmann weights $\mathbf{s}$ for the separating loops. It can happen that some dominant terms in the first situation contain no singularity with respect to $\mathbf{s}$, so we need to consider in the second situation terms which were subleading in the first situation, see \textit{e.g.} Theorem~\ref{CoscrF}. The saddle point analysis here is facilitated as similar handlings already appeared for cylinder generating series in \cite{BBD}, and the technical aspects of the present article rather focus on the combinatorics of maps of higher topology.

\section{Properties of the general $O(n)$ loop model}

\label{S2}

\subsection{Definitions}

We start by reminding the definition of the model, following the
presentation of \cite{BBG12a,BBG12b}.

\subsubsection{Loop models}

A \emph{map} is a finite connected graph (possibly with loops and
multiple edges) drawn on a closed orientable compact surface, in such
a way that the edges do not cross and that the connected components of
the complement of the graph (called \emph{faces}) are simply
connected. Maps differing by a homeomorphism of their underlying
surfaces are identified; thus there are countably many maps. The map is \emph{planar} if the underlying surface is topologically a sphere.
The \emph{degree} of a vertex or a face is its number of incident edges (around a face, we count incident edges with multiplicity). To each map we may associate its \emph{dual map} which,
roughly speaking, is obtained by exchanging the roles of vertices and
faces. For $k \geq 1$, a \emph{map with $k$ boundaries} is a map with
$k$ marked faces, \emph{pairwise distinct} and labeled from $1$ to $k$. By convention all the boundary faces are rooted,
that is to say for each boundary face $f$ we pick an oriented edge
(called a \emph{root}) having $f$ on the right.  The \emph{perimeter} of
a boundary is the degree of the corresponding face. Non-boundary faces
are called \emph{inner} faces. We do not impose any simplicity condition on any of the faces of the map. For $k' \geq 1$, a \emph{map with $k'$ marked points} is a map with
marked vertices, labeled from $1$ to $k'$, and by convention we \emph{do not} assume that the $k'$ marked points sit on pairwise distinct vertices. We call \emph{marked element} either a marked point or a marked face.

A \emph{triangulation with $k$
  boundaries} (resp.\ a \emph{quadrangulation with $k$ boundaries}) is
a map with $k$ boundaries such that each inner face has degree $3$
(resp.\ $4$).

\begin{figure}[htpb]
  \centering
  \includegraphics[width=.7\textwidth]{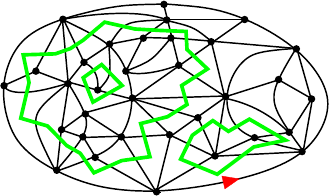}
  \caption{A planar triangulation with a boundary of perimeter $8$ (with
    root in red, the distinguished face being the outer face), endowed
    with a loop configuration (drawn in green).}
  \label{fig:myloopconfig_all}
\end{figure}

Given a map, a \emph{loop} is an undirected simple closed path on the
dual map (\textit{i.e.}\ it covers edges and vertices of the dual map, and
hence visits faces and crosses edges of the original map). This is not
to be confused with the graph-theoretical notion of loop (edge
incident twice to the same vertex), which plays no role here. A
\emph{loop configuration} is a collection of disjoint loops, and may
be viewed alternatively as a collection of \emph{crossed} edges such
that every face of the map is incident to either $0$ or $2$ crossed
edges. When considering maps with boundaries, we assume that the
boundary faces are not visited by loops. Finally, a
\emph{configuration} of the $O(n)$ loop model on random maps is a map
endowed with a loop configuration, see
Figure~\ref{fig:myloopconfig_all} for an example. For sake of clarity, we call \emph{usual map} a map without a loop configuration.

\begin{remark}
  In the original formulation of \cite{GaudinKostov,KOn,KSOn,EKOn},
  the loops cover vertices and edges of the map itself. Our motivation
  for drawing them on the dual map is that it makes our combinatorial
  decompositions easier to visualize.
\end{remark}

\subsubsection{The nesting graphs}
\label{Markm}

Given a configuration $C$ of the $O(n)$ loop model on a connected map $M$ of genus $\mathsf{g}$, we may cut the underlying surface
along every loop, which splits it into several connected components
$c_1,\ldots,c_N$. Let $\Gamma_0$ be the graph on the vertex set
$\{c_1,\ldots,c_N\}$ in which there is an edge between $c_i$ and $c_j$ if and only if they have a common boundary, \textit{i.e.}\ they touch each other
along a loop (thus the edges of $\Gamma_0$ correspond to the loops of $C$)\footnote{Note that multiple edges and edges which are incident twice to the same vertex can occur when building the nesting graphs.}. We assign to each vertex $\mathsf{v}$ the genus $\mathsf{h}(\mathsf{v})$ of the corresponding connected component and for each marked element in $M$ belonging to a connected component $c_i$, we put a mark on the  corresponding vertex of $\Gamma_0$. If the map is planar, $\Gamma_0$ is a tree and all its vertices carry genus $0$. We call $\Gamma_0$ the \emph{primary nesting graph} of $M$.

Let us consider an ensemble of maps with $k''=k+k'\geq 1$ marked elements, where $k$ is the number of boundaries and $k'$ the number of marked points. One can define the \emph{nesting graph} $\Gamma$ from $\Gamma_0$ by repeatedly performing the following first step until it leaves the graph unchanged and then applying the second step:
\begin{itemize}
\item[$(i)$] erasing all vertices that correspond to connected components which, in the complement of all loops in $M$, are homeomorphic to disks, and the edge incident to each of them (except for the case of an isolated vertex carrying a mark); 
\item[$(ii)$] replacing any maximal simple path of the form $v_0 - v_1 - \cdots  - v_{P}$ with $P\geq 2$, where the vertices $(v_i)_{i = 1}^{P-1}$ represent connected components homeomorphic to cylinders, by a single edge
$$
v_0 \mathop{-}^{P} v_{P}
$$
carrying a length $P$. By convention, edges which are not obtained in this way carry a length $P = 1$.
\end{itemize}
The outcome is $(\Gamma,\star,\mathbf{P})$ where $\Gamma$ is the \emph{nesting graph}, which is connected and has vertices labeled by genera such that
$$  
\mathsf{g} = b_1(\Gamma) + \sum_{\mathsf{v} \in V(\Gamma)} \mathsf{h}(\mathsf{v}),
$$
where $b_1(\Gamma)=|E(\Gamma)|-|V(\Gamma)|+1$ is the first Betti number which is equal to the number of cycles of $\Gamma$, where $E(\Gamma)$ and $V(\Gamma)$ denote the set of edges and the set of vertices of $\Gamma$.

The sequence of lengths $\mathbf{P}$ records the number of consecutive ``separating'' loops $P(\mathsf{e})$ for each edge $\mathsf{e}$. We call every $P(\mathsf{e})$ the \emph{depth} of the corresponding edge $\mathsf{e}$ in the nesting graph or of the corresponding arm in the original map. By construction, given the total genus $\mathsf{g}$ and a finite set of marked elements, one can only obtain finitely many inequivalent nesting graphs. $\star$ is the assignment of the marked elements of $M$ to the vertices of $\Gamma$. The valency $d(\mathsf{v})$ of vertices $\mathsf{v}$ of the nesting graph $\Gamma$ with no marked elements must satisfy:
$$
2\mathsf{h}(\mathsf{v}) - 2 + d(\mathsf{v}) > 0
$$
because we erased the vertices that corresponded to unmarked connected components of the map with the topologies of disks and cylinders, \textit{i.e.} with $\mathsf{h}(\mathsf{v})=0$, and $d(\mathsf{v})=1$ and $d(\mathsf{v})=2$, respectively, which are the only two possibilities of $(\mathsf{h}(\mathsf{v}),d(\mathsf{v}))$ that would not satisfy the equality.

\begin{figure}[h!]
\begin{center}
\def\svgwidth{\columnwidth}
 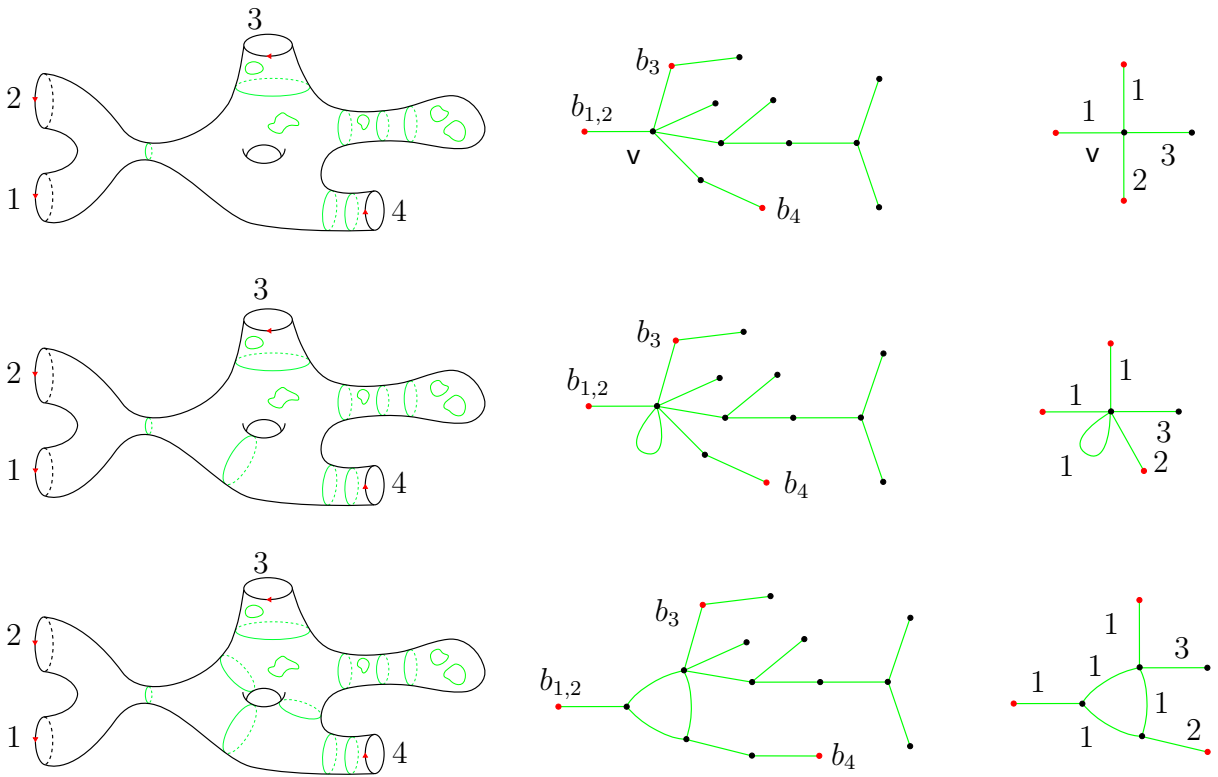
 \caption{  \label{fig:nestinggraph} Left: schematic representations of loop configurations on a
    map of genus $1$ with $4$ boundaries. Center: associated primary nesting graphs, where every
    red vertex carries the marks of the boundaries which belong to the corresponding
    connected component in the map. Right: associated nesting graphs,
    where every edge is labelled with its depth. All vertices carry genus $0$, except
    $\mathsf{v}$ in the first case which has $\mathsf{h}(\mathsf{v})=1$.}
    \end{center}
\end{figure}

This articles studies the distribution of nesting graphs and nesting variables $(\Gamma,\star,\mathbf{P})$ in ensembles of random maps of the $O(n)$ model that we now define.

\subsubsection{Statistical weights}
\label{bendI}

The $O(n)$ loop model is a statistical ensemble
of configurations in which $n$ plays the role of a fugacity per
loop. In addition to this ``nonlocal'' parameter, we need also some
``local'' parameters, controlling in particular the size of the maps
and of the loops. Precise instances of the model can be defined in
various ways.

The simplest instance is the \emph{$O(n)$ loop model on random
  triangulations} \cite{GaudinKostov,KOn,KSOn,EKOn}: here we require
the underlying map to be a triangulation, possibly with boundaries and marked points.
There are two local parameters $g$ and $h$, which are the weights per
face (triangle) distinct from a boundary and which is, respectively, not visited and visited
by a loop. The Boltzmann weight attached to a configuration $C$ with $k\geq 1$ boundaries is
thus $w(C)=n^{\mathcal{L}} g^{T} h^{T'}$, with $\mathcal{L}$ the number of loops of $C$,
$T$ its number of unvisited triangles and $T'$ its number of
visited triangles.

\begin{figure}[htpb]
  \centering
  \includegraphics[width=.5\textwidth]{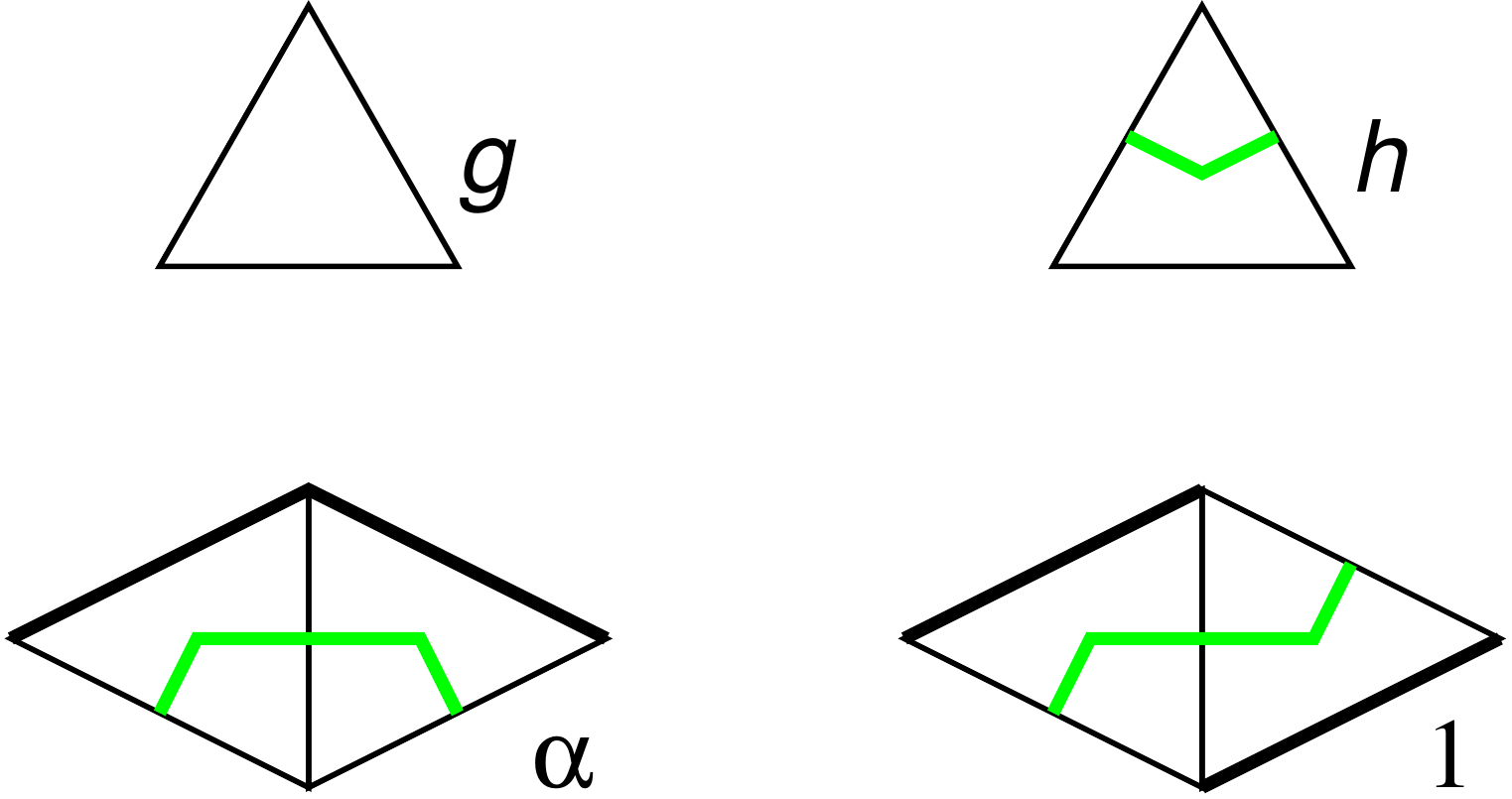}
  \caption{Top row: local weights for the $O(n)$ loop model on random
    triangulations. Bottom row: in the bending energy model, an extra
    weight $\alpha$ is attached to each segment of a loop between two
    successive turns in the same direction.}
  \label{fig:curv}
\end{figure}

A slight generalization of this model is the \emph{bending energy
  model} \cite{BBG12b}, where we incorporate in the Boltzmann weight
$w(C)$ an extra factor $\alpha^B$, where $B$ is the number of pairs of
successive loop turns in the same direction, see
Figure~\ref{fig:curv}. Another variant is the $O(n)$ loop model on
random quadrangulations considered in \cite{BBG12a} (and its ``rigid''
specialization).

In the general $O(n)$ loop model, the Boltzmann weight of a configuration is:
$$
w(C) = \frac{1}{|{\rm Aut}\,C|}\,n^{\mathcal{L}} \prod_{l \geq 3} g_{l}^{N_{l}} \prod_{\substack{\{l_1,l_2\} \\ l_1 + l_2 \geq 1}} g_{l_1,l_2}^{N_{l_1,l_2}},
$$
where $N_l$ is the number of unvisited faces of degree $l$, and $N_{l_1,l_2}$ is the number of visited faces of degree $(l_1 + l_2 + 2)$ whose boundary consists, in cyclic order with an arbitrary orientation, of $l_1$ uncrossed edges, $1$ crossed edge, $l_2$ uncrossed edges and $1$ crossed edge. As the loops are not oriented here, $N_{l_1,l_2} = N_{l_2,l_1}$ and we also assume $g_{l_1,l_2} = g_{l_2,l_1}$. ${\rm Aut}(C)$ is the subgroup of permutations of vertices and edges respecting the root edges and leaving $C$ invariant, which can be observed to be trivial if the number of boundaries $k\geq 1$.

\subsubsection{Generating series}

We now define the basic generating series of interest. Fixing three integers $k,k' \geq 0$ and $\mathsf{g} \geq 0$, we consider the ensemble of allowed configurations of the $O(n)$ model where the underlying map is a connected surface of genus $\mathsf{g}$, with $k$ boundaries of respective lengths $\ell_1,\ell_2,\ldots,\ell_k \geq 1$ (called perimeters) and $k'$ marked points. The corresponding generating series is then the sum of the Boltzmann weights $w(C)$ of all such
configurations. We find convenient to add an auxiliary weight $u$ per
vertex, and define
\begin{equation}
  \label{eq:Fdef}
  F^{(\mathsf{g},k,\bullet k')}_{\ell_1,\ldots,\ell_k} = \delta_{k,1}\delta_{\ell_1,0}\,u + \sum_C u^{|V(C)|} w(C),
\end{equation}
where the sum runs over all desired configurations $C$, and $|V(C)|$ denotes the number of vertices of the underlying map of $C$, also called \emph{volume}, which we will denote just $V$ if there is no possible confusion. We simply write $F^{(\mathsf{g},k)}$ when there are no marked points, and $F^{(\mathsf{g},k,\bullet)}$ when there is one marked point.
For planar maps, \textit{i.e.} $\mathsf{g} = 0$, we just write $F^{(k)}_{\ell_1,\ldots,\ell_k}$. We call \emph{cylinders} the planar maps with $k = 2$ boundaries, and \emph{disks} the planar maps with $k = 1$ boundary. By convention, the map consisting of a single vertex in the sphere is considered as a disk with a boundary of length $\ell_1 = 0$, accounting for the first term in \eqref{eq:Fdef}.

In the course of studying the $O(n)$ loop model, we will also need the generating series of usual maps. The Boltzmann weight of a configuration in this case is chosen to be
\beq
\label{usualmap} w(C) = \frac{1}{|{\rm Aut}\,C|}\,\prod_{l \geq 1} g_l^{N_{l}},
\eeq
and the generating series $\mathcal{F}^{(\mathsf{g},k,\bullet k')}_{\ell_1,\ldots,\ell_k}$ is defined as previously.

\subsection{Planar case: usual maps and the nested loop approach}

\label{SSub} 
In maps with the topology of a disk, there is a notion of inside and outside a loop, from the point of view of the boundary. Then, the nested loop approach \cite[Section 2]{BBG12b} puts in bijection disks $M$ with a loop configuration with triples $(\mathcal{M},\mathcal{R},M')$, where:
\begin{itemize}
\item[$\bullet$] $\mathcal{M}$ is a usual disk, called the \emph{gasket} of $M$. It is obtained as the connected component containing the boundary in the complement of all loops in $M$, filling the interior of each outermost loops by a face.
\item[$\bullet$] $\mathcal{R}$ is a disjoint union of sequences of faces visited by a single loop so as to form an annulus, which is rooted on its outer boundary. It is obtained as the collection of faces crossed by the outermost loops in $M$ -- from the point of view of the boundary -- and the root edge on the outer boundary of each ring (call it $B$) is conventionally defined to be the edge outgoing from the vertex in $B$ which is reached by the shortest leftmost geodesic between the origin of the root edge on the boundary of $M$, and $B$.
\item[$\bullet$] $M'$ is a disjoint union of disks carrying loop configurations. These are the inside of the outermost loops.
\end{itemize}

\begin{figure}[h!]
\begin{center}
  \includegraphics[width=.8\textwidth]{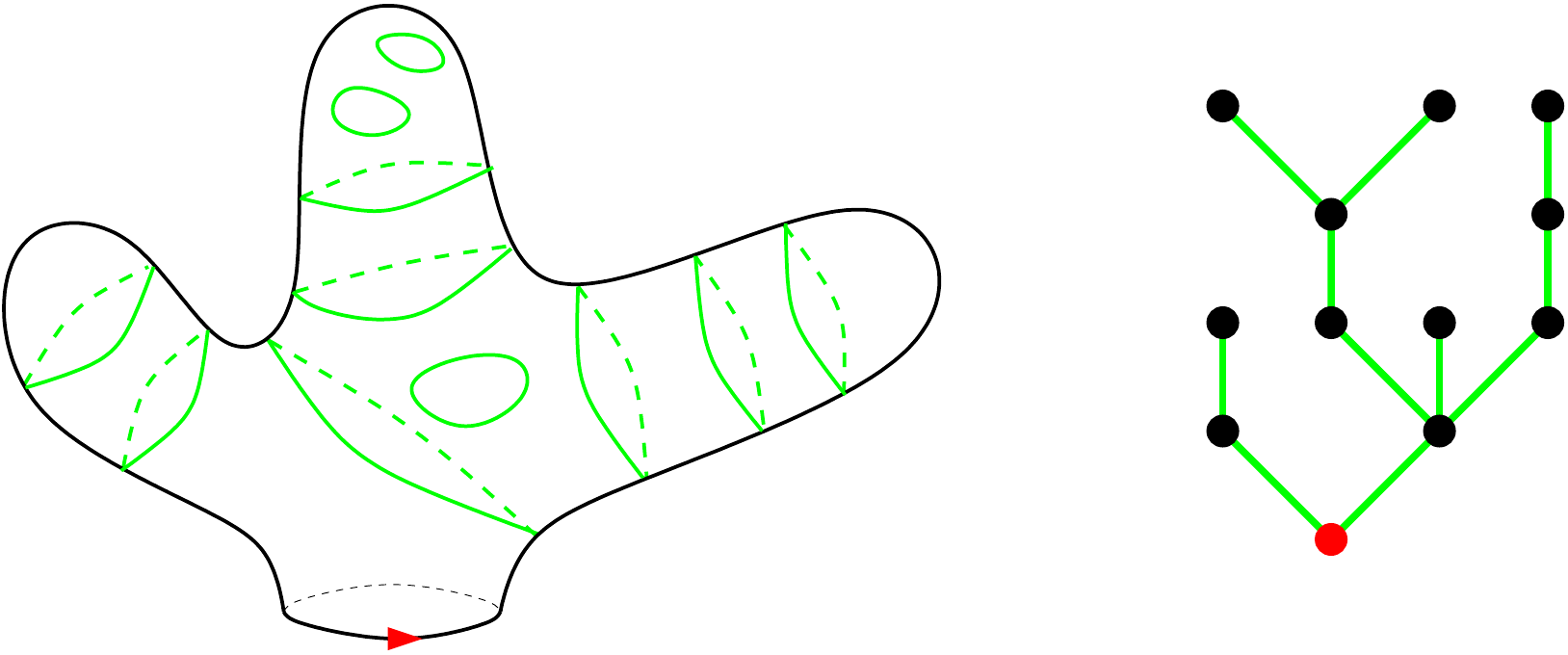}
  \caption{  \label{fig:nestingtree} Left: schematic representation of a loop configuration on a
    planar map with one boundary. Right: the associated primary nesting tree
    (the red vertex corresponds to the gasket).}
  \end{center}
\end{figure}

This translates into a functional relation for the generating series of disks:
\beq
\label{fixedpoint} F_{\ell} = \mathcal{F}_{\ell}(G_1,G_2,\ldots),
\eeq
where the weights $G_l$ of a face of degree $l$ must satisfy the following fixed point condition
\begin{equation}
  \label{eq:fixp}
  G_l = g_l + \sum_{\ell' \geq 0} A_{l,\ell'} \mathcal{F}_{\ell'}(G_1,G_2,\ldots) = g_l + \sum_{\ell' \geq 1} A_{l,\ell'}\,F_{\ell'},
\end{equation}
which was first established in \cite[Page 6]{BBG12b}.
We have denoted by $A_{l,\ell}$ the generating series of sequences of faces visited by a loop, which are glued together so as to form an annulus, in which the outer boundary is rooted and has length $l$, and the inner boundary is unrooted and has length $\ell$. Compared to the notations of \cite{BBG12b}, we decide to include in $A_{l,\ell}$ the weight $n$ for the loop crossing all faces of the annulus. We call $G_l$ the \emph{renormalized face weights}. Note that, although $g_l$ could be zero for $l = 1,2$ and for $l \geq l_0$, with $l_0<\infty$ a certain maximum allowed length, $G_1,G_2$ and $G_l$ for $l \geq l_0$ are a priori non-zero. For this reason, it was necessary to consider the model of usual maps with general face weights \eqref{usualmap}, while we could restrict e.g. to faces (visited or not) of perimeter larger of equal to $3$ in the definition of the general $O(n)$ loop model.

In all what follows, unless explicitly mentioned, the generating series of usual maps $\mathcal{F}^{(\mathsf{g},k)}$ will always be specialized to the renormalized face weights $(G_1,G_2,\ldots)$.

Functional relations for more general planar maps can be deduced from this fixed point equation. The operation of marking a boundary of length $\ell$ is realized by the operator $\ell\,\frac{\partial}{\partial g_{\ell}}$, while marking a vertex amounts to applying $u\,\frac{\partial}{\partial u}$. For instance:
\beq
\begin{array}{rclcrcl} \mathcal{F}^{(2)}_{\ell_1,\ell_2} & = & \ell_2\,\frac{\partial}{\partial g_{\ell_2}} \mathcal{F}_{\ell_1}, & \qquad & F_{\ell_1,\ell_2}^{(2)} & = & \ell_2\,\frac{\partial}{\partial g_{\ell_2}} F_{\ell_1}, \\
\label{rclcc} \mathcal{F}^{\bullet}_{\ell} & = & u\,\frac{\partial}{\partial u}\,\mathcal{F}_{\ell}, & \qquad & F^{\bullet}_{\ell} & = & u\,\frac{\partial}{\partial u}\,F_{\ell} .
\end{array}\nonumber
\eeq
By convention, the equation for $\mathcal{F}^{\bullet}$ assumes that the evaluation to renormalized face weights $G_l$ given by \eqref{eq:fixp} is done \emph{after} the derivative with respect to the vertex weight $u$. In other words, $\mathcal{F}^{\bullet}$ is the generating series of maps pointed in the gasket. Therefore, we deduce from \eqref{fixedpoint}-\eqref{eq:fixp}
\bea
\label{Cylfixed} F^{(2)}_{\ell_1,\ell_2} & = & \mathcal{F}_{\ell_1,\ell_2}^{(2)} + \sum_{l,l' \geq 1} \mathcal{F}_{\ell_1,l}^{(2)}\,R_{l,l'}\,F_{l',\ell_2}^{(2)}, \\
\label{Cyl2fixed} F^{\bullet}_{\ell} & = & \mathcal{F}_{\ell}^{\bullet} + \sum_{l' \geq 1, l'' \geq 0} \mathcal{F}_{\ell,l'}^{(2)}\,R_{l',l''} F_{l''}^{\bullet}, 
\eea
with $R_{l,\ell} = A_{l,\ell}/l$, where the factor $1/l$ amounts to removing the root from the first boundary of the annulus since it is glued to an object with a marked face which already takes care of killing the automorphisms. We will use $R_{l,\ell}$ every time we glue on both sides to objects carrying extra marked elements.

More generally, $\mathcal{F}_{\ell_1,\ldots,\ell_k}^{(\mathsf{g},k,\bullet k')}$ will denote the generating series of usual maps with $k'$ marked points, evaluated at renormalized face weights. According to the nested loop approach, it enumerates maps in the $O(n)$ model where the $k$ boundaries and the $k'$ marked points all belong to the same connected component after removal of all loops. As already remarked for $\mathcal{F}^{\bullet}_{\ell}$, due to the constraints on the relative position of the marked points and the loops,
$$
\mathcal{F}_{\ell_1,\ldots,\ell_k}^{(\mathsf{g},k,\bullet k')} \neq (u\partial_{u})^{k'}\big[\mathcal{F}_{\ell_1,\ldots,\ell_k}^{(\mathsf{g},k)}|_{\{g_l = G_l\}}\big].
$$
The difference comes from the order of differentiation/evaluation at $u$-dependent renormalized vertex weight $\{G_l\}$ from \eqref{eq:fixp}.

\subsection{Separating loops and refined enumeration}
\label{Srefff} In a map $M$ with a non empty set of marked elements $P$, a loop is \emph{separating}\footnote{With this definition, non-contractible loops in $M$ are separating, even though the name could be misleading in such a case.} if it is not contractible in $M \setminus P$. The separating loops (or sequences of separating loops) were encoded in the edges of the nesting graph. If the map is planar, an equivalent definition is saying that a loop is separating if it does not bound a disk in the underlying surface which contains no marked element.

Let us examine the simple case of two marked elements in a planar map. Then, either the two marked elements are not separated by a loop (the nesting graph consists of a single vertex carrying the two marks), or they are separated by $P \geq 1$ loops (the nesting graph consists of an edge of length $P$ between two vertices). To fix ideas, let us say that the first marked element is a boundary. Then, we can put such a map $M$ in bijection either with a cylinder having no separating loop, or a triple consisting of a cylinder with no separating loops, an annulus of faces visited by a single loop, and another map $M'$ with $p - 1$ separating loops. This is the combinatorial meaning of \eqref{Cylfixed}-\eqref{Cyl2fixed}, and it allows an easy refinement. Namely, let $F^{(2)}_{\ell_1,\ell_2}[s]$ (resp. $F^{\bullet}_{\ell}[s]$) be the generating series of cylinders (resp. pointed disks) where the Boltzmann weight includes an extra factor $s^{P}$ and $P$ is the number of separating loops. We obtain from the previous reasoning:
\bea
\label{Cylfixedref} F^{(2)}_{\ell_1,\ell_2}[s] & = & \mathcal{F}_{\ell_1,\ell_2}^{(2)} + s\sum_{l,l' \geq 1} \mathcal{F}_{\ell_1,l}^{(2)}\,R_{l,l'}\,F_{l',\ell_2}^{(2)}[s], \\
\label{Cyl2fixedref} F^{\bullet}_{\ell}[s] & = & \mathcal{F}_{\ell}^{\bullet} + s \sum_{l',l'' \geq 1} \mathcal{F}_{\ell,l'}^{(2)}\,R_{l',l''} F_{l''}^{\bullet}[s].
\eea

In full generality, we are interested in computing $\mathscr{F}^{(\mathsf{g},k,\bullet k')}_{\ell_1,\ldots,\ell_k}[\Gamma,\star;\mathbf{s}]$, the refined generating series of maps of the $O(n)$ model which are connected of genus $\mathsf{g}$, have $k$ boundaries and $k'$ marked points, achieve the nesting graph with its markings $(\Gamma,\star)$, and for which the usual Boltzmann weight contains an extra factor:
$$
\prod_{\mathsf{e} \in E(\Gamma)} s(\mathsf{e})^{P(\mathsf{e})}.
$$

The construction of the nesting graph provides a combinatorial decomposition of maps. Indeed, we can retrieve bijectively the original map from $(\Gamma,\star,\mathbf{P})$, by glueing together:
\begin{itemize}
\item[$\bullet$] for each vertex $\mathsf{v}$ of valency $d(\mathsf{v})$, a usual map (with renormalized weights) of genus $\mathsf{h}(\mathsf{v})$ with $k(\mathsf{v})$ labeled boundaries and $d(\mathsf{v})$ other unlabeled boundaries, and $k'(\mathsf{v})$ marked points;
\item[$\bullet$] for each edge $\mathsf{e}$ of length $1$, an annulus visited by a single loop;
\item[$\bullet$] for each each $\mathsf{e}$ of length $P(\mathsf{e}) \geq 2$, two annuli visited by a single loop capping a cylinder with $P(\mathsf{e}) - 2$ separating loops. 
\end{itemize}

Let us denote $E(\Gamma)$ the set of edges and $V(\Gamma)$ the set of vertices of the nesting graph $\Gamma$. At a given vertex $\mathsf{v}$, $\mathsf{e}(\mathsf{v})$ is the set of outcoming half-edges, and for a given edge $\mathsf{e}$, $\{\mathsf{e}_{+},\mathsf{e}_{-}\}$ is its set of half-edges. $\partial(\mathsf{v})$ the set of boundaries which are registered on marked elements on $\mathsf{v}$ -- if there are no marked elements on $\mathsf{v}$ or just $k(\mathsf{v}) = 0$, then $\partial(\mathsf{v}) = \emptyset$. Let $V_{0,2}(\Gamma)$ be the set of univalent vertices $\mathsf{v}$ of genus $0$ which carry exactly $1$ boundary; the outgoing half-edge (pointing towards the boundary) is then denoted $\mathsf{e}_+(\mathsf{v})$ and $\tilde{V}(\Gamma)=V(\Gamma) \setminus V_{0,2}(\Gamma)$. Let $E_{{\rm un}}(\Gamma)$ be the set of edges which are incident to vertices in $V_{0,2}(\Gamma)$, and $\tilde{E}(\Gamma) = E(\Gamma)\setminus E_{{\rm un}}(\Gamma)$. We define the set of glueing half-edges as follows: 
$$
E_{{\rm glue}}(\Gamma)= \bigcup_{\mathsf{e}\in \tilde{E}(\Gamma)}\{\mathsf{e}_{+},\mathsf{e}_{-}\} \cup \bigcup_{\mathsf{v}\in V_{0,2}(\Gamma)} \mathsf{e}_+(\mathsf{v}).
$$

Let us introduce the generating series of cylinders with one annulus (with unrooted outer boundary) glued to one of the two boundaries
\beq
\label{Fihat} \hat{F}_{\ell_1,\ell_2}^{(2)}[s] = s \sum_{l \geq 0} R_{\ell_1,l} F_{l,\ell_2}^{(2)}[s]
\eeq
and the generating series of cylinders cuffed with two annuli with unrooted outer boundaries
\beq
\label{Fiilde} \tilde{F}_{\ell_1,\ell_2}^{(2)}[s] = s R_{\ell_1,\ell_2} + s^2 \sum_{l,l' \geq 0} R_{\ell_1,l} F_{l,l'}^{(2)}[s]\,R_{l',\ell_2}.
\eeq
By convention, we included in the latter an extra term corresponding to a single annulus with its two boundaries unrooted.

We can determine the desired refined generating series of maps, whose corresponding nesting graph is fixed, using the decomposition of any such map  into the previously introduced pieces.
\begin{proposition}
\label{magi}
\bea
\mathscr{F}^{(\mathsf{g},k,\bullet k')}_{\ell_1,\ldots,\ell_k}[\Gamma,\star,\mathbf{s}] & = &  \sum_{l\,:\,E_{{\rm glue}}(\Gamma) \rightarrow \mathbb{N}}\,\prod_{\mathsf{v} \in \tilde{V}(\Gamma)} \frac{\mathcal{F}^{(\mathsf{h}(\mathsf{v}),k(\mathsf{v}) + d(\mathsf{v}),\bullet k'(\mathsf{v}))}_{\ell(\partial(\mathsf{v})),l(\mathsf{e}(\mathsf{v}))}}{|\mathrm{Aut}(\Gamma)|}
\nonumber \\
& & \qquad  \prod_{\mathsf{e} \in \tilde{E}(\Gamma)} \tilde{F}_{l(\mathsf{e}_-),l(\mathsf{e}_+)}^{(2)}[s(\mathsf{e})] \prod_{\mathsf{v} \in V_{0,2}(\Gamma)} \hat{F}^{(2)}_{l(\mathsf{e}_+(\mathsf{v})),\ell(\partial(\mathsf{v}))}[s(\mathsf{e}_+(\mathsf{v}))],
\eea
where $\ell : \bigcup_{\mathsf{v}\in V(\Gamma)}\partial(\mathsf{v}) \rightarrow \mathbb{N}$ is given by $\ell_1,\ldots, \ell_k$.
\end{proposition}

\section{Analytic properties of generating series}
\label{Sanal} 
So far, all the parameters of the model were formal. We now would like to assign them real values. In this section, we review the properties of generating series of maps obtained by recording all possible boundary perimeters at the same time.

\subsection{Usual maps}
\label{UsSection} 
In the context of usual maps (here not specialized to the renormalized face weigths), we say that $u$ and a sequence $(g_l)_{l \geq 1}$ of nonnegative real numbers are \emph{admissible} if $\mathcal{F}_{\ell}^{\bullet} < \infty$ for any $\ell$. By extension, we say that $u$ and a sequence $(g_l)_{l \geq 1}$ of real numbers are \emph{admissible} if $u$ and $(|g_l|)_{l \geq 1}$ are admissible. For admissible vertex and face weights, we can define
$$
\bs{\mathcal{F}}(x) = \sum_{\ell \geq 0} \frac{\mathcal{F}_{\ell}}{x^{\ell + 1}} \in \mathbb{Q}[[x^{-1}]].
$$

Then, $\bs{\mathcal{F}}(x)$ satisfies the one-cut lemma and a functional relation coming from Tutte's combinatorial decomposition of rooted disks.\begin{theorem} \cite[Section 6]{BBG12b}
\label{ponecut}If $(g_l)_{l \geq 1}$ is admissible, then the formal series $\bs{\mathcal{F}}(x)$ is the Laurent series expansion at $x = \infty$ of a function, still denoted $\bs{\mathcal{F}}(x)$, which is holomorphic for $x \in \mathbb{C}\setminus \gamma$, where $\gamma = [\gamma_-,\gamma_+]$ is a segment of the real line depending on the vertex and face weights. Its endpoints are characterized so that $\gamma_{\pm} = \mathfrak{s} \pm 2\mathfrak{r}$, and $\mathfrak{r}$ and $\mathfrak{s}$ are the evaluation at the chosen weights of the unique formal series in the variables $u$ and $(g_l)_{l \geq 1}$ such that
\bea
\oint_{\gamma} \frac{\dd y}{2{\rm i}\pi}\,\frac{\big(y - \sum_{l \geq 1} g_l\,y^{l - 1}\big)}{\sigma(z)} & = & 0, \\
u + \oint_{\gamma} \frac{\dd y}{2{\rm i}\pi}\,\frac{y\big(y - \sum_{l \geq 1} g_l\,y^{l - 1}\big)}{\sigma(z)} & = & 0,
\eea
where $\sigma(x) = \sqrt{x^2 - 2\mathfrak{s}x + \mathfrak{s}^2 - 4\mathfrak{r}}$. Besides, the endpoints satisfy $|\gamma_-| \leq \gamma_+$, with equality iff $g_{l} = 0$ for all odd $l$'s.
\end{theorem}

\begin{theorem}\cite[Section 6]{BBG12b}
\label{prop2} $\bs{\mathcal{F}}(x)$ is uniformly bounded for $x \in \mathbb{C}\setminus \gamma$. Its boundary values on the cut satisfy the functional relation:
\beq
\label{eq:funcnu}\forall x \in \gamma,\qquad \bs{\mathcal{F}}(x + {\rm i}0) + \bs{\mathcal{F}}(x - {\rm i}0) = x - \sum_{l \geq 1} g_l\,x^{l - 1},
\eeq
and $\bs{\mathcal{F}}(x) = u/x + O(1/x^2)$ when $x \rightarrow \infty$. These properties uniquely determine $\gamma_-,\gamma_+$ and $\bs{\mathcal{F}}(x)$.
\end{theorem}

Although \eqref{eq:funcnu} arise as a consequence of Tutte's equation and analytical continuation, it has itself not received a combinatorial interpretation yet. 

With Theorem~\ref{ponecut} at hand, the analysis of Tutte's equation for generating series of maps with several rooted boundaries and their analytical continuation has been performed (in a more general setting) in \cite{BEO13,Bstuff}. The first outcome is that, if $u$ and $(g_l)_{l \geq 1}$ are admissible, then $\mathcal{F}_{\ell_1,\ldots,\ell_k}^{(\mathsf{g},k,\bullet k')} < \infty$, for all $\mathsf{g}$, $k$ and $k'$, so that we can define
$$
\bs{\mathcal{F}}^{(\mathsf{g},k,\bullet k')}(x_1,\ldots,x_k) = \sum_{\ell_1,\ldots,\ell_k \geq 0} \frac{\mathcal{F}^{(\mathsf{g},k,\bullet k')}_{\ell_1,\ldots,\ell_k}}{x_1^{\ell_1 + 1} \cdots x_{k}^{\ell_k + 1}} \in \mathbb{Q}[[x_1^{-1},\ldots,x_k^{-1}]].
$$
The second outcome is that these are as well Laurent series expansions at $\infty$ of functions, still denoted $\bs{\mathcal{F}}^{(\mathsf{g},k,\bullet k')}(x_1,\ldots,x_k)$, which are holomorphic for $x_i \in \mathbb{C}\setminus\gamma$, with the same $\gamma$ as in Theorem~\ref{ponecut}, and which have upper/lower boundary values when $x_i$ approaches $\gamma$ while $(x_j)_{j \neq i} \in (\mathbb{C}\setminus \gamma)^{k - 1}$ are fixed. More specifically, for cylinders:
\begin{theorem}\cite[Section 3.2]{Eynardbook}
$\sigma(x_1)\sigma(x_2)\bs{\mathcal{F}}^{(2)}(x_1,x_2)$ remains uniformly bounded for $x_1,x_2 \in \mathbb{C}\setminus\gamma$. We have the functional relation, for $x_1 \in (\gamma_-,\gamma_+)$ and $x_2 \in \mathbb{C}\setminus\gamma$:
$$
\bs{\mathcal{F}}^{(2)}(x_1 + {\rm i}0,x_2) + \bs{\mathcal{F}}^{(2)}(x_1 - {\rm i}0,x_2) = -\frac{1}{(x_1 - x_2)^2},
$$
and we have $\bs{\mathcal{F}}^{(2)}(x_1,x_2) \in O(x_1^{-2}x_2^{-2})$ when $x_1,x_2 \rightarrow \infty$. These properties uniquely determine $\bs{\mathcal{F}}^{(2)}(x_1,x_2)$.
\end{theorem}

Once $\gamma_{\pm}$ have been obtained, the formula for the generating series of usual cylinders is well-known (see \textit{e.g.} \cite[Section 3.2]{Eynardbook}):
\beq
\label{F2solep} \bs{\mathcal{F}}^{(2)}(x_1,x_2) = \frac{1}{2(x_1 - x_2)^2}\bigg\{-1 + \frac{x_1x_2 - \frac{\gamma_- + \gamma_+}{2}(x_1 + x_2) + \gamma_-\gamma_+}{\sigma(x_1)\sigma(x_2)}\bigg\}.
\eeq

The generating series for usual pointed disks is also particularly simple (see \textit{e.g.} \cite[Equation (6.6)]{BBG12b}):
\beq
\label{diskpointed} \bs{\mathcal{F}}^{\bullet}(x) = \frac{1}{\sigma(x)}.\eeq

\begin{theorem} \cite[Section 4]{Bstuff}\cite{Eynardbook,BEO13}
Let $2\mathsf{g} - 2 + k > 0$. There exists $r(\mathsf{g},k) > 0$ such that
$$
\sigma(x_1)^{r(\mathsf{g},k)}\bs{\mathcal{F}}^{(\mathsf{g},k)}(x_1,\ldots,x_k)
$$
remains bounded when $x_1$ approaches $\gamma$ while $(x_i)_{i = 2}^k$ are kept fixed away from $\gamma$. \\
$\bs{\mathcal{F}}^{(\mathsf{g},k)}(x_1,\ldots,x_k)$ has upper/lower boundary values for $x_1 \in (\gamma_-,\gamma_+)$ and $x_I = (x_i)_{i = 2}^k$ fixed away from $\gamma$; it satisfies under the same conditions:
$$
\bs{\mathcal{F}}^{(\mathsf{g},k)}(x_1 + {\rm i}0,x_I) + \bs{\mathcal{F}}^{(\mathsf{g},k)}(x_1 - {\rm i}0,x_I) = 0,
$$
and $\bs{\mathcal{F}}^{(\mathsf{g},k)}(x_1,x_I) \in O(x_1^{-2})$ when $x_1 \rightarrow \infty$.
\end{theorem}

\subsection{In the $O(n)$ loop model}
\label{ONsection}

In the context of the $O(n)$ model, we say that $u$ and the two sequences of real numbers $(g_{l})_{l \geq 3}$ and $(A_{l_1,l_2})_{l_1,l_2}$ are admissible if $u$ and the corresponding sequence of renormalized face weights $(G_1,G_2,\ldots)$ computed by \eqref{eq:fixp} are admissible. For admissible weights, we can define:
$$
\mathbf{F}(x) = \sum_{\ell \geq 0} \frac{F_{\ell}}{x^{\ell + 1}} \in \mathbb{Q}[[x^{-1}]].
$$
In the remaining of the article, we always assume admissible weights.

As consequence of \eqref{fixedpoint}, $\mathbf{F}(x)$ satisfies the one-cut property (the analogue of Theorem~\ref{ponecut}), and we still denote $\gamma_{\pm}$ the endpoints of the cuts, which now depend on face weights $(g_l)_{l \geq 3}$ and annuli weights $(A_{l,l'})_{l,l' \geq 0}$. Admissibility also implies that the annuli generating series
\bea
\mathbf{R}(x,y) & = & \sum_{l + l' \geq 1} R_{l,l'}x^{l}y^{l'} \ \  \text{ and} \nonumber \\
\mathbf{A}(x,y) & = & \sum_{l \geq 1} \sum_{l' \geq 0} A_{l,l'}\,x^{l - 1}y^{l'} = \partial_{x}\mathbf{R}(x,y) \nonumber 
\eea
are holomorphic in a neighborhood of $\gamma\times\gamma$. And, $\mathbf{F}(x)$'s boundary values on the cut satisfy the following functional relation:
\begin{theorem}\label{onecutloop}\cite[Section 2]{BBG12b}
\label{funcF} $\mathbf{F}(x)$ is uniformly bounded for $x \in \mathbb{C}\setminus\gamma$ and has upper/lower boundary values on $\gamma$. For $x \in \gamma$, we have:
\beq
\label{eq:funcF}\mathbf{F}(x + {\rm i}0) + \mathbf{F}(x - {\rm i}0) + \oint_{\gamma} \frac{\dd z}{2{\rm i}\pi}\,\mathbf{A}(x,z)\,\mathbf{F}(z) = x - \sum_{k \geq 1} g_k\,x^{k - 1}
\eeq
and $\mathbf{F}(x) = u/x + O(1/x^2)$ when $x \rightarrow \infty$. These properties uniquely determine $\mathbf{F}(x)$ and $\gamma_{\pm}$.
\end{theorem}

Now with Theorem~\ref{onecutloop} at hand, the analysis of Tutte's equation for the partition functions of maps having several boundaries in the loop model, and their analytical continuation, has also been performed in \cite{BEO13,Bstuff}. The outcome is that
$$
\mathbf{F}^{(\mathsf{g},k,\bullet k')}(x_1,\ldots,x_k) = \sum_{\ell_1,\ldots,\ell_k \geq 0} \frac{F^{(\mathsf{g},k,\bullet k')}_{\ell_1,\ldots,\ell_k}}{x_1^{\ell_1 + 1} \cdots x_{k}^{\ell_k + 1}} \in \mathbb{Q}[[x_1^{-1},\ldots,x_k^{-1}]]
$$
are also well-defined and Laurent series expansions at infinity of functions, still denoted $\mathbf{F}^{(\mathsf{g},k,\bullet k')}(x_1,\ldots,x_k)$, which are holomorphic for $x_i \in \mathbb{C}\setminus\gamma$, with the same $\gamma$ independently of $\mathsf{g}$, $k$ and $k'$, and admit upper/lower boundary values for $x_i \in \gamma$ while $(x_j)_{j \neq i} \in (\mathbb{C}\setminus \gamma)^{k - 1}$ are kept fixed. Besides:
\begin{theorem}
\label{prop4loop} \cite[Section 3]{Bstuff}
$\sigma(x_1)\sigma(x_2)\mathbf{F}^{(2)}(x_1,x_2)$ remains uniformly bounded for $x_1,x_2 \in \mathbb{C}\setminus\gamma$. For $x_1 \in (\gamma_-,\gamma_+)$ and $x_2 \in \mathbb{C}\setminus\gamma$, we have the following functional relation:
$$
\mathbf{F}^{(2)}(x_1 + {\rm i}0,x_2) + \mathbf{F}^{(2)}(x_1 - {\rm i}0,x_2) + \oint_{\gamma} \frac{\dd y}{2{\rm i}\pi}\,\mathbf{A}(x_1,y) \mathbf{F}^{(2)}(y,x_2) = -\frac{1}{(x_1 - x_2)^2},
$$
and $\mathbf{F}^{(2)}(x_1,x_2) \in O(x_1^{-2}x_2^{-2})$ when $x_1,x_2 \rightarrow \infty$. These properties uniquely determine $\mathbf{F}^{(2)}(x_1,x_2)$.
\end{theorem}
\begin{theorem}\label{2g2m}\cite[Section 4]{Bstuff} \cite{Eynardbook,BEO13}
Let $2\mathsf{g} - 2 + k > 0$. There exists $r(\mathsf{g},k) > 0$ such that
$$
\sigma(x_1)^{r(\mathsf{g},k)}\mathbf{F}^{(\mathsf{g},k)}(x_1,\ldots,x_k)
$$
remains bounded when $x_1$ approaches $\gamma$ while $(x_i)_{i = 2}^k$ are kept fixed away from $\gamma$. \\
$\mathbf{F}^{(\mathsf{g},k)}(x_1,\ldots,x_k)$ has upper/lower boundary values for $x_1 \in (\gamma_-,\gamma_+)$ and $x_I = (x_i)_{i = 2}^k$ fixed away from $\gamma$, and it satisfies under the same conditions:
$$
\mathbf{F}^{(\mathsf{g},k)}(x_1 + {\rm i}0,x_I) + \mathbf{F}^{(\mathsf{g},k)}(x_1 - {\rm i}0,x_I) + \oint_{\gamma} \frac{\dd y}{2{\rm i}\pi}\,\mathbf{A}(x,y)\,\mathbf{F}^{(\mathsf{g},k)}(y,x_I) = 0
$$
and $\mathbf{F}^{(\mathsf{g},k)}(x_1,x_I) \in O(x_1^{-2})$ when $x_1 \rightarrow \infty$.
\end{theorem}

\subsection{Refined generating series}

We now recall the results of \cite{BBD} for the refined generating series of pointed disks and cylinders. First of all, for admissible weights and $s \in \mathbb{R}$ at least in a neighborhood of $[-1,1]$,
\bea
\mathbf{F}_{s}^{\bullet}(x_1) & = & \sum_{\ell \geq 0} \frac{F_{\ell}^{\bullet}[s]}{x_1^{\ell_1 + 1}} \in \mathbb{Q}[[x_1^{-1}]], \nonumber \\
\mathbf{F}_{s}^{(2)}(x_1,x_2) & = & \sum_{\ell_1,\ell_2 \geq 1} \frac{F_{\ell_1,\ell_2}^{(2)}[s]}{x_1^{\ell_1 + 1}x_2^{\ell_2 + 1}} \in \mathbb{Q}[[x_1^{-1},x_2^{-1}]]
\eea
are well-defined, and are Laurent series expansions at infinity of functions, still denoted $\mathbf{F}_{s}^{(2)}(x_1,x_2)$ and $\mathbf{F}_{s}^{\bullet}(x_1)$, which are holomorphic of $x_i \in \mathbb{C}\setminus \gamma$, for the same $\gamma$ appearing in Section~\ref{ONsection}, independently of $s$. Besides, we have linear functional relations very similar to those satisfied by the unrefined generating series:
 
\begin{theorem}\cite[Section 4]{BBD}
\label{prop76} $\sigma(x_1)\sigma(x_2)\mathbf{F}_{s}^{(2)}(x_1,x_2)$ is uniformly bounded for $x_1,x_2 \in \mathbb{C}\setminus\gamma$. For any $x_1 \in (\gamma_-,\gamma_+)$ and $x_2 \in \mathbb{C}\setminus\gamma$ fixed, we have:
$$
\mathbf{F}_{s}^{(2)}(x_1 + {\rm i}0,x_2) + \mathbf{F}^{(2)}_{s}(x_1 - {\rm i}0,x_2) +  s \oint_{\gamma} \frac{\dd y}{2{\rm i}\pi}\,\mathbf{A}(x_1,y)\,\mathbf{F}^{(2)}_{s}(y,x_2) = - \frac{1}{(x_1 - x_2)^2}
$$
and $\mathbf{F}_{s}^{(2)}(x_1,x_2) \in O(x_1^{-2}x_2^{-2})$ when $x_1,x_2 \rightarrow \infty$. These properties uniquely determine $\mathbf{F}_{s}^{(2)}(x_1,x_2)$.
\end{theorem}

\begin{theorem}\cite[Section 4]{BBD}
\label{prop77} $\sigma(x)\mathbf{F}_{s}^{\bullet}(x)$ is uniformly bounded when $x \in \mathbb{C}\setminus\gamma$. For $x \in (\gamma_-,\gamma_+)$, we have:
$$
\mathbf{F}^\bullet_{s}(x + {\rm i}0) + \mathbf{F}^\bullet_{s}(x - {\rm i}0) + s \oint_{\gamma} \frac{\dd y}{2{\rm i}\pi}\,\mathbf{A}(x,y)\,\mathbf{F}^\bullet_{s}(y) = 0
$$
and $\mathbf{F}_{s}^{\bullet}(x) = u/x + O(1/x^2)$ when $x \rightarrow \infty$. These properties uniquely determine $\mathbf{F}_{s}^{\bullet}(x)$.
\end{theorem}
From the analytic properties of $\mathbf{F}_{s}^{(2)}$ and $\mathbf{R}$, it follows that
\beq
\hat{\mathbf{F}}_{s}^{(2)}(x_1,x_2) = \sum_{\ell_1,\ell_2 \geq 0} \hat{F}_{\ell_1,\ell_2}^{(2)}[s]\,\frac{x_1^{\ell_1}}{x_2^{\ell_2 + 1}} = s \oint_{\gamma} \frac{\dd y}{2{\rm i}\pi}\,\mathbf{R}(x_1,y) \mathbf{F}^{(2)}_{s}(y,x_2) \nonumber
\eeq
is the series expansion when $x_1 \rightarrow 0$ and $x_2 \rightarrow \infty$ of a function denoted likewise, which is holomorphic for $x_1$ in a neighborhood of $\gamma$ and $x_2$ in $\mathbb{C}\setminus\gamma$. And,
\bea
\label{Ftilded}\tilde{\mathbf{F}}_{s}^{(2)}(x_1,x_2) & = & \sum_{\ell_1,\ell_2 \geq 0} \tilde{F}_{\ell_1,\ell_2}^{(2)}[s]\,x_1^{\ell_1}x_2^{\ell_2} \\
& = &  s\,\mathbf{R}(x_1,x_2) + s^2 \oint_{\gamma} \frac{\dd y_1}{2{\rm i}\pi}\,\frac{\dd y_2}{2{\rm i}\pi}\,\mathbf{R}(x_1,y_1) \mathbf{F}^{(2)}_{s}(y_1,y_2) \mathbf{R}(y_2,x_2) \nonumber
\eea
is the series expansion at $x_i \rightarrow 0$ of a function denoted likewise, which is holomorphic for $x_i$ in a neighborhood of $\gamma$. This fact and the analytic properties of $\bs{\mathcal{F}}^{(\mathsf{g},k,\bullet k')}$ for any $\mathsf{g},k,k'$ described in Section~\ref{ONsection} imply, together with the formula of Proposition~\ref{magi}:

\begin{proposition}
\label{P212} If $u$, $(g_{l})_{l \geq 3}$ and $(A_{l,\ell})_{l,\ell}$ are admissible, then at least for $s(\mathsf{e}) \in \mathbb{R}$ in a neighborhood of $[-1,1]$ for each $\mathsf{e} \in E(\Gamma)$, the generating series for fixed nesting graph
$$
\pmb{\mathscr{F}}^{(\mathsf{g},k,\bullet k')}_{\Gamma,\star,\mathbf{s}}(x_1,\ldots,x_k) = \sum_{\ell_1,\ldots,\ell_k \geq 0} \frac{\mathscr{F}^{(\mathsf{g},k,\bullet k')}_{\ell_1,\ldots,\ell_k}[\Gamma,\star,s]}{x_1^{\ell_1 + 1}\cdots x_k^{\ell_k + 1}}
$$
are well-defined, and are the Laurent expansions at $\infty$ of functions, denoted with same symbol, which are holomorphic in $(x_1,\ldots,x_k) \in (\mathbb{C}\setminus\gamma)^k$ for the same segment $\gamma$ appearing in Section~\ref{ONsection}. If $I$ is a finite set, $(x_i)_{i \in I}$ a collection of variables and $J$ a subset of $I$, we denote $x_{J} = (x_j)_{j \in J}$. The formula of Proposition~\ref{magi} can be translated into
\bea
\pmb{\mathscr{F}}^{(\mathsf{g},k)}_{\Gamma,\star,\mathbf{s}}(x_1,\ldots,x_k) & = &  \oint_{\gamma^{E_{{\rm glue}}(\Gamma)}} \prod_{\mathsf{e} \in E_{{\rm glue}}(\Gamma)}  \frac{\dd y_{\mathsf{e}}}{2{\rm i}\pi} \prod_{\mathsf{v} \in \tilde{V}(\Gamma)} \frac{\bs{\mathcal{F}}^{(\mathsf{h}(\mathsf{v}),k(\mathsf{v}) + d(\mathsf{v}),\bullet k'(\mathsf{v}))}(x_{\partial(\mathsf{v})},y_{\mathsf{e}(\mathsf{v})})}{d(\mathsf{v})!} \nonumber\\
&& \qquad \prod_{\mathsf{e} \in \tilde{E}(\Gamma)} \tilde{\mathbf{F}}^{(2)}_{s(\mathsf{e})}(y_{\mathsf{e}_+},y_{\mathsf{e}_-}) \prod_{\mathsf{v} \in V_{0,2}(\Gamma)}\hat{\mathbf{F}}^{(2)}_{s(\mathsf{e}_+(\mathsf{v}))}(y_{\mathsf{e}_+(\mathsf{v})},x_{\partial(\mathsf{v})}). 
\eea
\end{proposition}

\subsection{Topological recursion}

\begin{theorem}\cite{Eynardbook,BEO13}
The generating series $\mathbf{F}^{(\mathsf{g},k)}$ for arbitrary topologies can be obtained from the generating series of disks $\mathbf{F}^{(0,1)} = \mathbf{F}$ and of cylinders $\mathbf{F}^{(0,2)} = \mathbf{F}^{(2)}$ by the topological recursion of \cite{EOFg}. This is a universal recursion on $2\mathsf{g} - 2 + k > 0$. By specialization, the generating series of usual maps at renormalized face weights $\bs{\mathcal{F}}^{(\mathsf{g},k)}$ is also given by the topological recursion: the initial data of the recursion is then $\bs{\mathcal{F}}(x) = \mathbf{F}(x)$ and $\bs{\mathcal{F}}^{(2)}$ given by \eqref{F2solep}.
\end{theorem}

The general statement of the topological recursion formula demands more notions of complex analysis than desirable here, so we refer in general to \cite{EOFg,BEO13}. We shall describe its somewhat simpler application to the bending energy model in the next Section.

For the general $O(n)$ model, we cannot go much further at present. Let us summarize the logic of computation of $\pmb{\mathscr{F}}_{\Gamma,\star,\mathbf{s}}^{(\mathsf{g},k)}$, which is the main quantity of interest in this article. Firstly, one tries to solve for $\mathbf{F}(x)$ the linear equation of Theorem~\ref{funcF}, as a function of $\gamma_{\pm}$, only exploiting that $\sigma(x)\mathbf{F}(x)$ remains uniformly bounded for $x \rightarrow \gamma_{\pm}$ -- for the moment, we do not use the stronger fact that $\mathbf{F}(x)$ is bounded. This problem is known a priori to have a unique solution for any choice of $\gamma_{\pm}$, but is hardly amenable to an explicit solution. Secondly, imposing that $\mathbf{F}(x)$ is actually uniformly bounded for $x \in \mathbb{C}\setminus \gamma$ gives two non-linear equations which determine $\gamma_{\pm}$. These equations may not have a unique solution, but we look for the unique solution such that $\gamma_{\pm}$ are evaluations at the desired weights of formal power series of $\sqrt{u}$, $g_l$, $n$ and $A_{k,l}$. Thirdly, now knowing $\gamma_{\pm}$ -- or assuming to know them -- one tries to solve for $\mathbf{F}^{(2)}_{s}(x_1,x_2)$ the linear equation of Theorem~\ref{prop76}, in a uniform way for any $s$. This problem is as difficult as the first step\footnote{As a matter of fact, there exists a general and explicit linear formula to extract $\mathbf{F}(x)$ (resp. $\mathbf{F}^{\bullet}_{s}$) from the knowledge of $\mathbf{F}^{(2)}_{s = 1}(x_1,x_2)$ (resp. $\mathbf{F}^{(2)}_{s}(x_1,x_2)$), which we will not need here.}. In a fourth step, if $\gamma_{\pm}$, $\mathbf{F}(x)$, and $\mathbf{F}^{(2)}(x_1,x_2)$ are known or assumed so, the topological recursion allows the explicit computation of $\mathbf{F}^{(\mathsf{g},k)}(x_1,\ldots,x_k)$ by induction on $2\mathsf{g} - 2 + k$. We now have all the ingredients to compute in a fifth step the generating series $\pmb{\mathscr{F}}_{\Gamma,\star,\mathbf{s}}^{(\mathsf{g},k)}$ in absence of marked points. 

\subsection{Adding marked points}
\label{addinm}
The computation of generating series of maps with marked points is done a posteriori. For the generating series of maps with loops where the position of the marked points is not constrained, we simply have
$$
\mathbf{F}^{(\mathsf{g},k,\bullet k')} = (u\partial_u)^{k'} \mathbf{F}^{(\mathsf{g},k)}.
$$
To force marked points and boundaries to be all together, not separated by loops, \textit{i.e.} to compute $\bs{\mathcal{F}}^{(\mathsf{g},k,\bullet k')}$, we proceed differently.

Consider a usual map of genus $\mathsf{g}$ with $k$ boundaries of perimeters $(\ell_i)_{i = 1}^k$. Denote $V$ the number of vertices, $E$ the number of edges, and $(N_m)_{m \geq 1}$ the number of (non-marked) faces of degree $m$. We have the Euler relation
$$
2 - 2\mathsf{g} - k = V - E + \sum_{m \geq 1} N_m,
$$
and counting half-edges gives
$$
2E = \sum_{m \geq 1} mN_{m} + \sum_{i = 1}^k \ell_i.
$$
Then, the number of vertices is
$$
V = 2 - 2\mathsf{g} - k + \sum_{m \geq 1} (\tfrac{m}{2} - 1)N_m + \sum_{i = 1}^k \tfrac{1}{2} \ell_i.
$$

Therefore, the operation of marking a point is realized at the level of generating series by application of the operator
$$
2 - 2\mathsf{g} - k + \sum_{m \geq 1} (\tfrac{1}{2} - \tfrac{1}{m}) mg_m \partial_{g_m} - \sum_{i = 1}^k \tfrac{1}{2}\partial_{x_i} x_i.
$$
In particular, if we denote $\mathbf{V}(x) = \tfrac{1}{2}x^2 - \sum_{m \geq 1}\tfrac{g_m}{m}x^m$, the generating series of usual maps with marked points and (non-renormalized) face weights $\{g_m\}_{m \geq 1}$ satisfies, for all $k' \geq 1$
\bea
\bs{\mathcal{F}}^{(\mathsf{g},k,\bullet k')}_{{\rm bare}}(x_1,\ldots,x_k) & = &  \Big(2 - 2\mathsf{g} - k - \sum_{i = 1}^k \tfrac{1}{2}\,\partial_{x_i} x_i\Big)\bs{\mathcal{F}}_{{\rm bare}}^{(\mathsf{g},k,\bullet (k' - 1))}(x_1,\ldots,x_k) \nonumber \\
&& - \oint_{\gamma} \frac{\dd y}{2{\rm i}\pi}\,\big(\tfrac{y}{2}\mathbf{V}'(y) - \mathbf{V}(y)\big)\bs{\mathcal{F}}^{(\mathsf{g},k + 1,\bullet (k' - 1))}_{{\rm bare}}(y,x_1,\ldots,x_k). \nonumber
\eea
For renormalized face weights $\{G_m\}_{m\geq 1}$, we have to take into account the shift from $g_m$ to $G_m$ \eqref{eq:fixp}, resulting in
\bea
\bs{\mathcal{F}}^{(\mathsf{g},k,\bullet k')}(x_1,\ldots,x_k) & = & \Big(2 - 2\mathsf{g} - k - \sum_{i = 1}^k \tfrac{1}{2}\,\partial_{x_i} x_i\Big)\bs{\mathcal{F}}^{(\mathsf{g},k,\bullet (k' - 1))}(x_1,\ldots,x_k) \nonumber \\
\label{renFmarked} &&- \oint_{\gamma} \frac{\dd y}{2{\rm i}\pi}\,\big(\tfrac{y}{2}\tilde{\mathbf{V}}'(y) - \tilde{\mathbf{V}}(y)\big)\bs{\mathcal{F}}^{(\mathsf{g},k + 1,\bullet (k' - 1))}(y,x_1,\ldots,x_k), 
\eea
where
$$
\tilde{\mathbf{V}}(x) = \mathbf{V}(x) - \oint_{\gamma} \frac{\dd z}{2{\rm i}\pi}\,\mathbf{R}(x,z)\,\mathbf{F}(z).
$$

\section{The bending energy model}
\label{S4}
\subsection{Definition}

We shall focus on the class of loop models with bending energy on triangulations studied in \cite{BBG12b}, for which the computations can be explicitly carried out. On top of the loop fugacity $n$ and the vertex weight $u$, it features a weight $g$ per unvisited triangle, $h$ per visited triangle, and $\alpha$ per consecutive pair of visited triangles pointing in the same direction. The annuli generating series in this model are:
\bea
\label{rin}\mathbf{R}(x,z) & = &  n\ln\left(\frac{1}{1 - \alpha h(x + z) - (1 - \alpha^2)h^2xz}\right) \\
&  = &  n\ln\left(\frac{1}{z - \varsigma(x)}\right) + \frac{n}{2}\ln\left(\frac{\varsigma'(x)}{-h^2}\right), \nonumber \\
\mathbf{A}(x,z) & = & \partial_{x}\mathbf{R}(x,z) \nonumber \\
& = & n\left(\frac{\varsigma'(x)}{z - \varsigma(x)} + \frac{\varsigma''(x)}{2\varsigma'(x)}\right), \nonumber
\eea
where
\beq
\label{varsigma}\varsigma(x) = \frac{1 - \alpha hx}{\alpha h + (1 - \alpha^2)h^2x}
\eeq
is a rational involution. We assume that the weights are admissible, and thus all relevant generating series of maps with boundaries have a cut $[\gamma_-,\gamma_+]$.

Technically, the fact that $\mathbf{A}(x,y)$ is a rational function with a single pole allows for an explicit solution of the linear equation for $\mathbf{F}(x)$ and $\mathbf{F}_{s}^{(2)}(x_1,x_2)$, assuming $\gamma_{\pm}$ are known (see Section~\ref{slili}). Then, $\gamma_{\pm}$ are determined implicitly by two complicated equations -- cf. \eqref{determinab} below. This is nevertheless explicit enough to analyze the critical behavior of the model (see Section~\ref{critit}).

\subsection{Solving the linear equation}
\label{elparam}

\subsubsection{Preliminaries} If $f$ is a holomorphic function in $\mathbb{C}\setminus\gamma$ such that $f(x) \sim c_{f}/x$ when $x \rightarrow \infty$, we can evaluate the contour integral:
\beq
\label{contueq}\oint_{\gamma} \frac{\dd y}{2{\rm i}\pi}\,\mathbf{A}(x,y)\,f(y) = -n \varsigma'(x)\,f(\varsigma(x)) + n c_{f}\,\frac{\varsigma''(x)}{2\varsigma'(x)},
\eeq
where we notice that
$$
\frac{\varsigma''(x)}{2\varsigma'(x)} = -\frac{1}{x - \varsigma(\infty)}.$$
Therefore, a linear equation of the form
$$
f(x + {\rm i}0) + f(x - {\rm i}0) + s\oint_{\gamma} \frac{\dd y}{2{\rm i}\pi}\,\mathbf{A}(x,y)\,f(y) = \phi(x),\qquad \forall x \in (\gamma_-,\gamma_+),
$$
becomes
\beq
\label{fhomo} f(x + {\rm i}0) + f(x - {\rm i}0) - ns\,\varsigma'(x)f(\varsigma(x)) = \tilde{\phi}(x) \coloneqq \phi(x)  -ns c_{f}\,\frac{\varsigma''(x)}{2\varsigma'(x)},\ \  \forall x \in (\gamma_-,\gamma_+).
\eeq
When $ns \neq \pm 2$, which is assumed here, 
\beq
\label{fhom} f^{{\rm hom}}(x) = f(x) - \frac{2\tilde{\phi}(x) + ns \varsigma'(x)\tilde{\phi}(\varsigma(x))}{4 - n^2s^2},
\eeq
with $f$ a solution of (\ref{fhomo}), satisfies the following homogeneous linear equation:
\beq
\label{fhomoh} \forall x \in (\gamma_-,\gamma_+),\qquad f^{{\rm hom}}(x + {\rm i}0) + f^{{\rm hom}}(x - {\rm i}0) - ns\varsigma'(x)f^{{\rm hom}}(\varsigma(x)) = 0.
\eeq
If we assume that $\phi(x)$ is a given rational function with poles $q$ away from $\gamma$, $f^{{\rm hom}}(x)$ acquires poles at the same points, and we have:
$$
f^{{\rm hom}}(x) = \delta_{q,\infty}\,\frac{c_{f}}{x} - \frac{2\tilde{\phi}(x) + ns \varsigma'(x)\tilde{\phi}(\varsigma(x))}{4 - n^2s^2} + O(1),\qquad x \rightarrow q.
$$
So, we are left with the problem of solving \eqref{fhomoh} with vanishing right-hand side, but admitting rational singularity with prescribed divergent part at a finite set of points $q \in \mathbb{C}\setminus \gamma$.

\begin{figure}
\begin{center}
\includegraphics[width=0.7\textwidth]{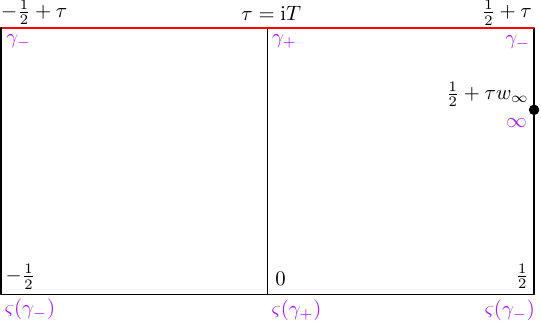}
\end{center}
\caption{\label{ParamF} The fundamental rectangle in the $v$-plane. We indicate the image of special values of $x$ in purple, and the image of the cut $\gamma$ in red. The left (resp. right) panel is the image of ${\rm Im}\,x > 0$ (resp. ${\rm Im}\,x < 0$).}
\end{figure}

The key to the solution is the use of an elliptic parametrization $x = x(v)$. It depends on a parameter $\tau = {\rm i}T$ which is completely determined by the data of $\gamma_{\pm}$ and $\varsigma(\gamma_{\pm})$. The domain $\mathbb{C}\setminus\big(\gamma\cup\varsigma(\gamma)\big)$ is mapped to the fundamental rectangle (Figure~\ref{ParamF})
\beq
\big\{v \in \mathbb{C},\qquad 0 < \mathrm{Re}\,v < 1/2,\quad |\mathrm{Im}\,v| < T\big\},
\eeq
with values at the corners:
\beq
\begin{array}{lcl}
x(\tau) = x(-\tau) = \gamma_{+}, & \qquad &  x(\tau + 1/2) = x(-\tau + 1/2) = \gamma_{-}, \\ 
x(0) = \varsigma(\gamma_{+}), & \qquad & x(1/2) = \varsigma(\gamma_{-}). \end{array} 
\eeq
Besides, when $x$ is in the physical sheet, $$v(\varsigma(x)) = \tau - v(x).$$ Since the involution $\varsigma$ is decreasing, $\varsigma(\gamma_-)$ belongs to the union $(\varsigma(\gamma_+),+\infty) \sqcup (-\infty,\gamma_-)$, and therefore $x = \infty$ is mapped to $v_{\infty} = \frac{1}{2} + \tau w_{\infty}$ with $0 < w_{\infty} < 1/2$. When $\alpha = 1$, by symmetry we must have $w_{\infty} = 1/2$.

The function $v \mapsto x(v)$ is analytically continued for $v \in \mathbb{C}$ by the relations:
\beq
x(-v) = x(v + 1) = x(v + 2\tau) = x(v).
\eeq
This parametrization allows the conversion \cite{EKOn,BBG12b} of the functional equation
\beq
\forall x \in \mathring{\gamma},\qquad f(x + {\rm i}0) + f(x - {\rm i}0) - n\,\varsigma'(x)\,f(\varsigma(x)) = 0
\eeq
for an analytic function $f(x)$ in $\mathbb{C}\setminus\gamma$, into the functional equation:
\beq
\label{reuh}\forall v \in \mathbb{C},\qquad \tilde{f}(v + 2\tau) + \tilde{f}(v) - n\,\tilde{f}(v - \tau) = 0, \qquad \text{with } \tilde{f}(v) = \tilde{f}(v + 1) = -\tilde{f}(-v),
\eeq
for the analytic continuation of the function $\tilde{f}(v) = f(x(v))x'(v)$. The second condition in \eqref{reuh} enforces the continuity of $f(x)$ on $\mathbb{R}\setminus\gamma$. We set:
\beq
b = \frac{\mathrm{arccos}(n/2)}{\pi}.
\eeq
The new parameter $b$ ranges from $\tfrac{1}{2}$ to $0$ when $n$ ranges from $0$ to $2$. Solutions of the first equation of \eqref{reuh} with prescribed meromorphic singularities can be build from a fundamental solution $\Upsilon_b$, defined uniquely by the properties:
\beq
\label{propoup}\Upsilon_{b}(v + 1) = \Upsilon_{b}(v),\qquad \Upsilon_{b}(v + \tau) = e^{{\rm i}\pi b}\Upsilon_{b}(v),\qquad \Upsilon_{b}(v) \mathop{\sim}_{v \rightarrow 0} \frac{1}{v}.
\eeq
Its expression and main properties are reminded in Appendix~\ref{AppUp}.

\subsubsection{Elementary generating series}\label{slili} We present the solution for the generating series of disks, and of refined disks and cylinders.
\label{SectionG}
Let $\mathbf{G}(v)$ be the analytic continuation of 
\beq
\label{EquationG} x'(v)\mathbf{F}(x(v)) - \partial_{v}\Bigg(\frac{2\mathbf{V}(x(v)) + n\mathbf{V}(\varsigma(x(v)))}{4 - n^2} - \frac{nu \ln \big[\varsigma'(x(v))\big]}{2(2 + n)}\Bigg),
\eeq
where $\mathbf{V}(x) = \tfrac{1}{2}x^2 - \sum_{k \geq 1} \tfrac{g_l}{l}x^{l}$ collects the weights of empty faces. In the model we study, empty faces are triangles counted with weight $g$ each, so $\mathbf{V}(x) = \tfrac{1}{2}x^2 - \tfrac{g}{3}x^3$. Let us introduce $(\tilde{g}_l)_{l \geq 1}$ as the coefficients of expansion:
$$
\label{deftildeg} \frac{\partial}{\partial v}\Big(-\frac{2\mathbf{V}(x(v))}{4 - n^2} + \frac{2 \ln x(v)}{2 + n}\Big) \mathop{=}_{v \rightarrow v_{\infty}} \sum_{k \geq 1} \frac{\tilde{g}_{k - 1}}{(v - v_{\infty})^k} + O(1).
$$
Note that there are only finitely many $\tilde{g}_l\neq 0$, since $V$ is polynomial and $x(v)$ has a single pole at $v=v_{\infty}$.
Their expressions for the model where all faces are triangles are recorded in Appendix~\ref{Appgdeter}.

\begin{proposition}[Disks] \cite{BBG12b}
\label{theimdisk}We have that
$$
\mathbf{G}(v) = \sum_{l \geq 0} \frac{1}{2}\,\frac{(-1)^l\tilde{g}_l}{l!}\,\partial_{v_{\infty}}^{l}\big[\Upsilon_{b}(v + v_{\infty}) + \Upsilon_{b}(v - v_{\infty}) - \Upsilon_{b}(-v + v_{\infty}) - \Upsilon_{b}(-v - v_{\infty})\big].
$$
The endpoints $\gamma_{\pm}$ are determined by the two conditions:
\beq
\label{determinab}\mathbf{G}(\tau + \varepsilon) = 0,\qquad \varepsilon = 0,\tfrac{1}{2},
\eeq
which follow from the fact that $\mathbf{F}(x)$ remains bounded when $x \rightarrow \gamma_{\pm}$.
\end{proposition}

For use in refined generating series, let us define
$$
b(s) = \frac{{\rm arccos}(ns/2)}{\pi}.
$$

\begin{proposition} \label{their}\cite{BBD} Define $\mathbf{G}^\bullet_{s}(v)$ as the analytic continuation of
$$
\label{second}x'(v)\mathbf{F}_{s}^\bullet(x(v)) + \partial_{v}\Bigg(\frac{nsu\,\ln[\varsigma'(x(v))]}{2(2 + ns)}\Bigg).
$$
We have:
$$
\mathbf{G}_{s}^\bullet(v) = \frac{u}{2 + ns}\Big[-\Upsilon_{b(s)}(v + v_{\infty}) - \Upsilon_{b(s)}(v - v_{\infty}) + \Upsilon_{b(s)}(-v + v_{\infty}) + \Upsilon_{b(s)}(-v - v_{\infty})\Big].
$$
\end{proposition}

\begin{proposition} \label{p15}\cite{BEOn,BBD} Define $\mathbf{G}^{(2)}_{s}(v_1,v_2)$ as the analytic continuation of
$$
x'(v_1)x'(v_2)\mathbf{F}^{(2)}_{s}(x(v_1),x(v_2)) + \frac{\partial}{\partial v_1}\frac{\partial}{\partial v_2}\Bigg(\frac{2\ln\big[x(v_1) - x(v_2)\big] + ns\ln\big[\varsigma(x(v_1)) - x(v_2)\big]}{4 - n^2s^2}\Bigg).
$$
We have:
$$
\mathbf{G}^{(2)}_{s}(v_1,v_2) = \frac{1}{4 - n^2s^2}\Big[\Upsilon_{b(s)}'(v_1 + v_2) - \Upsilon_{b(s)}'(v_1 - v_2) - \Upsilon_{b(s)}'(-v_1 + v_2) + \Upsilon_{b(s)}'(-v_1 - v_2)\Big].
$$
\end{proposition}

\begin{remark} When there is no bending energy, i.e.~$\alpha = 1$, the $4$-terms expression of Propositions~\ref{theimdisk}-\ref{their} can be reduced to $2$ terms using $\tau - v_{\infty} = v_{\infty}\,\,{\rm mod}\,\,\mathbb{Z}$ and the pseudo-periodicity of the special function $\Upsilon_{b}$. 
\end{remark}

\subsection{Topological recursion}
\label{TRtr}
Theorem~\ref{2g2m} in the special case of the bending energy model shows that $\mathbf{F}^{(\mathsf{g},k)}(x_1,x_2,\ldots,x_k)$ for $2\mathsf{g} - 2 + k > 0$ satisfies the homogeneous linear equation with respect to $x_1$, for fixed $(x_i)_{i = 2}^k$. Following Section~\ref{elparam}, we can thus introduce a meromorphic function $\mathbf{G}^{(\mathsf{g},k)}(v_1,\ldots,v_k)$ as the analytical continuation of
\beq
\label{140}\mathbf{F}^{(\mathsf{g},k)}(x(v_1),\ldots,x(v_k))\,\prod_{i = 1}^k x'(v_i).
\eeq
It is also convenient to introduce a shift for the case of cylinders. We consider:
\bea
\overline{\mathbf{G}}^{(\mathsf{g},k)}(v_1,\ldots,v_k) & = & \mathbf{G}^{(\mathsf{g},k)}(v_1,\ldots,v_k) \nonumber \\
\label{H5s} & & + \delta_{\mathsf{g},0}\delta_{k,2}\bigg(\frac{2 - n^2}{4 - n^2}\,\frac{x'(v_1)x'(v_2)}{(x(v_1) - x(v_2))^2} - \frac{n}{4 - n^2}\,\frac{\varsigma'(x(v_1))x'(v_1)x'(v_2)}{(x(v_1 - \tau) - x(v_2))^2} \bigg).
\eea
While $\mathbf{G}^{(2)}(v_1,v_2)$ satisfied the homogeneous linear equation, $\overline{\mathbf{G}}^{(2)}(v_1,v_2)$ satisfies, with respect to $v_1$, the inhomogeneous version of equation \eqref{reuh} with right-hand side $1/(x(v_1) - x(v_2))^2$.

Our starting point is the topological recursion residue formula proved in \cite{BEOn} or \cite[Section 5]{BEO13}. Let us define the recursion kernel, for $\varepsilon \in \{0,1/2\}$:
\beq
\mathbf{K}_{\varepsilon}(v_0,v) = -\frac{\dd v}{2}\,\frac{\int_{2(\tau + \varepsilon)  - v}^{v} \dd v'\,\overline{\mathbf{G}}^{(2)}(v',v_0)}{\mathbf{G}(v) + \mathbf{G}(2\tau - v)}.
\eeq
If $k \geq 2$, let $I = \{2,\ldots,k\}$, and if $k = 1$, $I = \emptyset$. If $J$ is a set, we denote $v_{J} = (v_{j})_{j \in J}$. 

\begin{theorem}
\label{them1o} For $2\mathsf{g} - 2 + k > 0$, we have
\bea
\overline{\mathbf{G}}^{(\mathsf{g},k)}(v_1,v_I) & = & \sum_{\varepsilon \in \{0,1/2\}} \Res_{v \rightarrow \tau + \varepsilon} \mathbf{K}_{\varepsilon}(v_1,v) \Bigg[\overline{\mathbf{G}}^{(\mathsf{g} - 1,k + 1)}(v,2(\tau + \varepsilon) - v,v_{I}) \nonumber \\
&& + \sum_{\substack{h + h' = g \\ J \sqcup J' = I}}^{{\rm no}\,\,{\rm disks}} \overline{\mathbf{G}}^{(\mathsf{h},1 + |J|)}(v,v_{J})\overline{\mathbf{G}}^{(\mathsf{h}',1 + |J'|)}(2(\tau + \varepsilon) - v,v_{J'})\Bigg], \nonumber
\eea
where ``no disks'' means that we exclude the terms containing disk generating series, that is $(\mathsf{h},J)$ or $(\mathsf{h}',J')$ equal to $(0,\emptyset)$.
\end{theorem}

We are going to rewrite this recursion without involving residues. We first need to introduce some notations. Let us define the \emph{elementary blocks}:
\beq
\label{BB0}\varepsilon \in \{0,\tfrac{1}{2}\},\qquad \mathbf{B}_{\varepsilon,l}(v) = \frac{\partial^{2l}}{\partial v_2^{2l}} \overline{\mathbf{G}}^{(2)}(v,v_2)\Big|_{v_2 = \tau + \varepsilon}.
\eeq
Since $x(\tau + \varepsilon + w)$ is an even function of $w$, formula \eqref{BB0} is insensitive to replacing $\overline{\mathbf{G}}^{(2)}$ by $\mathbf{G}^{(2)}$. From the structure of $\mathbf{G}^{(2)} = \mathbf{G}^{(2)}_{s = 1}$ shown in Proposition~\ref{p15}, we see that
\beq
\label{BB1} \mathbf{B}_{\varepsilon,l}(v) = \partial_{v}^{2l} \mathbf{B}_{\varepsilon,0}(v).
\eeq

\begin{proposition}
\label{2g2mr}For $2\mathsf{g} - 2 + k > 0$, we have a decomposition
$$
\overline{\mathbf{G}}^{(\mathsf{g},k)}(v_1,\ldots,v_k) = \mathbf{G}^{(\mathsf{g},k)}(v_1,\ldots,v_k) = \sum_{\substack{l_1,\ldots,l_k \geq 0 \\ \varepsilon_1,\ldots,\varepsilon_{k} \in \{0,\frac{1}{2}\}}} \mathsf{C}^{(\mathsf{g},k)}\bigl[{}^{l_1}_{\varepsilon_1} \cdots {}^{l_k}_{\varepsilon_k}\bigr]  \prod_{i = 1}^k \mathbf{B}_{\varepsilon_i,l_i}(v_i),
$$
where the sum contains only finitely many non-zero terms.
\end{proposition}

As a consequence of Theorem~\ref{them1o}, the coefficients $\mathsf{C}^{(\mathsf{g},k)}[{}^{l_I}_{\varepsilon_I}\bigr]$ satisfy the recursion given in Proposition~\ref{cocor1} below. Its proof appears right after Proposition~\ref{cocor1}.

\subsubsection{Initial conditions}
\label{Seinia} We denote $y_{\varepsilon,1}$ and $y_{\varepsilon,2}$ the first two coefficients in the Taylor expansion at $w \rightarrow 0$:
\beq
\label{316} \Delta_{\varepsilon} \mathbf{G}(w) \coloneqq \mathbf{G}(w + \tau + \varepsilon) + \mathbf{G}(-w + \tau + \varepsilon) = y_{\varepsilon,1}\,w^2 + \frac{y_{\varepsilon,2}}{6}\,w^4 + O(w^6).
\eeq
We also need the constants
\beq
\label{ububub} \upsilon_{b,2m+1} = \lim_{w \rightarrow 0} \Big(\Upsilon_{b}^{(2m+1)}(w) + \frac{(2m+1)!}{w^{2m+2}}\Big),
\eeq
introduced in Appendix~\ref{AppUp}. The initial conditions for the recursion concern $(\mathsf{g},k) = (0,3)$ and $(1,1)$:
$$
\boxed{\mathsf{C}^{(0,3)}\big[{}^{l_1}_{\varepsilon_1}\,{}^{l_2}_{\varepsilon_2}\,{}^{l_3}_{\varepsilon_3}\bigr] = -\frac{2\,\delta_{l_1,l_2,l_3,0}\,\delta_{\varepsilon_1,\varepsilon_2,\varepsilon_3}}{y_{\varepsilon_1,1}},
\qquad \mathsf{C}^{(1,1)}\bigl[{}^{l}_{\varepsilon}\bigr] = \delta_{l,0}\Big(\frac{y_{\varepsilon,2}}{24y_{\varepsilon,1}^2} + \frac{\upsilon_{b,1}}{y_{\varepsilon,1}}\Big) -\frac{\delta_{l,1}}{24 y_{\varepsilon,1}}.}
$$

\subsubsection{The recursion coefficients}
We first define
\beq
\label{Khoubi} \boxed{K\bigl[{}^{l}_{\varepsilon}\,{}^{m}_{\,\sigma}\,{}^{m'}_{\,\sigma'}\bigr] = \Res_{w \rightarrow 0} \frac{-w^{2l + 1}\dd w}{(2l + 1)!\,\Delta_{\varepsilon}\mathbf{G}(w)}\,\mathbf{B}_{\sigma,m}(w + \tau + \varepsilon)\,\mathbf{B}_{\sigma',m'}(-w + \tau + \varepsilon).}
\eeq
Since $\Delta_{\varepsilon}\mathbf{G}(w)$ is even, we have the symmetry
\beq
K\bigl[{}^{l}_{\varepsilon}\,{}^{m}_{\,\sigma}\,{}^{m'}_{\,\sigma'}\bigr] = K\bigl[{}^{l}_{\varepsilon}\,{}^{m'}_{\,\sigma'}\,{}^{m}_{\,\sigma}\bigr].
\eeq
By counting the degree of the integrand at $w = 0$, we find selection rules. There are finitely many indices for which $K$ does not vanish:
\bea
\label{selvK} &&\,\,\, \big\{\varepsilon = \sigma = \sigma' \,\,\mathrm{and}\,\, l \leq m + m ' + 2\big\}, \nonumber \\
& \mathrm{or}& \,\,\,\big\{\varepsilon = \sigma \neq \sigma'\,\,\mathrm{and}\,\,l \leq m + 1\big\}, \nonumber \\
& \mathrm{or}& \,\,\,\big\{\varepsilon \neq \sigma = \sigma'\,\,\mathrm{and}\,\,l = 0\big\}. \nonumber
\eea
We also define
\beq
\label{Ktildeou}\boxed{\tilde{K}\bigl[{}^{l}_{\varepsilon}\,{}^{l'}_{\varepsilon'}\,{}^{m}_{\,\sigma}\bigr] =\,\frac{-\delta_{\varepsilon,\varepsilon'}}{(2l + 1)!\,(2l')!}\Res_{w \rightarrow 0} \dd w\,\frac{w^{2(l + l')+1}}{\Delta_{\varepsilon}\mathbf{G}(w)}\,\mathbf{B}_{\sigma,m}(\tau + \varepsilon + w).}
\eeq
There are finitely many values of the parameters for which $\tilde{K}$ does not vanish:
\bea
\label{selvKt} &&\,\,\,\big\{\varepsilon = \varepsilon' = \sigma\,\,\mathrm{and}\,\,l + l' \leq m + 1\big\}, \nonumber \\
& \mathrm{or}& \,\,\,\big\{\varepsilon = \varepsilon' \neq \sigma\,\,\mathrm{and}\,\,(l,l') = (0,0)\big\}. \nonumber
\eea

\subsubsection{The recursion formula}

\begin{proposition}
\label{cocor1}Assume $2\mathsf{g} - 2 + k \geq 2$, and denote $L = \{2,\ldots,k\}$. The coefficients of the decomposition in Proposition~\ref{2g2mr} satisfy:
\bea
\mathsf{C}^{(\mathsf{g},k)}\bigr[{}^{l_1}_{\varepsilon_1} \cdots\,{}^{l_k}_{\varepsilon_k}\big] & = & \sum_{\substack{m,m' \geq 0 \\\sigma,\sigma' \in \{0,1/2\}}} K\bigl[{}^{l_1}_{\varepsilon_1}\,{}^{m}_{\,\sigma}\,{}^{m'}_{\,\sigma'}\bigr]\,\mathsf{C}^{(\mathsf{g} - 1,k+1)}\bigl[{}^{m}_{\,\sigma}\,{}^{m'}_{\,\sigma'}\,{}^{l_{L}}_{\varepsilon_{L}}\bigr] \nonumber \\
& & + \sum_{\substack{\mathsf{h} + \mathsf{h}' = \mathsf{g} \\ J \sqcup J' = L \\ m,m' \geq 0,\,\,\sigma,\sigma' \in \{0,1/2\}}}^{{\rm stable}} K\bigl[{}^{l_1}_{\varepsilon_1}\,{}^{m}_{\,\sigma}\,{}^{m'}_{\,\sigma'}\bigr]\,\mathsf{C}^{(\mathsf{h},|J| + 1)}\big[{}^{m}_{\,\sigma}\,{}^{l_{J}}_{\varepsilon_J}\bigr]\,\mathsf{C}^{(\mathsf{h}',1 + |J'|)}\big[{}^{m'}_{\,\sigma'}\,{}^{l_{J'}}_{\varepsilon_{J'}}\bigr] \nonumber \\
\label{chack}& & + \sum_{\substack{i \in L,\,\,m \geq 0 \\ \sigma \in \{0,1/2\}}} 2\,\tilde{K}\bigl[{}^{l_1}_{\varepsilon_1}\,{}^{l_i}_{\varepsilon_i}\,{}^{m}_{\,\sigma}\bigr]\,\mathsf{C}^{(\mathsf{g},k - 1)}\bigl[{}^{m}_{\,\sigma}\,{}^{l_{L\setminus\{i\}}}_{\varepsilon_{L\setminus\{i\}}}\bigr],
\eea
where ``stable'' means that we exclude the terms involving disk or cylinder generating series, i.e.~for which $(\mathsf{h}, |J|+1)$ or $(\mathsf{h}', |J'|+1)$ belongs to $\{(0,1),(0,2)\}$.
\end{proposition}

Although this recursion gives a non symmetric role to the first boundary, the result ensuing from the initial conditions of \S~\ref{Seinia} is symmetric. This must be true by consistency, and this is in fact a general property of the topological recursion, cf. \cite[Theorem 4.6]{EOFg}.

\subsection{Proof of Propositions~\ref{2g2mr}-\ref{cocor1}}

\subsubsection{Properties of the elementary blocks}

We have called elementary blocks the following functions:
\beq
\label{Bled}\mathbf{B}_{\varepsilon,l}(v) = \frac{\partial^{2l}}{\partial v_2^{2l}} \mathbf{G}^{(2)}(v,v_2)\Big|_{v_2 = \tau + \varepsilon}.
\eeq
\begin{lemma}
\label{regood} $\mathbf{B}_{\varepsilon,l}(\tau + \varepsilon' + w)$ is regular at $w = 0$ if $\varepsilon \neq \varepsilon'$, and behaves like $(2l + 1)!w^{-(2l + 2)} + O(1)$ when $w \rightarrow 0$ if $\varepsilon = \varepsilon'$.
\end{lemma}
\noindent \textbf{Proof.} We compute using Proposition~\ref{p15} and the properties \eqref{propoup} of $\Upsilon_{b}$:
\beq
\label{blbelb}\mathbf{B}_{\varepsilon,l}(\tau + \varepsilon' + w) = \frac{(e^{2{{\rm i}\pi b}} - 1)\Upsilon_{b}^{(2l + 1)}(\varepsilon + \varepsilon' + w) + (e^{-2{\rm i}\pi b} - 1)\Upsilon_{b}^{(2l + 1)}(\varepsilon + \varepsilon' - w)}{4 - n^2}.
\eeq
We deduce its behavior when $w \rightarrow 0$. Since $\Upsilon_{b}$ is regular at the value $\tfrac{1}{2}$, \eqref{blbelb} is regular at $w = 0$ when $\varepsilon \neq \varepsilon'$. If $\varepsilon = \varepsilon'$, the simple pole of $\Upsilon_b$ produces the divergent behavior:
$$
\mathbf{B}_{\varepsilon,l}(\tau + \varepsilon + w) = \frac{(2l + 1)!}{w^{2l + 2}} + O(1).
$$
\hfill $\Box$

We shall need later in the computation of $\mathbf{G}^{(1,1)}$:
\begin{lemma}
\label{Lupb}
\beq
\label{upb2}\overline{\mathbf{G}}^{(2)}(\tau + \varepsilon + w,\tau + \varepsilon - w) \,\mathop{=}_{w \rightarrow 0}\,\, \frac{1}{4w^2} - \upsilon_{b,1} + o(1),
\eeq
where $\upsilon_{b,1}$ is the constant computed in \eqref{upb1def}.
\end{lemma}
\noindent\textbf{Proof.} We compute with \eqref{H5s}:
\bea
\label{G2taue}  && \overline{\mathbf{G}}^{(2)}(\tau + \varepsilon + w,\tau + \varepsilon - w)\\
& = & -\frac{2 - n^2}{4 - n^2}\bigg(\upsilon_{b,1} + \frac{S_{x}(\tau + \varepsilon + w)}{6}\bigg) \nonumber \\
&& + \frac{n}{4 - n^2}\,\frac{x'(\varepsilon + w)x'(\tau + \varepsilon + w)}{(x(\varepsilon + w) - x(\tau + \varepsilon + w))^2} - \frac{\Upsilon_{b}'(2w) + \Upsilon_{b}'(-2w)}{4 - n^2}, \nonumber
\eea
where we introduced the Schwarzian derivative:
\beq
S_{x}(v) = \frac{x'''(v)}{x'(v)} - \frac{3}{2}\Big(\frac{x''(v)}{x'(v)}\Big)^2.
\eeq
Since $x'(\tau + \varepsilon + w)$ is an odd function of $w$, the second term in \eqref{G2taue} is $o(1)$ when $w \rightarrow 0$. We also compute:\bea
\frac{1}{6}\,S_{x}(\tau + \varepsilon + w) & = & \frac{1}{6}\Bigg[\frac{x''''(\tau + \varepsilon)}{x''(\tau + \varepsilon)} - \frac{3}{2w^2}\Bigg(\frac{1 + \frac{x''''(\tau + \varepsilon)}{2x''(\tau + \varepsilon)}\,w^2}{1 + \frac{x''''(\tau + \varepsilon)}{6x''(\tau + \varepsilon)}\,w^2}\Bigg)^2 + O(w^2)\Bigg] \nonumber \\
& = & -\frac{1}{4w^2} + O(w^2) \nonumber
\eea
and
\beq
\Upsilon_b'(2w) + \Upsilon_{b}'(-2w) = 2\Big(-\frac{1}{4w^2} + \upsilon_b\Big) + o(w^2). \nonumber
\eeq
Collecting all terms in \eqref{G2taue} we find \eqref{upb2}.
\hfill $\Box$

\subsubsection{Computing the residues}

Now we are ready to examine the formula of Theorem~\ref{them1o}. In order to compute the residues at $v \rightarrow \tau + \varepsilon$, we should first compute the expansion of the recursion kernel near those points. If we set $v = (\tau + \varepsilon) + w$ and $\Delta_{\varepsilon}\mathbf{G}(w) = \mathbf{G}(\tau + \varepsilon + w) + \mathbf{G}(\tau + \varepsilon - w)$, we find
\bea
\mathbf{K}_{\varepsilon}(v_0,\tau + \varepsilon + w) & = & \frac{-1}{2\Delta_{\varepsilon}\mathbf{G}(w)}\,\int_{-w}^{w} \dd z\Bigg(\sum_{l \geq 0} \mathbf{B}_{\varepsilon,l}(v_0)\,\frac{z^{2l}}{(2l)!} + ({\rm odd}\,\,{\rm terms})\Bigg) \nonumber \\
\label{expKq}& = & - \sum_{l \geq 0} \frac{w^{2l + 1}}{(2l + 1)!\Delta_{\varepsilon}\mathbf{G}(w)}\,\mathbf{B}_{\varepsilon,l}(v_0),
\eea
in terms of the elementary blocks \eqref{Bled}. Since we consider a model with off-critical weights, $\Delta_{\varepsilon}\mathbf{G}(w)$ has exactly a double zero at $w \rightarrow 0$. Subsequently, $\mathbf{K}_{\varepsilon}(v_0,\tau + \varepsilon + w)$ has a simple pole at $w = 0$, and the term indexed by $l$ in the sum has a simple pole if $l = 0$, and has a zero of order $(2l - 1)$ if $l \geq 1$.

We prove Propositions~\ref{2g2mr}-\ref{cocor1} by induction on $\chi = 2\mathsf{g} - 2 + k > 0$. The first case to consider is $\chi = 1$, i.e.~ $(\mathsf{g},k) = (0,3)$ or $(1,1)$. For $(\mathsf{g},k) = (0,3)$, Theorem~\ref{them1o} yields
\bea
&&  \mathbf{G}^{(0,3)}(v_1,v_2,v_3) \nonumber \\
& = & \sum_{\varepsilon \in \{0,1/2\}} \Res_{w \rightarrow 0} \mathbf{K}_{\varepsilon}(v_1,\tau + \varepsilon + w)\Big[\overline{\mathbf{G}}^{(2)}(\tau + \varepsilon + w,v_2)\overline{\mathbf{G}}^{(2)}(\tau + \varepsilon - w,v_3) \nonumber \\
& & \qquad\qquad\qquad\qquad\qquad\qquad\qquad + \overline{\mathbf{G}}^{(2)}(\tau + \varepsilon + w,v_3)\overline{\mathbf{G}}^{(2)}(\tau + \varepsilon - w,v_2)\Big]. \nonumber
\eea
As one can check from Proposition~\ref{p15}, $\mathbf{G}^{(2)}(\tau + \varepsilon + w,v')$ is regular when $w \rightarrow 0$. Therefore, the residue picks up the term $l = 0$ in the expansion of the recursion kernel, and evaluates the function between brackets to $w = 0$. The result is thus of the form announced in Proposition~\ref{2g2mr}, with only non-zero coefficients:
\beq
\label{Coini}\mathsf{C}^{(0,3)}\bigl[{}^{0}_{\varepsilon}\,{}^{0}_{\varepsilon}\,{}^{0}_{\varepsilon}\bigr] = -\frac{2}{y_{\varepsilon,1}},\qquad \varepsilon \in \{0,\tfrac{1}{2}\},
\eeq
computed using also the expansion \eqref{316} of $\Delta_{\varepsilon}\mathbf{G}(w)$. \\
For $(\mathsf{g},k) = (1,1)$, Theorem~\ref{them1o} yields
$$
\mathbf{G}^{(1,1)}(v_1) = \sum_{\varepsilon \in \{0,1/2\}} \Res_{w \rightarrow 0} \mathbf{K}_{\varepsilon}(v_1, \tau + \varepsilon + w)\cdot \overline{\mathbf{G}}^{(2)}(\tau + \varepsilon + w,\tau + \varepsilon - w).
$$
We have seen in Lemma~\ref{Lupb} that the last factor has a double pole when $w \rightarrow 0$, with no simple pole and constant term $-\upsilon_{b,1}$ defined in \eqref{upb1def}. Then, we have to expand the recursion kernel up to $O(w^2)$ in order to obtain the final answer for $\mathbf{G}^{(1,1)}$. In other words, we only need to include the terms $l = 0$ and $l = 1$, and use the expansion \eqref{316} of the denominator to perform the computation:
$$
\mathbf{K}_{\varepsilon}(v_1,\tau + \varepsilon + w) = -\frac{\mathbf{B}_{\varepsilon,0}(v_1)}{y_{\varepsilon,1}}\,\frac{1}{w} + \Big(\frac{y_{\varepsilon,2}\mathbf{B}_{\varepsilon,0}(v_1)}{y_{\varepsilon,1}^2} - \frac{\mathbf{B}_{\varepsilon,1}(v_1)}{y_{\varepsilon,1}}\Big)\,\frac{w}{6} + o(w).
$$
We find eventually
$$
\mathbf{G}^{(1,1)}(v_1) = \sum_{\varepsilon \in \{0,1/2\}} \Big(\frac{y_{\varepsilon,2}}{24y_{\varepsilon,1}^2} + \frac{\upsilon_{b,1}}{y_{\varepsilon,1}}\Big)\mathbf{B}_{\varepsilon,0}(v_1)  - \frac{\mathbf{B}_{\varepsilon,1}(v_1)}{24y_{\varepsilon,1}}.
$$
The answer is of the form of Proposition~\ref{2g2mr}, with only non-zero coefficients:
\beq
\label{C1ini} \mathsf{C}^{(1,1)}\bigl[{}^{0}_{\varepsilon}\bigr] = \frac{y_{\varepsilon,2}}{24y_{\varepsilon,1}^2} + \frac{\upsilon_b}{y_{\varepsilon,1}},\qquad \mathsf{C}^{(1,1)}\bigl[{}^{1}_{\varepsilon}\bigr] = -\frac{1}{24y_{\varepsilon,1}}.
\eeq

Now, take $\chi \geq 2$, and assume the result is true for all $\overline{\mathbf{G}}^{(\mathsf{g}',k')} = \mathbf{G}^{(\mathsf{g}',k')}$, with $0 < 2\mathsf{g}' - 2 + k' < \chi$. We would like to compute $\mathbf{G}^{(\mathsf{g},k)}$ for a topology such that $2\mathsf{g} - 2 + k = \chi$. The residue formula of Theorem~\ref{them1o} involves $\mathbf{G}^{(\mathsf{g}',k')}$ for $0 < 2\mathsf{g}' - 2 + k' < \chi$, which we replace by the decomposition of Proposition~\ref{2g2mr}, as well as $\overline{\mathbf{G}}^{(2)}$.

The terms which do not contain $\overline{\mathbf{G}}^{(2)}$ give a contribution which is the sum over indices $(l_j,\varepsilon_j)_{j \in I}$ and indices $(m,\sigma),(m',\sigma')$ of terms containing the factor:
\bea
& & \Big[\prod_{j \in I} \mathbf{B}_{\varepsilon_j,l_j}(v_j)\Big]\,\Res_{w \rightarrow 0} \mathbf{K}_{\varepsilon}(v_0,\tau + \varepsilon + w)\,\mathbf{B}_{\sigma,m}(\tau + \varepsilon + w)\,\mathbf{B}_{\sigma',m'}(\tau + \varepsilon - w) \nonumber \\
\label{233} & = & \sum_{l \geq 0} \Big[\prod_{j \in I} \mathbf{B}_{\varepsilon_j,l_j}(v_j)\cdot\mathbf{B}_{\varepsilon,l}(v_0)\Big]\,K\bigl[{}^{l}_{\varepsilon}\,{}^{m}_{\sigma}\,{}^{m'}_{\sigma'}\bigr]. \nonumber
\eea
We computed the residue thanks to the expansion of $\mathbf{K}_{\varepsilon}$ given in \eqref{expKq}, and we introduced the coefficient \eqref{Khoubi}:
\beq
K\bigl[{}^{l}_{\varepsilon}\,{}^{m}_{\,\sigma}\,{}^{m'}_{\,\sigma'}\bigr] = \Res_{w \rightarrow 0} \frac{-\dd w\,w^{2l + 1}}{(2l + 1)!\Delta_{\varepsilon}\mathbf{G}(w)}\,\mathbf{B}_{\sigma,m}(\tau + \varepsilon + w)\,\mathbf{B}_{\sigma',m'}(\tau + \varepsilon - w) \nonumber. 
\eeq
These terms thus form a linear combination of products of elementary blocks in the variables $v_0,(v_j)_{j \in I}$, which contribute to $\mathsf{C}^{(\mathsf{g},k)}\big[{}^{l}_{\varepsilon}\,{}^{l_{I}}_{\varepsilon_{I}}\bigr]$ by the two first lines in \eqref{chack}.

Since $2\mathsf{g} - 2 + k \geq 2$, the contribution to $\mathbf{G}^{(\mathsf{g},k)}(v_0,v_I)$ containing $\overline{\mathbf{G}}^{(2)}$ is precisely the sum over $\varepsilon \in \{0,1/2\}$ and $i \in I = \{2,\ldots,k\}$ of
\beq
\label{resiJ}\Res_{w \rightarrow 0} \mathbf{K}_{\varepsilon}(v_0,\tau + \varepsilon + w)\Big[\overline{\mathbf{G}}^{(2)}(\tau + \varepsilon + w,v_i)\,\mathbf{G}^{(\mathsf{g},k - 1)}(\tau + \varepsilon - w,v_{I\setminus \{i\}}) + (w \rightarrow -w)\Big].
\eeq
The quantity in brackets can be decomposed using odd and even parts:
\bea
& & 2\Big[\overline{\mathbf{G}}^{(2)}_{{\rm even}}(\tau + \varepsilon + w,v_i)\,\mathbf{G}^{(\mathsf{g},k - 1)}_{{\rm even}}(\tau + \varepsilon + w,v_{I\setminus\{i\}})\nonumber \\
& &  - \overline{\mathbf{G}}^{(2)}_{{\rm odd}}(\tau + \varepsilon + w,v_i)\,\mathbf{G}^{(\mathsf{g},k - 1)}_{{\rm odd}}(\tau + \varepsilon + w,v_{I\setminus\{i\}})\Big]. \nonumber
\eea
When we insert in this expression the decomposition of Proposition~\ref{2g2mr} for $\mathbf{G}^{(\mathsf{g},k - 1)}$, we have to deal with the sum over indices $(l_{j},\varepsilon_{j})_{j \neq i}$ and $(m,\sigma)$ of terms of the form
\bea
& & 2\mathsf{C}^{(\mathsf{g},k - 1)}\bigl[{}^{m}_{\,\sigma}\,{}^{l_{I\setminus\{i\}}}_{\varepsilon_{I\setminus\{i\}}}\bigr]\prod_{j \in I\setminus\{i\}} \mathbf{B}_{\varepsilon_j,l_j}(v_j)\cdot
\Big[\overline{\mathbf{G}}^{(2)}_{{\rm even}}(\tau + \varepsilon + w,v_i)\,\mathbf{B}_{\sigma,m}^{{\rm even}}(\tau + \varepsilon + w) \nonumber \\ \label{235}& & \phantom{ 2\,\mathsf{C}_{(\mathsf{g},k-1)}\bigl[{}^{m}_{\sigma}\,{}^{l_{I\setminus\{i\}}}_{\varepsilon_{I\setminus\{i\}}}\bigr]\prod_{j \in I\setminus\{i\}} \mathbf{B}_{\varepsilon_j,l_j}(v_j)\cdot}\, - \overline{\mathbf{G}}^{(2)}_{{\rm odd}}(\tau + \varepsilon + w,v_i)\,\mathbf{B}_{\sigma,m}^{{\rm odd}}(\tau + \varepsilon + w)\Big]. 
\eea
According to Lemma~\ref{regood}, $\mathbf{B}_{\sigma,m}^{{\rm odd}}(\tau + \varepsilon + w) \in O(w)$ when $w \rightarrow 0$. Since $\overline{\mathbf{G}}^{(2)}(\tau + \varepsilon + w,v_i)$ is regular when $w \rightarrow 0$, this implies that the product of the odd parts does not contribute to the residue \eqref{resiJ}. Besides, the expansion at $w \rightarrow 0$ of the product of even parts in \eqref{235} can be expressed in terms of the elementary blocks. We thus obtain a contribution
\beq
\label{238}\sum_{l,l_i \geq 0} 2\,\mathsf{C}^{(\mathsf{g},k - 1)}\bigl[{}^{m}_{\,\varepsilon}\,{}^{l_{I\setminus\{i\}}}_{\varepsilon_{I\setminus\{i\}}}\bigr] \Bigg[\prod_{j \in I\setminus\{i\}} \mathbf{B}_{\varepsilon_j,l_j}(v_j) \cdot \mathbf{B}_{\varepsilon,l}(v_0)\mathbf{B}_{\varepsilon,l_i}(v_i) \Bigg]\,\tilde{K}\bigl[{}^{l}_{\varepsilon}\,{}^{l'}_{\varepsilon'}\,{}^{m}_{\,\sigma}\bigr]
\eeq
and we have defined
\beq
\tilde{K}\bigl[{}^{l}_{\varepsilon}\,{}^{l'}_{\varepsilon'}\,{}^{m}_{\,\sigma}\bigr] = \delta_{\varepsilon,\varepsilon'}\,\Res_{w \rightarrow 0} \frac{-\dd w\,w^{2l + 1}}{(2l + 1)!\Delta_{\varepsilon}\mathbf{G}(w)}\cdot \frac{w^{2l'}}{(2l')!} \cdot \mathbf{B}_{\sigma,m}(w + \tau + \varepsilon), \nonumber
\eeq
which is the coefficient announced in \eqref{Ktildeou}. Since the prefactor of $\mathbf{B}$ in the residue is an odd $1$-form in $w$, the residue picks up the even part of $\mathbf{B}$, so it did not change the result to replace $\mathbf{B}^{{\rm even}}$ by $\mathbf{B}$. Let us examine the cases for which $\tilde{K}$ does not vanish. If $\varepsilon = \sigma$, we take into account the behavior at $w \rightarrow 0$ of $\mathbf{B}_{\varepsilon,m}(\tau + \varepsilon + w)$ given by Lemma~\ref{regood}, and find
\beq
\label{fikii}\tilde{K}\bigl[{}^{l}_{\varepsilon}\,{}^{l'}_{\varepsilon'}\,{}^{m}_{\,\sigma}\bigr] = \delta_{\varepsilon,\varepsilon'}\,\frac{1}{(2l + 1)!(2l')!}\,\Res_{w \rightarrow 0} \frac{-\dd w\, w^{2(l + l')}}{\Delta_{\varepsilon}\mathbf{G}(w)}\left(\frac{ (2m + 1)! w^{-2m}}{w} + w\, \upsilon_{b,2m+1} \right),
\eeq
where $\upsilon_{b,2m+1}$ are the constants introduced in \eqref{expord1Upsilon}.
Since $\Delta_{\varepsilon}\mathbf{G}(w)$ has a double zero at $w = 0$, \eqref{fikii} vanishes if $l + l' \geq m + 2$. If $\varepsilon \neq \sigma$, Lemma~\ref{regood} tells us that $\mathbf{B}_{\sigma,m}(\tau + \varepsilon + w)$ is regular at $w = 0$, hence $\tilde{K}$ vanish unless $(l,l') = (0,0)$, and we have
$$
\tilde{K}\bigl[{}^{0}_{\varepsilon}\,{}^{0}_{\varepsilon}\,{}^{m}_{\,\sigma}\bigr] = - \frac{\mathbf{B}_{\sigma,m}(\tau + \varepsilon)
}{y_{\varepsilon,1}} = \frac{\Upsilon_{b}^{(2m + 1)}(\tfrac{1}{2})}{y_{\varepsilon,1}}.
$$
The last equality follows from \eqref{blbelb} and the properties of $\Upsilon_{b}$ described in Appendix~\ref{AppA}. We can study in a similar way the cases for which $K$ does not vanish: we leave this computation to the reader, which only uses Lemma~\ref{regood} and the double zero of $\Delta_{\varepsilon}\mathbf{G}(w)$ at $w = 0$.

Collecting all the terms from \eqref{233} and \eqref{238}, we arrive to Formula ~\eqref{chack} and conclude the recursive proof. \hfill $\Box$

\subsection{Diagrammatic representation}
\label{diagTR}
Unfolding the recursion yields a formula for $\mathbf{G}^{(\mathsf{g},k)}$ with $2\mathsf{g} - 2 + k > 0$ as a sum over the set $\mathcal{S}^{(\mathsf{g},k)}$ of graphs $\mathcal{G}$ with first Betti number $\mathsf{g}$, trivalent vertices equipped with a cyclic order of their incident edges, and legs (univalent vertices) labeled $\{1,\ldots,k\}$. With this definition, if there is an edge from a trivalent vertex to itself (a loop), the cyclic order is just the transposition of the two distinct incident edges. The weight given to a graph actually depends on the choice of an initial leg $i_0$, but the sum over graphs is independent of those choices \cite{EORev}.

Before stating the formula, we need a preliminary construction. If $\mathcal{G} \in \mathcal{S}^{(\mathsf{g},k)}$, we denote $V(\mathcal{G})$ the set of trivalent vertices and a $E(\mathcal{G})$ the set of edges. We also denote $V_o(\mathcal{G})$ the set of trivalent vertices with a loop. If $\mathsf{v}$ is a vertex, we denote $\mathsf{e}[\mathsf{v}]$ its set of incident edges. A simple counting gives:
\beq
|E(\mathcal{G})| = 3\mathsf{g} - 3 + 2k,\qquad |V(\mathcal{G})| = 2\mathsf{g} - 2 + k.
\eeq

\subsubsection{Exploration of a cyclically ordered graph}
\label{explodocus}
The choice of an initial leg and the data of the cyclic order determines a way to explore $\mathcal{G}$, i.e.~two bijections
$$\varphi\,:\,\{1,\ldots,|E(\mathcal{G})|\} \rightarrow E(\mathcal{G}),\qquad \eta\,:\,\{1,\ldots,|V(\mathcal{G})| + k\} \rightarrow V(\mathcal{G}) \cup \{1,\ldots,k\}$$
which record in which order the edges, and the vertices or legs, are visited. Let us describe how $\varphi$ and $\eta$ are constructed.

We declare that $\eta(1)$ is the initial leg, and $\phi(1)$ is the edge incident to the initial leg $i_0$. Since $2\mathsf{g} - 2 + k > 0$, $\mathcal{G}$ must have at least a trivalent vertex, so $\phi(1)$ is also incident to a trivalent vertex that we declare to be $\eta(2)$. We define a seed with initial value $(\phi(1),\eta(2))$. Then, we apply the following algorithm. Let $(\mathsf{e} = \phi(j_1),\mathsf{v} = \eta(j_2))$ be the seed. If $\mathsf{v}$ is not a leg, let $\mathsf{e}^+$ (resp. $\mathsf{e}^-$) be the edge following (resp. preceding) $\mathsf{e}$ in the cyclic order around $\mathsf{v}$.

\smallskip

\noindent $\bullet$ \textbf{First cases:} either $\mathsf{v}$ is a leg or, otherwise, $\mathsf{e}^+$ and $\mathsf{e}^-$ have already been explored (i.e.~are equal to $\phi(i^+)$ and $\phi(i^-)$ for some $i^{\pm} < j_1$). If actually all vertices have already been explored (i.e.~$j_2 = |V(\mathcal{G})| + k$), the algorithm terminates; otherwise, we consider the maximal $j_2' < j_2$ such that $\eta(j_2')$ is not a leg, and the maximal $j_1' \leq j_1$ such that $\phi(j_1')$ is incident to $\eta(j_2')$, and reset the seed to $(\phi(j_1'),\eta(j_2'))$.\smallskip

\noindent $\bullet$ \textbf{Second case:} $\mathsf{e}^+$ has not been explored. We define $\phi(j_1 + 1) = \mathsf{e}^+ = \{\mathsf{v},\mathsf{v}^+\}$ and $\eta(j_2 + 1) = \mathsf{v}^+$, and reset the seed to $(\mathsf{e}^+,\mathsf{v}^+)$.
\smallskip

\noindent $\bullet$ \textbf{Third case:} $\mathsf{e}^+$ has already been explored, but not $\mathsf{e}^-$. We define $\phi(j_1 + 1) = \mathsf{e}^- = \{\mathsf{v},\mathsf{v}^-\}$ and $\eta(j' + 1) = \mathsf{v}^-$, and reset the seed to $(\mathsf{e}^-,\mathsf{v}^-)$.

\smallskip

Now, at any trivalent vertex $\mathsf{v}$ which does not have a loop, we can label the incident edges $\mathsf{e}^0_{\mathsf{v}},\mathsf{e}^1_{\mathsf{v}},\mathsf{e}^2_{\mathsf{v}}$, starting from the edge such that $\phi^{-1}(\mathsf{e}_{\mathsf{v}}^0)$ is minimal among $\mathsf{e}[\mathsf{v}]$, and following the cyclic order. If a trivalent vertex $\mathsf{v}$ has a loop, we can just label $\mathsf{e}^0_{\mathsf{v}}$ the incident edge which is not a loop, and $\mathsf{e}_{\mathsf{v}}^1$ the other one; this definition also agrees with the order of exploration at $\mathsf{v}$.
\begin{definition}
A trivalent vertex $\mathsf{v}$ is \emph{bi-terminal} if $\mathsf{e}_{\mathsf{v}}^{1}$ and $\mathsf{e}_{\mathsf{v}}^{2}$ are incident to legs. It is \emph{terminal} if $\mathsf{e}_{\mathsf{v}}^{1}$ xor $\mathsf{e}_{\mathsf{v}}^{2}$ is incident to a leg. We denote $V_{t}(\mathcal{G})$ (resp. $V_{tt}(\mathcal{G})$) the set of (bi-)terminal vertices, and $V'(\mathcal{G})$ the set of trivalent vertices which are neither terminal, neither bi-terminal, nor have a loop.
\end{definition}
We stress that, for a given graph, all these notions depend on the choice of an initial leg.

\subsubsection{The unfolded formula}

Let ${\rm Col}(\mathcal{G};(\bs{l},\bs{\varepsilon}))$ be the set of colorings of edges by labels in $\mathbb{N}\times\{0,\tfrac{1}{2}\}$ such that
\begin{itemize}
\item[$\bullet$] the coloring of edges incident to legs agrees with the fixed coloring $(\bs{l},\bs{\varepsilon})$ of the legs;
\item[$\bullet$] the color of a loop is identical to the color of the other edge incident to the vertex where the loop is attached.
\end{itemize}
If $(\bs{m},\bs{\sigma})$ is such a coloring, and $\mathsf{v}$ is a trivalent vertex which does not have a loop, we define $\bs{m}[\mathsf{v}]$ to be the sequence $(m(\mathsf{e}_{\mathsf{v}}^0),m(\mathsf{e}_{\mathsf{v}}^{1}),m(\mathsf{e}_{\mathsf{v}}^{2}))$, and similarly for the sequence $\bs{\sigma}[\mathsf{v}]$. One proves by induction:

\begin{proposition}
\label{cosums}
For $2\mathsf{g} - 2 + k > 0$, we have
\bea
 && \mathsf{C}^{(\mathsf{g},k)}\bigl[{}^{l_1}_{\varepsilon_1}\,\cdots\,{}^{l_k}_{\varepsilon_k}\bigr] \nonumber \\
& = & \sum_{\substack{\mathcal{G} \in \mathcal{S}^{(\mathsf{g},k)} \\ (\bs{m},\bs{\sigma}) \\ \in {\rm Col}(\mathcal{G};(\bs{l},\bs{\varepsilon}))}} \prod_{\mathsf{v} \in V'(\mathcal{G})} K\bigl[{}^{\bs{m}[\mathsf{v}]}_{\,\bs{\sigma}[\mathsf{v}]}\bigr] \prod_{\mathsf{v} \in V_{t}(\mathcal{G})} \tilde{K}\bigl[{}^{\bs{m}[\mathsf{v}]}_{\,\bs{\sigma}[\mathsf{v}]}\bigr]\prod_{\mathsf{v} \in V_{tt}(\mathcal{G})} \mathsf{C}^{(0,3)}\bigl[{}^{\bs{m}[\mathsf{v}]}_{\,\bs{\sigma}[\mathsf{v}]}\bigr] \prod_{\mathsf{v} \in V_{o}(\mathcal{G})} \mathsf{C}^{(1,1)}\bigr[{}^{m(\mathsf{e}_{\mathsf{v}}^0)}_{\,\sigma(\mathsf{e}_{\mathsf{v}}^{0})}\bigr]. \nonumber
\eea
 \hfill $\Box$
\end{proposition}

\subsubsection{Usual maps with renormalized face weights}

$\mathbf{F}^{(\mathsf{g},k)}|_{n = 0}$ is the generating series of usual triangulations, with weight $g_3$ per triangle. This is different from $\bs{\mathcal{F}}^{(\mathsf{g},k)}$, which is by definition the generating series of usual maps with renormalized face weights \eqref{eq:fixp}, and still depends on $n$. 

Recall that $\mathbf{F}^{(\mathsf{g},k)}$ depends on $n$ in two ways. Firstly, $n$ appears as a proportionality coefficient in $\mathbf{A}(x,y)$ -- see \eqref{rin} -- in the linear functional relation of Theorem~\eqref{2g2m}. Secondly, the linear equation for $\mathbf{F}(x)$ gives two equations determining $\gamma_{\pm}$ as functions of $n$, and this data gives the interval $x \in (\gamma_{-},\gamma_+)$ on which the linear equation for $\mathbf{F}^{(\mathsf{g},k)}$ holds. For $(\mathsf{g},k) \neq (0,1)$, we can disentangle the two dependences in $n$: let us call $n_1$ the variable appearing linearly in the linear equation, and $n_2$ the variable on which $\gamma_{\pm}$ depends. We denote momentarily $\mathbf{F}^{(\mathsf{g},k)}_{n_1,n_2}$ the corresponding generating series. Note that the parametrization $x(v)$ only depends on $n_2$.

The previous remarks show that the generating series of maps in the $O(n)$ model is
$$
\mathbf{F}^{(\mathsf{g},k)} = \mathbf{F}^{(\mathsf{g},k)}_{n_1 = n,n_2 = n},
$$
while the generating series of usual maps with renormalized face weights is
$$
\bs{\mathcal{F}}^{(\mathsf{g},k)} = \mathbf{F}^{(\mathsf{g},k)}_{n_1 = 0,n_2 = n}.
$$
Note however that $\bs{\mathcal{F}}(x) = \mathbf{F}(x)$.

Let us use curly letters to denote the analogue, in the context of usual maps with renormalized face weights, of all quantities defined in the context of maps of the $O(n)$ model. We have
\bea
\overline{\bs{\mathcal{G}}}^{(2)}(v_1,v_2) & = & \frac{1}{4}\bigg[\Upsilon'_{1/2}(v_1 + v_2) - \Upsilon_{1/2}'(v_1 - v_2) - \Upsilon'_{1/2}(-v_1 + v_2) + \Upsilon'_{1/2}(-v_1 - v_2)\bigg] \nonumber \\
&& + \frac{x'(v_1)x'(v_2)}{2(x(v_1) - x(v_2))^2}, \nonumber
\eea
where $\Upsilon_{1/2}$ is a function of the elliptic modulus $\tau$, thus a function of $n_2$. The modified building block is defined as:
\bea
\bs{\mathcal{B}}_{\varepsilon,l}(v) = \partial_{v}^{2l}\bs{\mathcal{B}}_{\varepsilon,0}(v),\qquad \bs{\mathcal{B}}_{\varepsilon,0}(v) = \overline{\bs{\mathcal{G}}}^{(2)}(v,\tau + \varepsilon).
\eea
As the generating series of disks are $\mathbf{F}(x) = \bs{\mathcal{F}}(x)$ and the parametrization $x(v)$ only depends on $n_2$, we have
$$
\Delta_{\varepsilon}\bs{\mathcal{G}}(v) = \Delta_{\varepsilon}\mathbf{G}(v).
$$
The modified recursion coefficients (compare with \eqref{Khoubi}-\eqref{Ktildeou}) are
\bea
\mathcal{K}\bigl[{}^{l}_{\varepsilon}\,{}^{m}_{\sigma}\,{}^{m'}_{\sigma'}\bigr] & = & \Res_{w \rightarrow 0} \frac{-w^{2l + 1}\dd w}{(2l + 1)!\Delta_{\varepsilon}\mathbf{G}(w)}\,\bs{\mathcal{B}}_{\sigma,m}(w + \tau + \varepsilon) \bs{\mathcal{B}}_{\sigma',m'}(-w + \tau + \varepsilon), \nonumber \\
\tilde{\mathcal{K}}\bigl[{}^{l}_{\varepsilon}\,{}^{l'}_{\varepsilon'}\,{}^{m}_{\sigma}\bigr] & = & \frac{-\delta_{\varepsilon,\varepsilon'}}{(2l - 1)!\,(2l')!} \Res_{w \rightarrow 0} \frac{\dd w}{w}\,\frac{w^{2(l + l')}}{\Delta_{\varepsilon}\mathbf{G}(w)}\,\bs{\mathcal{B}}_{m,\sigma}(\tau + \varepsilon + w). \nonumber
\eea
Following the proof of Proposition~\ref{cocor1}, the non-zero modified initial data read:
$$
\boxed{\mathcal{C}^{(0,3)}\big[{}^{l_1}_{\varepsilon_1}\,{}^{l_2}_{\varepsilon_2}\,{}^{l_3}_{\varepsilon_3}\bigr] = -\frac{2\,\delta_{l_1,l_2,l_3,0}\,\delta_{\varepsilon_1,\varepsilon_2,\varepsilon_3}}{y_{\varepsilon_1,1}},
\qquad \mathcal{C}^{(1,1)}\bigl[{}^{l}_{\varepsilon}\bigr] = \delta_{l,0}\Big(\frac{y_{\varepsilon,2}}{24y_{\varepsilon,1}^2} + \frac{\upsilon_{1/2,1}}{y_{\varepsilon,1}}\Big) -\frac{\delta_{l,1}}{24 y_{\varepsilon,1}}.}$$
Compared to the initial conditions for $\mathsf{C}$'s, the only difference is the replacement of $\upsilon_{b,1}$ by $\upsilon_{1/2,1}$ (see \eqref{ububub} for their definition) in $\mathcal{C}^{(1,1)}\bigl[{}^{0}_{\varepsilon}\bigr]$. Then, the analogue of Propositions \ref{2g2mr}-\ref{cosums} is:
\begin{proposition}
\label{cosums2} For $2\mathsf{g} - 2 + k > 0$, we have a decomposition into a finite sum:
$$
\bs{\mathcal{G}}^{(\mathsf{g},k)}(v_1,\ldots,v_k) = \sum_{\substack{l_1,\ldots,l_k \geq 0 \\ \varepsilon_1,\ldots,\varepsilon_k \in \{0,\frac{1}{2}\}}} \mathcal{C}^{(\mathsf{g},k)}\bigl[{}^{l_1}_{\varepsilon_1}\,\cdots\,{}^{l_k}_{\varepsilon_k}\bigr]\prod_{i = 1}^k \bs{\mathcal{B}}_{\varepsilon_i,l_i}(v_i).
$$
The coefficients are given by the unfolded formula:
\bea
&& \mathcal{C}^{(\mathsf{g},k)}\bigl[{}^{l_1}_{\varepsilon_1}\,\cdots\,{}^{l_k}_{\varepsilon_k}\bigr] \nonumber \\
& = & \sum_{\substack{\mathcal{G} \in \mathcal{S}^{(\mathsf{g},k)} \\ \substack{(\bs{m},\bs{\sigma}) \\ \in {\rm Col}(\mathcal{G};(\bs{l},\bs{\varepsilon}))}}} \prod_{\mathsf{v} \in V'(\mathcal{G})} \mathcal{K}\bigl[{}^{\bs{m}[\mathsf{v}]}_{\,\bs{\sigma}[\mathsf{v}]}\bigr] \prod_{\mathsf{v} \in V_{t}(\mathcal{G})} \tilde{\mathcal{K}}\bigl[{}^{\bs{m}[\mathsf{v}]}_{\,\bs{\sigma}[\mathsf{v}]}\bigr]\prod_{\mathsf{v} \in V_{tt}(\mathcal{G})} \mathcal{C}^{(0,3)}\bigl[{}^{\bs{m}[\mathsf{v}]}_{\,\bs{\sigma}[\mathsf{v}]}\bigr] \prod_{\mathsf{v} \in V_{o}(\mathcal{G})} \mathcal{C}^{(1,1)}\bigr[{}^{m(\mathsf{e}_{\mathsf{v}}^0)}_{\,\sigma(\mathsf{e}_{\mathsf{v}}^{0})}\bigr]. \nonumber
\eea
\hfill $\Box$
\end{proposition}

\section{Critical behavior in the bending energy model (disregarding nesting)}
\label{critit}
\label{S5}

\subsection{Phase diagram}
\label{Secphase}
\begin{center}
\begin{figure}[h!]
\includegraphics[width=0.6\textwidth]{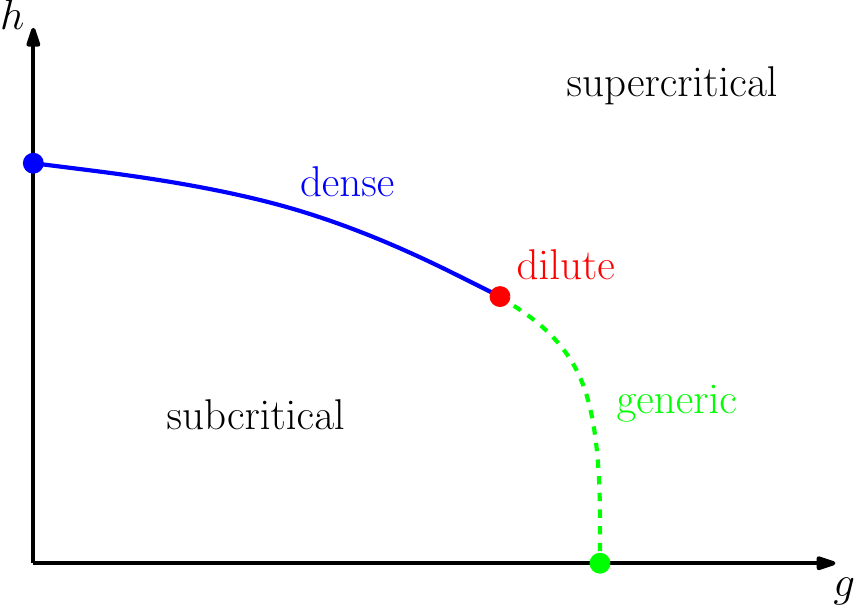}
\caption{\label{Qualiphas} The phase diagram of the model with bending energy is qualitatively insensitive to the value of $n \in (0,2)$ and $\alpha$ not too large.}
\end{figure}
\end{center}

For fixed values $(n,\alpha,g,h)$, we introduce
$$
u_{c} = \sup\{u \geq 0\,:\, F_{\ell}^{\bullet} < \infty\}
$$
in terms of the generating series of pointed disks defined in \eqref{Cyl2fixed}. If $u_{c} = 1$ (resp. $u_{c} < 1$, $u_{c} > 1$), we say that the model is at a critical (resp. subcritical, supercritical) point. At a critical point, the generating series $\bs{\mathcal{F}}(x) = \mathbf{F}(x)$ has a singularity when $u \rightarrow 1^{-}$, and the nature (universality class) of this singularity is characterized by some critical exponents. The phase diagram of the model with bending energy was rigorously determined in \cite{BBG12b,BBD}, and is plotted qualitatively in Figure~\ref{Qualiphas}, see also the early works \cite{KOn,GaudinKostov} for $\alpha = 1$. We now review the precise results obtained in \cite{BBG12b,BBD}.

Three universality classes can be found in the model with bending energy: \emph{generic}, non-generic \emph{dilute} and non-generic \emph{dense}. For $n > 0$, we find a dense critical line, which ends with a dilute critical point, and continues as a generic critical line. For $n = 0$, only the generic critical line remains. As the generic universality class is already present in maps without loops, we will not pursue its study. On the contrary, the non-generic universality class is specific to the loop model, and it corresponds to a regime where macroscopic loops continue to exist in maps of volume $V \rightarrow \infty$ \cite{KOn,BEThese}. The remaining of the text aims at describing our various generating series on the non-generic critical line.

A non-generic critical point occurs when $\gamma_+$ approaches the fixed point of $\varsigma$:
$$
\gamma_+^* = \varsigma(\gamma_+^*) = \frac{1}{h(\alpha + 1)}.
$$
In this limit, the two cuts $\gamma$ and $\varsigma(\gamma)$ merge at $\gamma_+^*$, and one can justify on the basis of combinatorial arguments \cite[Section 6]{BBG12b} that $\gamma_- \rightarrow \gamma_-^*$ with
$$
|\gamma_-^*| < |\gamma_+^*|\qquad {\rm and}\qquad \varsigma(\gamma_-^*) \neq \gamma_-^*.
$$
In terms of the parametrization $x(v)$, it amounts to letting $T \rightarrow 0$, and this is conveniently measured in terms of the parameter
$$
q = e^{-\frac{\pi}{T}} \rightarrow 0.
$$
After establishing the behavior of $x(v)$ and the special function $\Upsilon_{b}(v)$ in this regime (see Appendix~\ref{AppA} for a summary), one can prove:

\begin{theorem}\cite{BBG12b}
\label{thphase}
Assume $\alpha = 1$, and introduce the parameter
$$
\rho = 1 - 2h\gamma_-^* = 1 - \frac{\gamma_-^*}{\gamma_+^*}.
$$
There is a non-generic critical line, parametrized by $\rho \in (\rho_{\min},\rho_{\max}]$:
\bea
\frac{g}{h} & = & \frac{4(\rho b\sqrt{2 + n} - \sqrt{2 - n})}{\rho^2(b^2 - 1)\sqrt{2 - n} + 4\rho b\sqrt{2 + n} - 2\sqrt{2 - n}}, \nonumber \\
h^2 & = & \frac{\rho^2 b}{24\sqrt{4 - n^2}}\,\frac{\rho^2\,b(1 - b^2)\sqrt{2 + n}  - 4\rho\sqrt{2 - n} + 6b\sqrt{2 + n}}{-\rho^2(1 - b^2)\sqrt{2 - n} + 4\rho b\sqrt{2 + n} - 2\sqrt{2 - n}}.
\nonumber 
\eea
It realizes the dense phase of the model. The endpoint 
$$
\rho_{\max} = \frac{1}{b}\,\sqrt{\frac{2 - n}{2 + n}}
$$
corresponds to the fully packed model $g = 0$, with the critical value $h = \frac{1}{2\sqrt{2}\sqrt{2 + n}}$. The endpoint
$$
\rho_{\min} = \frac{\sqrt{6 + n} - \sqrt{2 - n}}{(1 - b)\sqrt{2 + n}}
$$
is a non-generic critical point realizing the dilute phase, and it has coordinates:
\bea
\frac{g}{h} & = & 1 + \sqrt{\frac{2 - n}{6 + n}}, \nonumber \\
h^2 & = &  \frac{b(2 - b)}{3(1-  b^2)(2 + n)}\Big(1 - \frac{1}{4\sqrt{(2 - n)(6 + n)}}\Big). \nonumber
\eea
\end{theorem}
The fact that the non-generic critical line ends at $\rho_{\max} < 2$ is in agreement with $|\gamma_-^*| < |\gamma_+^*|$.

\begin{theorem}\cite{BBG12b}
\label{alphanotlarge} There exists $\alpha_c(n) > 1$ such that, in the model with bending energy $\alpha < \alpha_c(n)$, the qualitatitive conclusions of the previous theorem still hold. For $\alpha = \alpha_c(n)$, only a non-generic critical point in the dilute phase exist, and for $\alpha > \alpha_c(n)$, non-generic critical points do not exist.
\end{theorem}

\begin{theorem}\cite{BBD}
\label{th38} Assume $(g,h)$ are chosen such that the model has a non-generic critical point for vertex weight $u = 1$. When $u < 1$ tends to $1$, we have
$$
q \sim \Big(\frac{1 - u}{q_*}\Big)^{c},
$$
with the universal exponent
$$
c = \left\{\begin{array}{lll} \frac{1}{1 - b} & & {\rm dense}, \\ 1 & & {\rm dilute}. \end{array}\right.
$$
The non-universal constant reads, for $\alpha = 1$:
$$
q_* = \left\{\begin{array}{lll} \frac{6(n + 2)}{b}\,\frac{\rho^2(1 - b)^2\sqrt{2 + n} + 2\rho(1 - b)\sqrt{2 - n} - 2\sqrt{2 + n}}{\rho^2b(1 - b^2)\sqrt{2 + n} - 4\rho(1 - b^2)\sqrt{2 - n} + 6b\sqrt{2 + n}} & & {\rm dense}, \\ \frac{24}{b(1 - b)(2 - b)} & & {\rm dilute}. \end{array}\right. $$
For $\alpha \neq 1$, its expression is much more involved, see \cite[Appendix E]{BBD}.
\end{theorem}

Actually, it is proved in \cite[Appendix~E and~J]{BBD} that:
\begin{lemma}\label{deltaanal}\cite{BBD}
The function $u\mapsto q$ is delta-analytic.
\end{lemma}

\subsection{Principles}
\label{D3crti}

\subsubsection{Small and large boundaries}

The generating series of connected maps of genus $\mathsf{g}$ in the $O(n)$ model with fixed volume $V$ and fixed boundary lengths $\ell_1,\ldots,\ell_k$ reads
$$
[u^{V}] F_{\ell_1,\ldots,\ell_k}^{(\mathsf{g},k)} = \oint \frac{\dd u}{2{\rm i}\pi\,u^{V + 1}} \oint \prod_{i = 1}^k \frac{\dd x_i\,x_i^{\ell_i}}{2{\rm i}\pi} \mathbf{F}^{(\mathsf{g},k)}(x_1,\ldots,x_k).
$$
The contour for integration of $x_i$ is originally around $\infty$ with negative orientation, but we can move it to surround $\gamma$. At a critical point, the asymptotics when $V \rightarrow \infty$ are dominated by the behavior of the generating series at $u = 1$. If we want to keep $\ell_i$ finite, we can leave the contour integral over $x_i$ in a neighborhood of $\infty$, and by setting $x_i = x(\tfrac{1}{2} + \tau w_i)$ we trade it for a contour surrounding $w_i = w_{\infty}^*$. If we want to let $\ell_i \rightarrow \infty$ at a rate controlled by $V \rightarrow \infty$, the asymptotics will be dominated by the behavior of the generating series for $x_i$ near the singularity $\gamma_+ \rightarrow \gamma_+^*$, i.e.~for $x_i = x(\tau w_i)$ with $w_i$ of order $1$. The same principle holds for any of the unrefined generating series $\bs{\mathcal{F}}$ and $\pmb{\mathscr{F}}_{\Gamma}$.

If $\mathbf{H}_s(x_1\ldots,x_k)$ is a refined generating series of maps with $k$ boundaries (with $s$ a Boltzmann weight for certain separating loops), we can compute the number of such maps having fixed volume $V$, fixed number $P$ of such separating loops, and fixed boundary perimeters, by$$
\Big[s^Pu^{V} \prod_{i} x_i^{-(\ell_i + 1)}\Big] \mathbf{H}_s(x_1,\ldots,x_k) = \oint \frac{\dd s}{2{\rm i}\pi}\,\oint \frac{\dd u}{2{\rm i}\pi} \oint \Big[\prod_{i = 1}^k \frac{x_i^{\ell_i}\dd x_i}{2{\rm i}\pi}\Big] \frac{\mathbf{H}_s(x_1,\ldots,x_k)}{s^{P + 1}u^{V + 1}}.$$
In the regime $P,V \rightarrow \infty$, the contour integral over $s$ will be determined by the behavior of the generating series near the dominant singularity in the variable $s$, and $u \rightarrow 1$.

To summarize, we need to study the behavior of generating series approaching criticality, i.e.~$q = e^{-\frac{\pi}{T}}$ with $\tau = {\rm i}T \rightarrow 0$, while $x = x(v)$ with $v = \varepsilon + \tau w$ and $w$ is in a fixed compact. With $\varepsilon = \frac{1}{2}$ we have access to the regime of finite (also called ``small'') boundaries, and with $\varepsilon = 0$ to the regime of large boundaries.

\subsection{Organization of the computations}

In the present Section~\ref{critit}, we will study maps without marked points. The modifications arising to include a number $k' > 0$ of marked points will be discussed in Section~\ref{crititmarked}. We will find, as can be expected, that marked points behave -- as far as critical exponents are concerned -- as small boundaries.

Our first goal is to determine the behavior of the generating series of maps $\mathbf{F}^{(\mathsf{g},k)}$ and of usual maps with renormalized face weights $\bs{\mathcal{F}}^{(\mathsf{g},k)}$. To obtain it, we first determine the behavior of the building blocks of Propositions ~\ref{cosums}-\ref{cosums2} in the next paragraph, and then study the behavior of the sum over colorings and graphs to derive the behavior of $\mathsf{C}^{(\mathsf{g},k)}$ and $\mathcal{C}^{(\mathsf{g},k)}$ (Lemma~\ref{Cbehavior}). This step is rather technical, and the result for the critical exponent for $\mathsf{C}$'s and $\mathcal{C}$'s is not particularly simple. Yet, the final result for the critical behavior of the generating series of maps themselves turns out to be much simpler (Theorem~\ref{ouqusf}). We recall that the $\mathsf{C}$'s do not have a combinatorial interpretation in terms of maps, so this technical part should only be seen as a (necessary) intermediate step to arrive to the $\mathbf{F}$'s and $\bs{\mathcal{F}}$'s.

\subsection{Preliminaries}\subsubsection{Building blocks}

\vspace{-0.7cm}

\label{Sbuildi}
We first examine the behavior at criticality, i.e.~$q = e^{-\frac{\pi}{T}} \rightarrow 0$, of the various bricks appearing in Proposition~\ref{cosums}.  Let us define
$$
\varepsilon,\varepsilon' \in \{0,\tfrac{1}{2}\},\qquad \varepsilon \oplus \varepsilon' \coloneqq \left\{\begin{array}{rcl} 0 & & {\rm if}\,\,\varepsilon = \varepsilon', \\ \tfrac{1}{2} & & {\rm if}\,\,\varepsilon \neq \varepsilon', \end{array}\right.
$$
and for $\varepsilon,\sigma,\sigma'\in \{0,\tfrac{1}{2}\}$
$$
f(\varepsilon,\sigma,\sigma'\vert B) \coloneqq  B\big[(\varepsilon \oplus \sigma) + (\varepsilon \oplus \sigma')\big] + \big(\mathfrak{d}\tfrac{b}{2} - 1\big)(1 - 2\varepsilon),
$$
with $\mathfrak{d} = 1$ in the dense phase, and $\mathfrak{d} = -1$ in the dilute phase. We give its table of values (dense on the left, dilute on the right) for $B=b$:
\begin{center}
\begin{tabular}{|c||c|c|c|}
\hline $\bs{\sigma + \sigma'}$ & $\bs{0}$ & $\bs{\tfrac{1}{2}}$ & $\bs{1}$ \\
\hline\hline
$\bs{\varepsilon = 0}$ & $\tfrac{b}{2} - 1$ & $b - 1$ & $\tfrac{3b}{2} - 1$ \\
\hline
$\bs{\varepsilon = \tfrac{1}{2}}$ & $b$ & $\tfrac{b}{2}$ & $0$ \\
\hline
\end{tabular}
\quad \begin{tabular}{|c||c|c|c|}
\hline $\bs{\sigma + \sigma'}$ & $\bs{0}$ & $\bs{\tfrac{1}{2}}$ & $\bs{1}$ \\
\hline\hline
$\bs{\varepsilon = 0}$ & $-\tfrac{b}{2} - 1$ & $-1$ & $\tfrac{b}{2} - 1$ \\
\hline
$\bs{\varepsilon = \tfrac{1}{2}}$ & $b$ & $\tfrac{b}{2}$ & $0$ \\
\hline
\end{tabular}
\end{center}

\begin{lemma}
\label{pieces}
In the critical regime $\tau = {\rm i}T$ with $T \rightarrow 0^+$, we have for the building blocks of the generating series of maps in the bending energy model
\bea
K\bigl[{}^{l}_{\varepsilon}\,{}^{m}_{\sigma}\,{}^{m'}_{\sigma'}\bigr] & = & \Big(\frac{\pi}{T}\Big)^{2(m + m' - l) + 1}\,q^{f(\varepsilon,\sigma,\sigma'\vert b)}\Big\{K^*\bigl[{}^{l}_{\varepsilon}\,{}^{m}_{\sigma}\,{}^{m'}_{\sigma'}\bigr] + O(q^{b})\Big\}, \nonumber \\
\tilde{K}\bigl[{}^{l}_{\varepsilon}\,{}^{l'}_{\varepsilon}\,{}^{m}_{\sigma}\bigr] & = &  \Big(\frac{\pi}{T}\Big)^{2(m - l - l') - 1}\,q^{f(\varepsilon,\varepsilon,\sigma\vert b)}\Big\{\tilde{K}^*\bigl[{}^{l}_{\varepsilon}\,{}^{m}_{\sigma}\,{}^{m'}_{\sigma'}\bigr] + O(q^{b})\Big\}, \nonumber \\
\mathsf{C}^{(0,3)}\bigl[{}^{0}_{\varepsilon}\,{}^{0}_{\varepsilon}\,{}^{0}_{\varepsilon}\bigr] & = &  \Big(\frac{\pi}{T}\Big)^{-3}\,q^{f(\varepsilon,\varepsilon,\varepsilon\vert b)}\Big\{\mathsf{C}^{(0,3)}_{*}\bigl[{}^{0}_{\varepsilon}\,{}^{0}_{\varepsilon}\,{}^{0}_{\varepsilon}\bigr] + O(q^{b})\Big\}, \nonumber \\
\mathsf{C}^{(1,1)}\bigl[{}^{l}_{\varepsilon}\bigr] & = &  \Big(\frac{\pi}{T}\Big)^{-(2l + 1)}\,q^{f(\varepsilon,\varepsilon,\varepsilon\vert b)}\Big\{\mathsf{C}^{(1,1)}_*\bigl[{}^{l}_{\varepsilon}\bigr] + O(q^{b})\Big\}, \nonumber \\
\mathbf{B}_{\varepsilon,l}(\tau\phi + \varepsilon') & = &  \Big(\frac{\pi}{T}\Big)^{2l + 2}\,q^{b(\varepsilon \oplus \varepsilon')}\Big\{\mathbf{B}_{\varepsilon\oplus \varepsilon',l}^{*,(2l + 1)}(\pi \phi) + O(q^{b})\Big\}. \nonumber 
\eea
And, for the building blocks of the generating series of usual maps with renormalized face weights
\bea
\mathcal{K}\bigl[{}^{l}_{\varepsilon}\,{}^{m}_{\sigma}\,{}^{m'}_{\sigma^{\prime}}\bigr] & = &  \Big(\frac{\pi}{T}\Big)^{2(m + m' - l) + 1}\,q^{f\left(\varepsilon,\sigma,\sigma'\vert \frac{1}{2}\right)}\Big\{\mathcal{K}^*\bigl[{}^{l}_{\varepsilon}\,{}^{m}_{\sigma}\,{}^{m'}_{\sigma'}\bigr] + O(q^{b})\Big\}, \nonumber \\
\tilde{\mathcal{K}}\bigl[{}^{l}_{\varepsilon}\,{}^{l'}_{\varepsilon}\,{}^{m}_{\sigma}\bigr] & = &  \Big(\frac{\pi}{T}\Big)^{2(m - l - l') - 1}\,q^{f\left(\varepsilon,\varepsilon,\sigma\vert \frac{1}{2}\right)}\Big\{\tilde{\mathcal{K}}^*\bigl[{}^{l}_{\varepsilon}\,{}^{m}_{\sigma}\,{}^{m'}_{\sigma'}\bigr] + O(q^{b})\Big\}, \nonumber \\
\mathcal{C}^{(0,3)}\bigl[{}^{0}_{\varepsilon}\,{}^{0}_{\varepsilon}\,{}^{0}_{\varepsilon}\bigr] & = &  \Big(\frac{\pi}{T}\Big)^{-3}\,q^{f\left(\varepsilon,\varepsilon,\varepsilon \vert \frac{1}{2}\right)}\Big\{\mathcal{C}^{(0,3)}_*\bigl[{}^{0}_{\varepsilon}\,{}^{0}_{\varepsilon}\,{}^{0}_{\varepsilon}\bigr] + O(q^{b})\Big\}, \nonumber \\
\mathcal{C}^{(1,1)}\bigl[{}^{l}_{\varepsilon}\bigr] & = &  \Big(\frac{\pi}{T}\Big)^{-(2l + 1)}\,q^{f\left(\varepsilon,\varepsilon,\varepsilon\vert \frac{1}{2}\right)}\Big\{\mathcal{C}^{(1,1)}_*\bigl[{}^{l}_{\varepsilon}\bigr] + O(q^{b})\Big\}, \nonumber \\
\bs{\mathcal{B}}_{\varepsilon,l}(\tau\phi + \varepsilon') & = &  \Big(\frac{\pi}{T}\Big)^{2l + 2}\,q^{\frac{1}{2}(\varepsilon \oplus \varepsilon')}\Big\{\bs{\mathcal{B}}_{\varepsilon\oplus \varepsilon',l}^{*,(2l + 1)}(\pi \phi) + O(q^{\frac{1}{2}})\Big\}. \nonumber 
\eea 
\end{lemma}
We will do many computations just for the $\mathsf{C}^{(\mathsf{g},k)}$'s, but they will work analogously for the $\mathcal{C}^{(\mathsf{g},k)}$'s, just by specializing $b=\frac{1}{2}$ in all the critical exponents of $K, \tilde{K}, \mathsf{C}^{(0,3)}, \mathsf{C}^{(1,1)}$ given by some $f(\varepsilon,\sigma,\sigma'\vert b)$, and $b = \tfrac{1}{2}$ in the exponent of $\mathbf{B}$ to obtain $\bs{\mathcal{B}}$.

The expressions for the leading order coefficients -- here denoted with $*$ -- are provided in Appendix~\ref{AppBB}. They are non-zero and satisfy the same selection rules as the unstarred quantities on the left-hand side. An interesting feature of the result is that, in the formula of Proposition~\ref{cosums} (resp. Proposition~\ref{cosums2}), the contribution to $\mathsf{C}^{(\mathsf{g},k)}$ (resp. $\mathcal{C}^{(\mathsf{g},k)}$) of a colored graph $(\mathcal{G},\bs{\sigma})$ has order of magnitude $q^{f(\mathcal{G},\bs{\sigma})}$ with
$$
f(\mathcal{G},\bs{\sigma}) = \sum_{\mathsf{v} \in V(\mathcal{G})} f(\bs{\sigma}[\mathsf{v}]\,\vert\, B), \text{ with } B=b \text{ (resp. } B=1/2 \text{)}.
$$
We remark that $f(\bs{\sigma}[\mathsf{v}]\,\vert\, B)$ does not depend on the vertex being terminal, bi-terminal, having a loop or not. Since $q = e^{-\frac{\pi}{T}} \rightarrow 0$ when $T \rightarrow 0^+$, the leading term in $\mathsf{C}^{(\mathsf{g},k)}$ and $\mathcal{C}^{(\mathsf{g},k)}$ are given by the colored graphs minimizing $f(\mathcal{G},\bs{\sigma})$. We will study the minimizing graphs and their exponent in Section~\ref{NextS}.

\subsubsection{Minimization over colorings}

\begin{lemma}
\label{lemanun}
For a given graph $\mathcal{G}$ of genus $\mathsf{g}$ with $k$ legs, the coloring assigning $0$ to each edge realizes the minimum of $f(\mathcal{G};\bs{\sigma})$, which is 
$$
(2\mathsf{g} - 2 + k)\left(\mathfrak{d}\tfrac{b}{2}-1\right).
$$
\end{lemma}
\noindent \textbf{Proof.} Every $f(\varepsilon,\sigma,\sigma')$ realizes its minimum $\left(\mathfrak{d}\frac{b}{2}-1\right)$ at $(\varepsilon,\sigma,\sigma')=(0,0,0)$, and the coloring with $\bs{\sigma}[\mathsf{v}] = (0,0,0)$ for all $\mathsf{v} \in V(\mathcal{G})$ receives a non-zero contribution at this order.  \hfill $\Box$

\subsubsection{Study of a critical exponent}

Let $\lfloor x \rfloor$ denote the unique integer such that $\lfloor x \rfloor \leq x < \lfloor x \rfloor + 1$. Let us define
\bea
\beta_1(i_{1/2}) & \coloneqq & \lfloor \tfrac{i_{1/2}}{2}\rfloor+2\delta_{i_{1/2},1}, \nonumber \\
\beta_2(\mathsf{g},k,i_0) & \coloneqq & 2 \mathsf{g} -2 +\lfloor\tfrac{k}{2}\rfloor+\lfloor\tfrac{i_0 + (k \,{\rm mod}\, 2)}{2}\rfloor  .\nonumber\eea
We then define a function of three integers $\mathsf{g},i_0,i_{1/2}$ such that $2\mathsf{g} - 2 + i_0 + i_{1/2} \geq 1$:
\beq
\label{betadef} \beta(\mathsf{g},i_{0},i_{1/2}\vert B) = \left\{\begin{array}{lll} \beta_1(i_{1/2})\tfrac{B}{2}+ \beta_2(\mathsf{g},i_{0}+i_{1/2},i_0)(\mathfrak{d} \tfrac{b}{2}-1) && {\rm if}\,\,\beta_2( \mathsf{g},i_0 + i_{1/2},i_0) > 0, \\ 0 & & {\rm otherwise}. \end{array}\right.
\eeq
It will be useful later to know what happens when we decrement $i_0$ and increment $i_{1/2}$.
\begin{lemma}
\label{betadecret} For $i_0>0$, we have $\beta(\mathsf{g},i_0,i_{1/2}\vert B)+\Delta=\beta(\mathsf{g},i_0-1,i_{1/2}+1\vert B)$, where 
\beq
\Delta =  \left\{\begin{array}{rl}
2 \frac{B}{2} -\left( \mathfrak{d}\frac{b}{2} -1 \right), &\text{ if } i_{1/2}=0, \\
-\frac{B}{2}, &\text{ if } i_{1/2}=1, \\
-\left( \mathfrak{d}\frac{b}{2} -1 \right), &\text{ if } i_{1/2}>0 \text{ even}, \\
\frac{B}{2}, &\text{ if } i_{1/2}>1 \text{ odd},
\end{array} \right.  \nonumber
\eeq
except for the exceptional cases $(\mathsf{g},k)=(0,3)$, $(0,4)$ and $(\mathsf{g},k,i_0)=(1,1,1), (0,5,1)$. In the last cases, we obtain
\beq
\Delta =  \left\{\begin{array}{rl}
 -\left( \mathfrak{d}\frac{b}{2} -1 \right), &\text{ if } (\mathsf{g},k)=(1,1), \, i_0=1, \\
-2\frac{B}{2}-\left( \mathfrak{d}\frac{b}{2} -1 \right), &\text{ if } (\mathsf{g},k)=(0,5), \, i_0=1,
\end{array} \right.  \nonumber
\eeq
and, in the other exceptional cases, where some configurations $(i_0, i_{1/2})$ give $\mathsf{C}^{(\mathsf{g},k)}=0$, we only record the variations between configurations giving non-zero $\mathsf{C}$'s:
\begin{itemize}
\item $\beta(0,3,0) -\left(\mathfrak{d}\frac{b}{2}-1\right)=\beta(0,0,3)=0$,
\item $\beta(0,4,0) +\frac{B}{2}-\left(\mathfrak{d}\frac{b}{2}-1\right)=\beta(0,2,2)$,
\item $\beta(0,2,2)-\frac{B}{2} -\left(\mathfrak{d}\frac{b}{2}-1\right)=\beta(0,0,4)=0$.
\end{itemize}
\end{lemma}
\noindent \textbf{Proof.} The exceptional cases can be easily checked with the expression for $\beta$. For the general situation, we separate cases according to the parity of $i_0$ and $k$, and we check first how $\beta_2(\mathsf{g},i_0+i_{1/2},i_0)$ varies depending on the parity of $i_{1/2}$:
\begin{itemize}
\item If $i_{1/2}$ is even, then $\beta_2(\mathsf{g},k,i_0-1)=\beta_2(\mathsf{g},k,i_0)-1$.
\item If $i_{1/2}$ is odd, then $\beta_2(\mathsf{g},k,i_0-1)=\beta_2(\mathsf{g},k,i_0)$.
\end{itemize}
For the variation of $\beta_1(i_{1/2})$, we distinguish four cases:
\begin{itemize}
\item $\beta_1(1)=\beta_1(0)+2=2$.
\item $\beta_1(2)=\beta_1(1)-1=1$.
\item If $i_{1/2}>0$ is even, then $\beta_1(i_{1/2}+1)=\beta_1(i_{1/2})$.
\item If $i_{1/2}>1$ is odd, then $\beta_1(i_{1/2}+1)=\beta_1(i_{1/2})+1$.
\end{itemize}
\hfill $\Box$

\subsection{Coefficients $\mathsf{C}$ and $\mathcal{C}$}
\label{NextS}

\begin{lemma}
\label{Cbehavior}
Let $\mathsf{g} \geq 0$ and $k \geq 1$ such that $2\mathsf{g} - 2 + k > 0$. Let $\varepsilon_1, \ldots, \varepsilon_k \in \{0,\tfrac{1}{2}\}$ be fixed, and denote $i_0$ (resp. $i_{1/2}$) be the number of $\varepsilon_i = 0$ (resp. $=\tfrac{1}{2}$). Then, in the critical regime $\tau = {\rm i}T$ with $T \rightarrow 0^+$ we obtain
\bea
\mathsf{C}^{(\mathsf{g},k)}\bigl[{}^{l_1}_{\varepsilon_1}\,\cdots\,{}^{l_k}_{\varepsilon_k}\bigr] & = &  \Big(\frac{\pi}{T}\Big)^{- \sum_{i = 1}^k (2l_i + 1)}\,q^{\beta(\mathsf{g},i_{0},i_{1/2}\vert b)}\Big(\mathsf{C}^{(\mathsf{g},k)}_*\bigl[{}^{l_1}_{\varepsilon_1}\,\cdots\,{}^{l_k}_{\varepsilon_k}\bigr] + O(q^{\frac{b}{2}})\Big),\nonumber \\
\mathcal{C}^{(\mathsf{g},k)}\bigl[{}^{l_1}_{\varepsilon_1}\,\cdots\,{}^{l_k}_{\varepsilon_k}\bigr] & = &  \Big(\frac{\pi}{T}\Big)^{-\sum_{i = 1}^k (2l_i + 1)}\,q^{\beta\left(\mathsf{g},i_{0},i_{1/2}\vert \frac{1}{2}\right)}\Big(\mathcal{C}^{(\mathsf{g},k)}_*\bigl[{}^{l_1}_{\varepsilon_1}\,\cdots\,{}^{l_k}_{\varepsilon_k}\bigr] + O(q^{\frac{b}{2}})\Big), \nonumber
\eea
where the leading coefficients indicated with $*$ are non-zero.
\end{lemma}
 
\noindent\textbf{Proof.} We shall do the reasoning for $\mathsf{C}^{(\mathsf{g},k)}$, i.e.~for $B=b$, but all the comparisons we do will work also for the special case of $B=\frac{1}{2}$, so the final scaling exponent will be the same for $\mathcal{C}^{(\mathsf{g},k)}$ specifying $B=\frac{1}{2}$, instead of $B=b$. For simplicity, we will write $\beta(\mathsf{g},i_0,i_{1/2})\equiv\beta(\mathsf{g},i_0,i_{1/2}\vert b)$ in this proof. The determination of the exponent of $\tfrac{\pi}{T}$ will be addressed in the third part of the proof. For the moment, we only focus on the powers of $q$. Since we know $\mathsf{C}^{(\mathsf{g},k)}\bigl[{}^{l_1}_{\varepsilon_1}\,\cdots\,{}^{l_k}_{\varepsilon_k}\bigr]$ is invariant by permutation of the pairs $(l_i,\varepsilon_i)_{i = 1}^k$, the scaling exponent will only depend on $\mathsf{g}$, $i_0$ and $i_{1/2}$. In the case $i_{1/2} = 0$, we have:
$$
\beta(\mathsf{g},k,0)=\left(2 \mathsf{g} -2 +\lfloor\tfrac{k}{2}\rfloor+\lfloor\tfrac{k + (k \,{\rm mod}\, 2)}{2}\rfloor\right)\left(\mathfrak{d} \tfrac{b}{2}-1\right) = (2 \mathsf{g} -2 +k)\left(\mathfrak{d} \tfrac{b}{2}-1\right),
$$ 
so the claim is correct according to Lemma \ref{lemanun}.

We prove all the other cases by induction on $2 \mathsf{g} -2 +k$, starting by the two base cases with $2 \mathsf{g} -2 +k=1$. In both base cases there is only one graph with a single vertex.

\smallskip

\noindent $\bullet$ $(\mathsf{g},k)=(0,3)$. Remember $\mathsf{C}^{(0,3)}\big[{}^{l_1}_{\varepsilon_1}\,{}^{l_2}_{\varepsilon_2}\,{}^{l_3}_{\varepsilon_3}\bigr] = 0$ in case we do not have $\varepsilon_1=\varepsilon_2=\varepsilon_3$. So the only case to consider is $i_{1/2}=3$ and we have $f(\frac{1}{2},\frac{1}{2},\frac{1}{2})=0$, which is equal to $\beta(0,0,3)$ since $\beta_2(0,3,0)=-1 < 0$.

\smallskip

\noindent $\bullet$ $(\mathsf{g},k)=(1,1)$. Remember the color of a loop should be identical to the color of the other edge. So in the case $i_{1/2}=1$, we get $f(\frac{1}{2},\frac{1}{2},\frac{1}{2})=0$, which is equal to $\beta(1,0,1)$ since $\beta_2(1,1,0)=0$.

\smallskip

Now we will prove the result for cases with $2 \mathsf{g} -2 +k$, supposing it is true for all cases $( \overline{\mathsf{g}},\overline{k})$ with $2 \overline{\mathsf{g}} -2 +\overline{k} <2 \mathsf{g} -2 +k$. We can decompose graphs $\mathcal{G}\in\mathcal{S}^{(\mathsf{g},k)}$ in terms of a graph $\mathcal{P}$ which consists of only one trivalent vertex $\mathsf{v}_0$ without loops, and either one graph $\tilde{\mathcal{G}}\in\mathcal{S}^{(\mathsf{g} - 1,k+1)}$, or two graphs $\mathcal{G}'\in \mathcal{S}^{(\mathsf{g}',k'+1)}$ and $\mathcal{G}''\in\mathcal{S}^{(\mathsf{g}'',k''+1)}$, with $\mathsf{g}' + \mathsf{g}'' = \mathsf{g}$ and $k'+k''=k-1$, excluding the cases $(\mathsf{g}',k')=(0,0)$ and $(\mathsf{g}'',k'')=(0,0)$.

The two last legs of $\mathcal{P}$ are shared either with two legs of $\tilde{\mathcal{G}}$, or with one in $\mathcal{G}'$ and one in $\mathcal{G''}$. Consider the following decompositions $\tilde{k}=\tilde{i}_0+\tilde{i}_{1/2}$, $k'=i_0'+i_{1/2}'$ and $k''=i_0''+i_{1/2}''$, with $\tilde{k}+2=k+1$ and $(k'+1)+(k''+1)=k+1$, where $\tilde{k},k'$ and $k''$ correspond to the number of legs which are not shared with $\mathcal{P}$ in the respective subgraphs $\tilde{\mathcal{G}}$, $\mathcal{G}'$ and $\mathcal{G''}$.

In order to extend a coloring for the corresponding subgraph $\tilde{\mathcal{G}}$, or $\mathcal{G}'$ and $\mathcal{G''}$ to a coloring of the whole $\mathcal{G}$, we will pick $\bs{\sigma}[\mathsf{v}_0]=(\sigma_0,\sigma_1,\sigma_2)$ in a compatible way, i.e.~the colorings $\sigma_1$ and $\sigma_2$ of the two legs of $\mathcal{P}$ which are shared with the corresponding subgraphs will coincide with the given ones for these legs on the subgraphs. We will make these choices to minimize $f(\mathcal{G};\bs{\sigma})$, which will be $f(\bs{\sigma}[\mathsf{v}_0])+\sum_{\mathsf{v}\in V(\tilde{\mathcal{G}})}f(\tilde{\bs{\sigma}}[\mathsf{v}])$ or $f(\bs{\sigma}[\mathsf{v}_0])+\sum_{\mathsf{v}\in V(\mathcal{G}')}f(\bs{\sigma}'[\mathsf{v}])+\sum_{\mathsf{v}\in V(\mathcal{G}'')}f(\bs{\sigma}''[\mathsf{v}])$.

For every configuration $i_0+i_{1/2}=k$, we will first build a graph with a coloring which is compatible with the fixed colorings of the legs $(\mathcal{G},\bs{\sigma})\in \mathcal{S}^{(\mathsf{g},k)}\times\mathsf{Col}(\mathcal{G};(\bs{l},\bs{\varepsilon}))$ from the ones from previous induction steps such that $f(\mathcal{G};\bs{\sigma})=\beta(\mathsf{g}, i_0,i_{1/2})$, i.e.~a graph realizing the desired value. Secondly, we will have to prove that there is no other $(\overline{\mathcal{G}},\overline{\bs{\sigma}})\in \mathcal{S}^{(\mathsf{g},k)}\times\mathsf{Col}(\mathcal{G};(\bs{l},\bs{\varepsilon}))$ with $f(\overline{\mathcal{G}};\overline{\bs{\sigma}})<\beta(\mathsf{g}, i_0,i_{1/2})$, i.e.~$\beta(\mathsf{g}, i_0,i_{1/2})$ is actually the minimum.

Remember that the cases with $i_0=k$ and $i_{1/2}=0$ were already checked, so we do not consider them in the following.

\bigskip

\begin{center}
\textit{First part: special cases}
\end{center}

\bigskip

We will deal first with the two special cases $(\mathsf{g},k)=(0,4), (0,5)$.

\medskip

\noindent $\bullet$ $(\mathsf{g},k)=(0,4)$. 
The graphs in $\mathcal{S}^{(0,4)}$ have only two vertices, one terminal and one bi-terminal. This implies that the only options with $\mathsf{C}^{(0,4)}\big[{}^{l_1}_{\varepsilon_1}\,\cdots\,{}^{l_4}_{\varepsilon_4}\bigr]\neq 0$ are $i_0=0, 2, 4$.
We show in Figure~\ref{fig:graphs(0,4)} the graphs with a suitable coloring which realize the desired value in every remaining case. 
\begin{center}
\begin{figure}[h!]
\includegraphics[width=.6\textwidth]{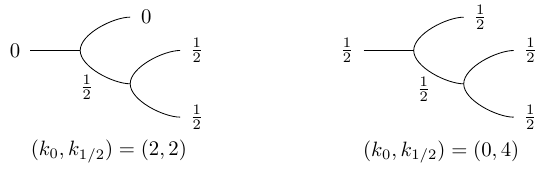}
  \caption{\label{fig:graphs(0,4)} $(\mathsf{g},k)=(0,4)$} 
\end{figure}
\end{center}
Observe that $\mathcal{G}''$ is the only graph in $\mathcal{S}^{(0,3)}$. Since $i_{1/2}''=2$ and the only vertex is biterminal, we have to set $\sigma_2=\tfrac{1}{2}$ here. We already checked that $f(\mathcal{G}'';\bs{\sigma}'')=\beta(0,0,3)=0$. Therefore, for $(i_0,i_{1/2})=(2,2)$ we obtain $f(\mathcal{G};\bs{\sigma})= f(0,0,\frac{1}{2}) = \mathfrak{d}\tfrac{b}{2} -1 + \tfrac{b}{2}=\beta(0,2,2)$ and for $(i_0,i_{1/2})=(0,4)$, $f(\mathcal{G};\bs{\sigma})=f(\frac{1}{2},\frac{1}{2},\frac{1}{2})=0= \beta(0,0,4)$, as we wanted.

\medskip

\noindent $\bullet$ $(\mathsf{g},k)=(0,5)$. For every possible $(i_0,i_{1/2})$ we choose the graph with the corresponding coloring shown in Figure~\ref{fig:graphs(0,5)}:

\begin{center}
\begin{figure}[h!]
\includegraphics[width=.9\textwidth]{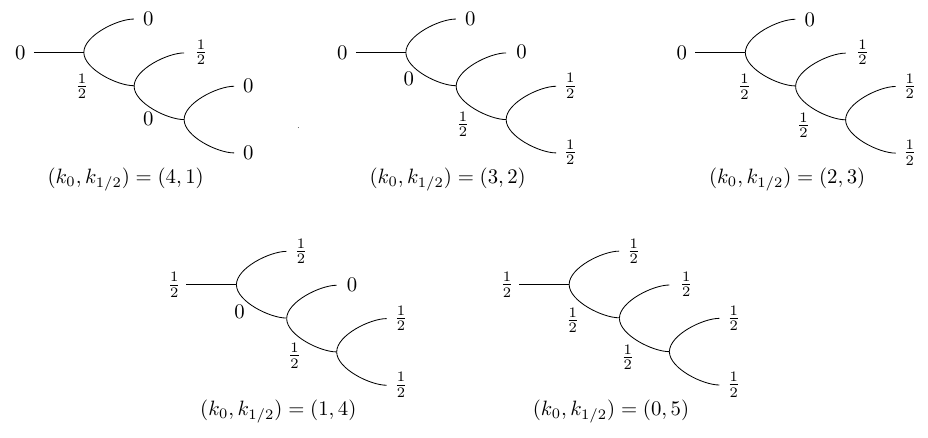}
  \caption{\label{fig:graphs(0,5)} $(\mathsf{g},k)=(0,5)$} 
\end{figure}
\end{center}
Observe that $\mathcal{G}'' \in \mathcal{S}^{(0,4)}$, which also makes the choice of $\sigma_2$ special. 

\smallskip 

\noindent $\diamond$ If $i_0=4$ and $i_{1/2}=1$, with the chosen graph we can only set $\sigma_2= \tfrac{1}{2}$. By induction hypothesis, we have $f(\mathcal{G}'';\bs{\sigma}'')=\beta(0,2,2)=\frac{b}{2}+\mathfrak{d}\frac{b}{2}-1$. Therefore
$$
f(\mathcal{G};\bs{\sigma})= f(0,0,\tfrac{1}{2}) + \beta(0,2,2) = (\mathfrak{d} + 1)\tfrac{b}{2}-1+(\mathfrak{d} + 1)\tfrac{b}{2} -1 = 2\tfrac{b}{2} + 2\left(\mathfrak{d}\tfrac{b}{2}-1\right)= \beta(0,4,1).
$$

\smallskip

\noindent $\diamond$ If $i_0=3$ and $i_{1/2}=2$, we can choose $\sigma_2=0$. By induction hypothesis, we have $f(\mathcal{G}'';\bs{\sigma}'')=\beta(0,2,2)=\mathfrak{d}\tfrac{b}{2} -1 + \tfrac{b}{2}$. Therefore$$
f(\mathcal{G};\bs{\sigma})= f(0,0,0) + \beta(0,2,2) = \mathfrak{d}\tfrac{b}{2} -1+ \tfrac{b}{2} +\mathfrak{d}\tfrac{b}{2}-1 = \beta(0,3,2).
$$

\smallskip

\noindent $\diamond$ If $i_0=2$ and $i_{1/2}=3$, we can only choose $\sigma_2=\frac{1}{2}$. By induction hypothesis, we have $f(\mathcal{G}'';\bs{\sigma}'')=\beta(0,0,4)=0$. Therefore
$$
f(\mathcal{G};\bs{\sigma})= f(0,0,\tfrac{1}{2}) = (\mathfrak{d} + 1)\tfrac{b}{2}-1= \beta(0,2,3).
$$

\smallskip

\noindent $\diamond$ If $i_0=1$ and $i_{1/2}=4$, we can only choose $\sigma_2=0$. Therefore $f(\mathcal{G};\bs{\sigma})= f(\frac{1}{2},\frac{1}{2},0)+ \beta(0,2,2)= \frac{b}{2}+\frac{b}{2}+\mathfrak{d}\frac{b}{2}-1= \beta(0,1,4)$.

\smallskip

\noindent $\diamond$ If $i_0=0$ and $i_{1/2}=5$, we can only choose $\sigma_2= \frac{1}{2}$. Thus $f(\mathcal{G};\bs{\sigma})= f(\frac{1}{2},\frac{1}{2},\frac{1}{2})+ \beta(0,0,4)= 0= \beta(0,0,5)$.

\bigskip 

\begin{center}
\textit{First part: general cases}
\end{center}

\bigskip

For the general cases with $(\mathsf{g},k)\neq (0,4), (0,5)$ we will consider four cases. If $k>2$, we will be automatically in one of the first two cases.

\noindent \textbf{Case I: $\mathbf{i_0 \geq 2}$.} We will choose the graph $\mathcal{G}$ constructed from $\mathcal{G}'$ and $\mathcal{G}''$, with $k'=i_0'=1$ and $\mathsf{g}=0$. Observe that $\mathcal{G}'$ has $2$ legs, both with coloring $0$.
In this case, $\mathsf{v}_0$ is a terminal vertex, so for the contribution to be non-zero, we have $\sigma_0=\sigma_1$ and we know that $\sigma_1=0$ because the leg is shared with $\mathcal{G}'$. Note that $i_0=i_0''+2$ and $i_{1/2}=i_{1/2}''$.
\begin{center}
\begin{figure}[h!]
\includegraphics[width=.4\textwidth]{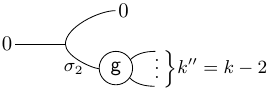} 
\caption{\label{fig:11} $i_0 \geq 2$} 
\end{figure}
\end{center}
In the general case, we can always choose $\sigma_2=0$. By the induction hypothesis, we can choose $(\mathcal{G}'',\bs{\sigma}'')$ such that $f(\mathcal{G}'';\bs{\sigma}'')= \beta(\mathsf{g},i_0''+1,i_{1/2}'')$. Therefore 
\bea
f(\mathcal{G};\bs{\sigma})
 & =& \mathfrak{d}\tfrac{b}{2}-1 + \beta_1(i_{1/2}'')\tfrac{b}{2} + \beta_2(\mathsf{g}, k''+1, i_0''+1)\left(\mathfrak{d}\tfrac{b}{2}-1\right) \nonumber\\
 &=& \beta_1(i_{1/2})\tfrac{b}{2} + (\beta_2(\mathsf{g}, k-1, i_0-1)+1)\left(\mathfrak{d}\tfrac{b}{2}-1\right)=\beta(\mathsf{g},i_0,i_{1/2}). \nonumber
\eea
The last step is a simple computation separating the cases where $k-1$ is even and odd.

\medskip

\noindent  \textbf{Case II: $\mathbf{i_{1/2}\geq 2}$.} Again we choose the graph $\mathcal{G}$ constructed from $\mathcal{G}'$ and $\mathcal{G}''$, with $k'=1$ and $\mathsf{g}=0$, but with $i'_{1/2}=1$ because in this case we have no assumption on $i_0$. And again $\mathsf{v}_0$ is a terminal vertex, so for the contribution to be non-zero, we have $\sigma_0=\sigma_1$, but here $\sigma_1= \frac{1}{2}$. Note that $i_0=i_0''$ and $i_{1/2}=i_{1/2}''+2$. 
\begin{center}
\begin{figure}[h!]
\includegraphics[width=.38\textwidth]{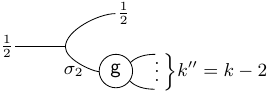}  
\caption{\label{fig:12} $i_{1/2}\geq 2$} 
\end{figure}
\end{center}
It will minimize to choose $\sigma_2=\frac{1}{2}$, if $i_{1/2}''=1$, and $\sigma_2=0$, otherwise. If $i_{1/2}''=1$, we have
\bea
f(\mathcal{G};\bs{\sigma})
 & =& 0 + \beta_1(2)\tfrac{b}{2} + \beta_2(\mathsf{g}, k''+1, i_0'')\left(\mathfrak{d}\tfrac{b}{2}-1\right)\nonumber\\
 &=& \beta_1(i_{1/2})\tfrac{b}{2} + \beta_2(\mathsf{g}, k-1, i_0)\left(\mathfrak{d}\tfrac{b}{2}-1\right)=\beta(\mathsf{g},i_0,i_{1/2}). \nonumber
\eea
In the last step we separate the cases where $k-1$ is even and odd, and we use $i_0+3=k$ to deduce the parity of $i_0$ in every case. If $i_{1/2}''\neq 1$, we have:
\bea
f(\mathcal{G};\bs{\sigma}) & =& \tfrac{b}{2} + \beta_1(i_{1/2}'')\tfrac{b}{2} + \beta_2(\mathsf{g}, k''+1, i_0''+1)\left(\mathfrak{d}\tfrac{b}{2}-1\right) \nonumber\\
 &=& \beta_1(i_{1/2}''+2)\tfrac{b}{2} + \beta_2(\mathsf{g}, k-1, i_0+1)\left(\mathfrak{d}\tfrac{b}{2}-1\right)=\beta(\mathsf{g},i_0,i_{1/2}). \nonumber
\eea
The last step is again a simple computation separating cases according to the parity of $k-1$.

\medskip

\noindent \textbf{Case III: $\mathbf{i_0=1, i_{1/2}=1}$.}  This is the remaining case of $k=2$. Observe that here $\mathsf{g}>0$ so that $2\mathsf{g}-2+2>0$. We distinguish two cases:

\smallskip

\noindent $\bullet$ $\mathsf{g} = 1$. We choose the graph $\mathcal{G}$ constructed from $\tilde{\mathcal{G}}$ with $\tilde{i}=\tilde{i}_{1/2}=1$. Since the vertex of $\tilde{\mathcal{G}}$ is terminal in $\mathcal{G}$, at least $\sigma_1$ or $\sigma_2$ should be $\frac{1}{2}$ for the contribution of the graph to be non-zero. Actually if we set $\sigma_1=\sigma_2=\frac{1}{2}$, we get $f(\mathcal{G};\bs{\sigma})=f(0,\frac{1}{2},\frac{1}{2})+f\left(\frac{1}{2},\frac{1}{2},\frac{1}{2}\right)= \mathfrak{d}\tfrac{b}{2} - 1 +  b = \beta(1,1,1)$.

\begin{center}
\begin{figure}[h!]
\includegraphics[width=.3\textwidth]{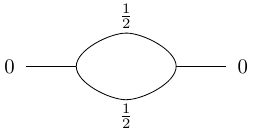}
\caption{\label{fig:(1,2)131} $(\mathsf{g},i_0,i_{1/2}) = (1,1,1)$}    \end{figure}
\end{center}

\smallskip

\noindent $\bullet$ $\mathsf{g} > 1$. We build $\mathcal{G}$ from $\mathcal{G}'$ and $\mathcal{G}''$ with $k'=i_{1/2}'=1$, $\mathsf{g}'>0$, $k''=0$ and $\mathsf{g}''> 0$, which we can choose because $\mathsf{g}>1$. Observe that if $\overline{\mathsf{g}}>0$, then $2 \overline{\mathsf{g}} -2+1 >0$ and hence in our cases we will have $\beta_2 >0$. Since $i_0', i_0'' =0$ and $i_0=1$, $\sigma_0=0$ and we can choose $\sigma_1=\sigma_2=0$.
\begin{center}
\begin{figure}[h!]
\includegraphics[width=.25\textwidth]{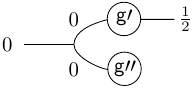}
\caption{\label{fig:132} $\mathsf{g}>1, (i_0,i_{1/2}) = (1,1)$ } 
\end{figure}
\end{center}
Therefore 
\bea
f(\mathcal{G};\bs{\sigma}) &=& \mathfrak{d}\tfrac{b}{2}-1 + \beta_1(1)\tfrac{b}{2}+\beta_2(\mathsf{g}',2,1)\left(\mathfrak{d}\tfrac{b}{2}-1\right) + \beta_1(0)\tfrac{b}{2}+\beta_2(\mathsf{g}'',1,1)\left(\mathfrak{d}\tfrac{b}{2}-1\right) \nonumber \\
&=& b + (2\mathsf{g}-2+1)\left(\mathfrak{d}\tfrac{b}{2}-1\right) = \beta(\mathsf{g},1,1). \nonumber
\eea

\medskip

\noindent \textbf{Case IV: $\mathbf{k=1}$.} The case $(1,1)$ was already a base one, so here we suppose $\mathsf{g}>1$. We consider the case $k=i_{1/2}=1$. So $\sigma_0=\varepsilon_1=\frac{1}{2}$ and we construct $\mathcal{G}$ from $\mathcal{G}'$ and $\mathcal{G}''$ with $k'=k''=0$ and $\mathsf{g}', \mathsf{g}'' >0$ . We can choose $\sigma_1=\sigma_2=0$.
\begin{center}
\begin{figure}[h!]
\includegraphics[width=.2\textwidth]{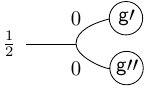}
\caption{\label{fig:14} $k=1$} 
\end{figure}
\end{center}
Therefore 
\bea
f(\mathcal{G};\bs{\sigma}) &=& b +\beta_1(0)\tfrac{b}{2}+\beta_2(\mathsf{g}',1,1)\left(\mathfrak{d}\tfrac{b}{2}-1\right) + \beta_1(0)\tfrac{b}{2}+\beta_2(\mathsf{g}'',1,1)\left(\mathfrak{d}\tfrac{b}{2}-1\right) \nonumber \\
&=& b + (2\mathsf{g}-2)\left(\mathfrak{d}\tfrac{b}{2}-1\right) = \beta(\mathsf{g},0,1). \nonumber
\eea

\medskip

\begin{center}
\textit{Second part: disconnected cases}
\end{center}

\medskip

For the second part of the proof, we will check that all other possible graphs and colorings for every case do not give a smaller exponent. We first discuss the disconnected case, i.e.~the case where $\mathcal{G}$ is constructed from $\mathcal{G}'$ and $\mathcal{G}''$  so that $\mathcal{G}', \mathcal{G}''\not\in\mathcal{S}^{(0,2)}$. The cases with $\mathcal{G}'$ or $\mathcal{G}''$ in $\mathcal{S}^{(0,3)}$ or $\mathcal{S}^{(0,4)}$ will be considered apart because they will have some extra restrictions to choose $\sigma_1$ and $\sigma_2$ and will be called the exceptional cases in this part.
\begin{center}
\begin{figure}[h!]
\includegraphics[width=.26\textwidth]{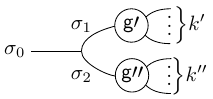}
\caption{\label{fig:21} $\mathcal{G}$ from $\mathcal{G}'$ and $\mathcal{G}''$} 
\end{figure}
\end{center}

Moreover, remember that a graph in $\mathcal{S}^{(0,5)}$ and with $i_0=0, i_{1/2}=5$ was also giving the special value of $\beta_2(0,5,0)=0$ and hence $\beta(0,0,5)=0$ automatically. However, observe that when one of the pieces $\mathcal{G}'$ or $\mathcal{G}''$ is in $\mathcal{S}^{(0,5)}$, we will not have any exceptional situation here because a graph in $\mathcal{S}^{(0,5)}$ with $i_{1/2}'=4$ and $\sigma_1=\tfrac{1}{2}$, or $i_{1/2}''=4$ and $\sigma_2=\tfrac{1}{2}$ will never be chosen to minimize; it will always be better to choose $\sigma_1$ or $\sigma_2$ to be $0$, which in this case is possible.

\smallskip

\noindent \textbf{Case} $\mathbf{\sigma_0=0}$. Let us check that choosing $\sigma_1=0$, if possible, always minimizes. Observe that $f(0,0,\sigma_2)=f\left(0,\frac{1}{2},\sigma_2\right)-\frac{b}{2}$. Then, making use of Lemma~\ref{betadecret}, we get\beq
f(0,0,\sigma_2)  +\beta(\mathsf{g}',i_0'+1,i_{1/2}') \leq f(0,\tfrac{1}{2},\sigma_2)+\beta(\mathsf{g}',i_0',i_{1/2}'+1). \nonumber
\eeq
Indeed, it is clear if $i_{1/2}'\neq 1$, and if $i_{1/2}'= 1$, we always have an equality because
\beq
f(0,0,\sigma_2)  +\beta(\mathsf{g}',i_0'+1,1) = f(0,\tfrac{1}{2},\sigma_2)-\tfrac{b}{2}+\beta(\mathsf{g}',i_0',2)+\tfrac{b}{2}. \nonumber
\eeq
The same argument works for $\sigma_2$. Now, we should check that the exceptional cases, where we cannot choose $\sigma_1=\sigma_2=0$, do not minimize further.

\smallskip

\noindent $\bullet$ If $\sigma_1=\tfrac{1}{2}, \sigma_2=0, (\mathsf{g}',k'+1)=(0,3), i_{1/2}'=2$, we have
$$
f(\mathcal{G};\bs{\sigma}) = \big((\mathfrak{d} + 1)\tfrac{b}{2} - 1\big) + 0 + \beta(\mathsf{g},i_0,i_{1/2} - 1) = \delta_{i_{1/2},3}\,b + \beta(\mathsf{g},i_0,i_{1/2}),
$$
where we have used that here $i_{1/2}\geq 2$.

\smallskip

\noindent $\bullet$ If $\sigma_1=\frac{1}{2}, \sigma_2=0, (\mathsf{g}',k'+1)=(0,4), i_{1/2}'=1$, we have
\bea
f(\mathcal{G};\bs{\sigma})
& = &(\mathfrak{d} + 1)\tfrac{b}{2}-1 + \tfrac{b}{2}+\left(\mathfrak{d}\tfrac{b}{2}-1\right) + \beta_1(i_{1/2}-1)\tfrac{b}{2}+\beta_2(\mathsf{g},k-4,i_0-2)\left(\mathfrak{d}\tfrac{b}{2}-1\right) \nonumber \\
&=& (\beta_1(i_{1/2}-1)+2)\tfrac{b}{2}+(\beta_2(\mathsf{g},k-4,i_0-2)+2)\left(\mathfrak{d}\tfrac{b}{2}-1\right) \nonumber \\
&=& (\beta_1(i_{1/2}-1)+2)\tfrac{b}{2}+(\beta_2(\mathsf{g},k,i_0)-1)\left(\mathfrak{d}\tfrac{b}{2}-1\right) \geq \beta(\mathsf{g},i_0,i_{1/2}). \nonumber
\eea
In the last step we have used that $\beta_1(i_{1/2}-1)+2>\beta(i_{1/2})$.\smallskip

\noindent $\bullet$ If $\sigma_1=\frac{1}{2}, \sigma_2=0, (\mathsf{g}',k'+1)=(0,4), i_{1/2}'=3$, we have
\bea
f(\mathcal{G};\bs{\sigma})
&=&\big(\mathfrak{d}\tfrac{b}{2}-1 + \tfrac{b}{2}\big)+ 0 + \beta_1(i_{1/2}-3)\tfrac{b}{2}+\beta_2(\mathsf{g},k-4,i_0)\left(\mathfrak{d}\tfrac{b}{2}-1\right) \nonumber \\
&=& (\beta_1(i_{1/2}-3)+1)\tfrac{b}{2}+(\beta_2(\mathsf{g},k-4,i_0)+1)\left(\mathfrak{d}\tfrac{b}{2}-1\right) \nonumber \\
&=& (\beta_1(i_{1/2}-3)+1)\tfrac{b}{2}+(\beta_2(\mathsf{g},k,i_0)-1)\left(\mathfrak{d}\tfrac{b}{2}-1\right)  \nonumber \\
&\geq& (\beta_1(i_{1/2})-1)\tfrac{b}{2}+(\beta_2(\mathsf{g},k,i_0)-1)\left(\mathfrak{d}\tfrac{b}{2}-1\right) \nonumber \\
&=& \beta(\mathsf{g},i_0,i_{1/2}) -\big(\mathfrak{d}\tfrac{b}{2}-1 + \tfrac{b}{2}\big) \geq \beta(\mathsf{g},i_0,i_{1/2}). \nonumber
\eea

\smallskip

\noindent $\bullet$ The remaining cases with $\sigma_1=\frac{1}{2}, \sigma_2=\frac{1}{2}$ consist of $\mathcal{G}', \mathcal{G}'' \in\mathcal{S}^{(0,4)}$; $\mathcal{G}'\in\mathcal{S}^{(0,3)}$ and $\mathcal{G}'\in\mathcal{S}^{(0,4)}$, and symmetric ones by exchanging the role of $\sigma_1$ and $\sigma_2$. They can be checked easily from the results for the base cases $(0,3)$ and $(0,4)$. 

\medskip
Choosing $\sigma_1=\sigma_2=0$ for the non-exceptional cases, we obtain:
\bea
f(\mathcal{G};\bs{\sigma})&=& \big(\mathfrak{d}\tfrac{b}{2} - 1\big) + \beta(\mathsf{g}',i_0'+1,i_{1/2}')+\beta(\mathsf{g}-\mathsf{g}',i_0''+1,i_{1/2}'')\nonumber \\ 
&=&(\beta_1(i_{1/2}')+\beta_1(i_{1/2}''))\tfrac{b}{2} \nonumber \\
& & + \big(\beta_2(\mathsf{g}',k'+1,i_{0}'+1)+\beta_2(\mathsf{g}'',k''+1,i_{0}''+1)+1\big)\left(\mathfrak{d}\tfrac{b}{2}-1\right).\nonumber
\eea
On the one hand, separating cases according to the parity of $k'$ and $k''$, and the parity of $i_0'$ and $i_0''$, we check that
\bea
\beta_2(\mathsf{g}',k'+1,i_{0}'+1)+\beta_2(\mathsf{g}'',k''+1,i_{0}''+1)+1 \nonumber \\
\label{ineq} \leq \beta_2(\mathsf{g}'+\mathsf{g}'',k'+k''+1,i_0'+i_0''+1)=\beta_2(\mathsf{g},k,i_0) 
\eea
and hence with this part we cannot minimize further.

On the other hand, distinguishing cases according to the parity of $i'_{1/2}$, $i_{1/2}''$ and $i_{1/2}=i'_{1/2}+i_{1/2}''$, and considering the special cases with some of them equal to $1$, we see that $\beta_1(i_{1/2}')+\beta_1(i_{1/2}'')=\beta_1(i_{1/2})-1$, if both $i_{1/2}'$ and $i_{1/2}''$ are odd and $>1$, and $\beta_1(i_{1/2}')+\beta_1(i_{1/2}'')\geq \beta_1(i_{1/2})$, otherwise.

Finally, we check easily that in the case of odd $i_{1/2}',i_{1/2}''>1$, where we have minimized $\beta_1$ by $1$, we lie in the cases with 
$\beta_2(\mathsf{g}',k'+1,k_{0}'+1)+\beta_2(\mathsf{g}'',k''+1,k_{0}''+1)+1 <\beta_2(\mathsf{g},k,i_0)$. Therefore, we also do not minimize globally, i.e.
\beq
f(\mathcal{G};\bs{\sigma}) \geq \beta(\mathsf{g},i_0,i_{1/2}), \nonumber
\eeq
because $-\frac{b}{2} >\mathfrak{d}\frac{b}{2}-1$ and thus, with a minimizing purpose, we prefer $\beta_2+1$ to $\beta_1-1$.

\medskip

\noindent \textbf{Case $\mathbf{\sigma_0=\frac{1}{2}}$.} As in the previous cases, it can be checked first that the exceptional cases do not minimize further. Let us check now which $\sigma_1$ we should choose to minimize, making use of Lemma~\ref{betadecret}.
\beq
f(\tfrac{1}{2},0,\sigma_2)+\beta_1(\mathsf{g}',i_0'+1,i_{1/2}')=f(\tfrac{1}{2},\tfrac{1}{2},\sigma_2)+ \tfrac{b}{2}+\beta_1(\mathsf{g}',i_0',i_{1/2}'+1)-\Delta. \nonumber
\eeq
If $i_{1/2}'\neq 1$, $\Delta \geq \frac{b}{2}$ and hence $\sigma_1=0$ minimizes. But, if $i_{1/2}'= 1$, $\Delta = -\frac{b}{2}$ and hence $\sigma_1=\frac{1}{2}$ is the minimizing choice. By the symmetry of the situation, the same argument works for the choice of $\sigma_2$ depending on $i_{1/2}''$.

\smallskip

\noindent $\bullet$ $i_{1/2}'=i_{1/2}''=1$. Using the inequality \eqref{ineq}, we have
\bea
f(\mathcal{G};\bs{\sigma})&=& 0 +\beta(\mathsf{g}',i_0',i_{1/2}'+1)+\beta(\mathsf{g}'',i_0',i_{1/2}'+1)\nonumber \\ 
&=& \beta_1(2) b +(\beta_2(\mathsf{g}',k'+1,i_{0}')+\beta_2(\mathsf{g}'',k''+1,i_{0}''))\left(\mathfrak{d}\tfrac{b}{2}-1\right) \nonumber \\
& \geq & \beta_1(3)\tfrac{b}{2}+(\beta_2(\mathsf{g},k,i_0'+i_0''-1)-1)\left(\mathfrak{d}\tfrac{b}{2}-1\right)  \nonumber\\
& \geq & \beta_1(3)\tfrac{b}{2}+\beta_2(\mathsf{g},k,i_0'+i_0'')\left(\mathfrak{d}\tfrac{b}{2}-1\right) = \beta(\mathsf{g},i_0,i_{1/2}). \nonumber
\eea

\smallskip

\noindent $\bullet$ $i_{1/2}'=1, i_{1/2}''\neq 1$ (and the analogous case $i_{1/2}'\neq1, i_{1/2}''=1$). Again using \eqref{ineq}, we obtain
\bea
f(\mathcal{G};\bs{\sigma})&=& \tfrac{b}{2} +\beta(\mathsf{g}',i_0',i_{1/2}'+1)+\beta(\mathsf{g}'',i_0''+1,i_{1/2}'')\nonumber \\ 
&=&\tfrac{b}{2}+(\beta_1(2)+\beta_1(i_{1/2}''))\tfrac{b}{2}+(\beta_2(\mathsf{g}',k'+1,i_{0}')+\beta_2(\mathsf{g}'',k''+1,i_{0}''+1))\left(\mathfrak{d}\tfrac{b}{2}-1\right) \nonumber \\
& \geq &\tfrac{b}{2}+ \beta_1(i_{1/2}''+2)\tfrac{b}{2}+(\beta_2(\mathsf{g},k,i_0'+i_0'')-1)\left(\mathfrak{d}\tfrac{b}{2}-1\right) \nonumber\\
& \geq & \beta_1(i_{1/2})\tfrac{b}{2}+\beta_2(\mathsf{g},k,i_0'+i_0'')\left(\mathfrak{d}\tfrac{b}{2}-1\right) = \beta(\mathsf{g},i_0,i_{1/2}). \nonumber
\eea

\smallskip

\noindent $\bullet$ $i_{1/2}'\neq 1, i_{1/2}''\neq 1$.
\bea
f(\mathcal{G};\bs{\sigma})&=& b +\beta(\mathsf{g}',i_0'+1,i_{1/2}')+\beta(\mathsf{g}'',i_0''+1,i_{1/2}'') \nonumber \\
&=& b+(\beta_1(i_{1/2}')+\beta_1(i_{1/2}''))\tfrac{b}{2} \nonumber \\
&& + (\beta_2(\mathsf{g}',k'+1,i_{0}'+1)+\beta_2(\mathsf{g}'',k''+1,i_{0}''+1))\left(\mathfrak{d}\tfrac{b}{2}-1\right) \nonumber \\
& \geq &(2+\beta_1(i_{1/2}')+\beta_1(i_{1/2}''))\tfrac{b}{2}+(\beta_2(\mathsf{g},k,i_0'+i_0''+1)-1)\left(\mathfrak{d}\tfrac{b}{2}-1\right) \nonumber \\
& \geq & \beta_1(i_{1/2}'+i_{1/2}''+1)\tfrac{b}{2}+(\beta_2(\mathsf{g},k,i_0'+i_0'')+1-1)\left(\mathfrak{d}\tfrac{b}{2}-1\right) = \beta(\mathsf{g},i_0,i_{1/2}). \nonumber
\eea
	
\medskip

\begin{center}
\textit{Second part: connected case}
\end{center}

\medskip

Now let us examine the case in which $\mathcal{G}$ is constructed from $\tilde{\mathcal{G}}$. Firstly it can be easily checked apart that special cases with $\tilde{\mathcal{G}}\in \mathcal{S}^{(0,3)}, \mathcal{S}^{(0,4)}$ do not minimize further.

\medskip

\begin{center}
\begin{figure}[h!]
\includegraphics[width=.26\textwidth]{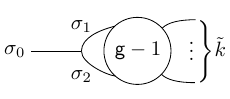}
\caption{\label{fig:22}  $\mathcal{G}$ from $\tilde{\mathcal{G}}$} 
\end{figure}
\end{center}
\noindent \textbf{Case $\mathbf{\sigma_0=0}$.} When we are not in the exceptional cases, we can always choose $\sigma_1=\sigma_2=0$ to minimize.
\bea
f(\mathcal{G};\bs{\sigma})&=& \big(\mathfrak{d}\tfrac{b}{2} - 1\big) +\beta_1(\tilde{i}_{1/2})\tfrac{b}{2}+\beta_2(\tilde{\mathsf{g}},\tilde{k}+2,\tilde{i}_0+2)\left(\mathfrak{d}\tfrac{b}{2}-1\right)\nonumber \\ 
&=&\left(\mathfrak{d}\tfrac{b}{2}-1\right)+\beta_1(i_{1/2})\tfrac{b}{2}+\beta_2(\mathsf{g}-1,k+1,i_0+1)\left(\mathfrak{d}\tfrac{b}{2}-1\right) \nonumber \\
&=&  \beta_1(i_{1/2})\tfrac{b}{2}+(\beta_2(\mathsf{g},k+1,i_0+1)-1)\left(\mathfrak{d}\tfrac{b}{2}-1\right)= \beta(\mathsf{g},i_0,i_{1/2}), \nonumber
\eea
where for the last computation we distinguish cases according to the parity of $k$.

\medskip

\noindent \textbf{Case $\mathbf{\sigma_0=\frac{1}{2}}$.} By the symmetry argument at the beginning of the proof, the only case remaining to be checked is the one corresponding to $i_0=0$ and $i_{1/2}=k$. By a computation similar to the one in the previous case with $\sigma_0=\frac{1}{2}$, we get that if $\tilde{i}_{1/2}=1$, $(\sigma_1,\sigma_2)=(\frac{1}{2},0)$ or $(\sigma_1,\sigma_2)=(0,\frac{1}{2})$ are the minimizing choices and if if $\tilde{i}_{1/2}\neq 1$, then we choose $(\sigma_1,\sigma_2)=(0,0)$ to minimize.

\smallskip

\noindent $\bullet$ $\tilde{i}_{1/2}=1\,\, (k=i_{1/2}=2)$.
\bea
f(\mathcal{G};\bs{\sigma})&=& \tfrac{b}{2} +\beta_1(2)\tfrac{b}{2}+\beta_2(\mathsf{g}-1,3,1)\left(\mathfrak{d}\tfrac{b}{2}-1\right)\nonumber \\ &=&\tfrac{b}{2}+\beta_1(2)\tfrac{b}{2}+(\beta_2(\mathsf{g},2,0)-1)\left(\mathfrak{d}\tfrac{b}{2}-1\right) \nonumber \\
&\geq&\beta_1(2)\tfrac{b}{2}+\beta_2(\mathsf{g},2,0)\left(\mathfrak{d}\tfrac{b}{2}-1\right)= \beta(\mathsf{g},0,2). \nonumber \nonumber
\eea

\smallskip

\noindent $\bullet$ $\tilde{i}_{1/2}\neq 1$.
\bea
f(\mathcal{G};\bs{\sigma})&=& b +\beta_1(\tilde{i}_{1/2})\tfrac{b}{2}+\beta_2(\mathsf{g}-1,k+1,2)\left(\mathfrak{d}\tfrac{b}{2}-1\right)\nonumber \\ 
&=&b+\beta_1(i_{1/2}-1)\tfrac{b}{2}+\beta_2(\mathsf{g}-1,k+1,2)\left(\mathfrak{d}\tfrac{b}{2}-1\right)\left(\mathfrak{d}\tfrac{b}{2}-1\right) \nonumber \\
&\geq&\beta_1(i_{1/2})\tfrac{b}{2}+\beta_2(\mathsf{g},k,0)\left(\mathfrak{d}\tfrac{b}{2}-1\right)= \beta(\mathsf{g},0,i_{1/2}), \nonumber \nonumber
\eea
where the last inequality is simple to check distinguishing the usual cases.

This exhausts all possible graphs and shows that the colored graphs constructed in the \textit{first part} for each $(\mathsf{g},i_0,i_{1/2})$ achieve the minimal value for the exponent, and this value is given by $\beta(\mathsf{g},i_0,i_{1/2})$ of \eqref{betadef}.

\medskip

\begin{center}
\textit{Third part}
\end{center}

\medskip

We show by induction on $\chi = 2\mathsf{g} - 2 + k \geq 1$ that $\mathsf{C}^{(\mathsf{g},k)}\bigl[{}^{l_1}_{\varepsilon_1}\,\cdots\,{}^{l_k}_{\varepsilon_k}\bigr]$ receives a power $(\frac{\pi}{T})^{- \sum_{i = 1}^k (2\ell_i + 1)}$ as prefactor. It is already correct for $(\mathsf{g},k) = (0,3)$ and $(1,1)$ according to Lemma~\ref{pieces}. If it is true for all $(\overline{\mathsf{g}},\overline{k})$ such that $2\overline{\mathsf{g}} - 2 + \overline{k} < \chi$,  then one easily checks with the recursive formula of Proposition~\ref{cocor1}, the behavior of $K$ and $\tilde{K}$, and the induction hypothesis that it continues to holds for all $(\mathsf{g},k)$ such that $2\mathsf{g} - 2 + k = \chi$.\medskip

Together with the identification of the leading power of $q$ in the previous steps, this concludes the proof for the critical behavior of $\mathsf{C}^{(\mathsf{g},k)}$. The arguments are identical for $\mathcal{C}^{(\mathsf{g},k)}$.

\hfill $\Box$

\subsection{Generating series of maps}
\label{SGNAP}

We arrive to the final result for the generating series of maps in the $O(n)$ model.
\begin{theorem}
\label{ouqusf} Let $k= k_0+k_{1/2} \geq 1$ and $\mathsf{g} \geq 0$ such that $2\mathsf{g} - 2 + k > 0$. Let $x_j = x(\tfrac{1}{2} + \tau \phi_j)$ for $j \in \{1,\ldots,k_{1/2}\}$, i.e.~$x_j$ remains finite and away from $[\gamma_-^*,\gamma_+^*]$. Let $y_j = x(\tau \psi_j)$
 for $j \in \{1,\ldots,k_0\}$, i.e.~$y_j$ scales with $q \rightarrow 0$ such that $y_j - \gamma_+ \in O(q^{\frac{1}{2}})$. Then, we have in the critical regime $q \rightarrow 0$:
\bea
&& \mathbf{F}^{(\mathsf{g},k)}(x_1,\ldots,x_{k_{1/2}},y_1,\ldots,y_{k_0}) \nonumber \\
& = & \Big(\frac{\pi}{T}\Big)^{k}\,q^{(2\mathsf{g} - 2 + k)(\mathfrak{d}\frac{b}{2}-1)-\frac{k}{2} + \frac{b + 1}{2}k_{1/2}}\Big( \mathbf{F}^{(\mathsf{g},k)}_{*}(\phi_1,\ldots,\phi_{k_{1/2}},\psi_1,\ldots,\psi_{k_0}) + O(q^{\frac{b}{2}})\Big), \nonumber
\eea
and for the generating series of usual maps with renormalized face weights:
\bea
&& \bs{\mathcal{F}}^{(\mathsf{g},k)}(x_1,\ldots,x_{k_{1/2}},y_1,\ldots,y_{k_0}) \nonumber \\
& = & \Big(\frac{\pi}{T}\Big)^{k}\,q^{\widetilde{\beta}(\mathsf{g},k,k_{1/2})}\Big(\bs{\mathcal{F}}^{(\mathsf{g},k)}_{*}(\phi_1,\ldots,\phi_{k_{1/2}},\psi_1,\ldots,\psi_{k_0}) + O(q^{\frac{b}{2}})\Big), \nonumber 
\eea
with $\widetilde{\beta}(\mathsf{g},k,k_{1/2})=(2\mathsf{g} - 2 + k)(\mathfrak{d}\frac{b}{2}-1)-\frac{k}{2} + \frac{3}{4}k_{1/2}$. Recall that $\mathfrak{d} = 1$ in the dense phase and $\mathfrak{d} = -1$ in the dilute phase.
\end{theorem}
The result for $(\mathsf{g},k) = (0,2)$ is much easier to derive: this is done in Corollary~\ref{dgfsgg} below, and the outcome is that Theorem~\ref{ouqusf} is still valid for $(\mathsf{g},k) = (0,2)$. Remark that in this case, the first term in the critical exponent vanishes so the result is the same in the dense and dilute phase -- only the relation between $u$ and $q$ differ, according to Theorem~\ref{th38}.

\vspace{0.2cm}

\noindent \textbf{Proof.} First, we study the critical behavior of $\mathbf{G}^{(\mathsf{g},k)}(v_1,\ldots,v_k)$. From the decomposition of Proposition~\ref{2g2mr}, the critical behavior for its coefficients $\mathsf{C}^{(\mathsf{g},k)}$  from Lemma~\ref{Cbehavior} and the asymptotic behavior for $\mathbf{B}_{\varepsilon,l}(v)$ in the two regimes $v = \varepsilon' + \tau\phi$ with $\varepsilon'=0,\frac{1}{2}$ given in Lemma \ref{pieces}, it follows that each summand with $\varepsilon_1,\ldots,\varepsilon_k$ fixed behaves like $q^{\bar{\beta}(\mathsf{g},i_0,i_{1/2},j_0,j_{1/2}\vert b)}$, with
\beq
\label{bbar} \bar{\beta}(\mathsf{g},i_0,i_{1/2},j_0,j_{1/2}\vert b)=\beta(\mathsf{g},i_0,i_{1/2}\vert b) + (j_0+j_{1/2})\tfrac{b}{2}, 
\eeq
where $j_0+j_{1/2}=|\{j\in \{1,\ldots,k\} \,\,:\,\, \varepsilon_j \neq \varepsilon'_j\}|$ and, more concretely, 
\bea
j_{1/2} &\coloneqq &|\{ j\in \{1,\ldots,k_{1/2}\} \,\,:\,\, 0=\varepsilon_j \neq \varepsilon'_j=\tfrac{1}{2}\}|, \nonumber \\
j_0 &\coloneqq &|\{ j\in \{k_{1/2}+1,\ldots,k_{1/2}+k_0\} \,\,:\,\, \tfrac{1}{2}=\varepsilon_j \neq \varepsilon'_j=0\}|. \nonumber
\eea
Since we are interested in the dominant behavior of $\mathbf{G}^{(\mathsf{g},k)}\big((v_i)_{i = 1}^k\big)$ with fixed $k_0$ and $k_{1/2}$, we need to decide which $0\leq j_0\leq k_0$ and $0\leq j_{1/2}\leq k_{1/2}$ minimize $\bar{\beta}(\mathsf{g},i_0,i_{1/2},j_0,j_{1/2}\vert b)$. Observe that $i_0=k_0-j_0+j_{1/2}$ and $i_{1/2}=k_{1/2}-j_{1/2}+j_0$. We have to take into account the already known behavior of $\beta(\mathsf{g},i_0,i_{1/2}\vert b)$ varying $i_0$ and $i_{1/2}$ to find in the end the quadruple $(i_0,i_{1/2},j_0,j_{1/2})$ minimizing $\bar{\beta}(\mathsf{g},i_0,i_{1/2},j_0,j_{1/2}\vert b)$ for fixed $\mathsf{g}, k_0$ and $k_{1/2}$.

We will consider first the special base cases, where some configurations $(i_0, i_{1/2})$ give vanishing $\mathsf{C}$'s. 

\smallskip

\noindent $\bullet$ $(\mathsf{g},k) = (0,3)$.  Since the only configurations with $\mathsf{C}^{(\mathsf{g},k)}\neq 0$ are $(i_0, i_{1/2})=(3,0)$ and $(i_0, i_{1/2})=(0,3)$, we automatically have $(j_0, j_{1/2})=(0,k_{1/2})$ and $(j_0, j_{1/2})=(k_{0},0)$, respectively.
\beq
\bar{\beta}(0,3,0,0,k_{1/2}\vert b)=\beta(0,3,0\vert b)+k_{1/2}\tfrac{b}{2} \leq 0 \leq \beta(0,0,3\vert b)+k_{0}\tfrac{b}{2} = \bar{\beta}(0,0,3,k_0,0\vert b). \nonumber\eeq

\smallskip

\noindent $\bullet$ $(\mathsf{g},k) = (0,4)$. First, similarly to the previous case, we have
\beq
\bar{\beta}(0,4,0,0,k_{1/2}\vert b)=\beta(0,4,0\vert b)+k_{1/2}\tfrac{b}{2}  \leq 0 \leq \beta(0,0,4\vert b)+k_{0}\tfrac{b}{2} = \bar{\beta}(0,0,4,k_0,0\vert b). \nonumber
\eeq
The only possibility remaining to compare is $(i_0, i_{1/2})=(2,2)$. Here we use $\beta(0,2,2\vert b)=\beta(0,4,0\vert b)+1+(1-\mathfrak{d})\frac{b}{2}$ from Lemma ~\ref{betadecret}.
\begin{itemize}
\item[$\diamond$] $\bar{\beta}(0,2,2,2,k_{1/2}\vert b)=\beta(0,4,0\vert b)+1+(1-\mathfrak{d})\frac{b}{2}+(2+k_{1/2})\tfrac{b}{2} \geq \bar{\beta}(0,4,0,0,k_{1/2}\vert b)$.
\item[$\diamond$] $\bar{\beta}(0,2,2,1,k_{1/2}-1\vert b)=\beta(0,4,0\vert b)+1+(1-\mathfrak{d})\frac{b}{2}+k_{1/2}\tfrac{b}{2}\geq \bar{\beta}(0,4,0,0,k_{1/2}\vert b)$.
\item[$\diamond$] $\bar{\beta}(0,2,2,0,k_{1/2}-2\vert b)=\beta(0,4,0\vert b)+1+(1-\mathfrak{d})\frac{b}{2}+(k_{1/2}-2)\tfrac{b}{2}\geq \bar{\beta}(0,4,0,0,k_{1/2}\vert b)$.
\end{itemize}

\smallskip

Using again Lemma~\ref{betadecret}, observe that in all the remaining cases, for $i_0>0$, we have
$$
\beta(\mathsf{g},i_0,i_{1/2}\vert b)+\Delta=\beta(\mathsf{g},i_0-1,i_{1/2}+1\vert b),$$
with $\Delta \geq \frac{b}{2}$, except for $i_{1/2}=1$.

\smallskip

\noindent $\bullet$ We now justify that it is always better to decrease $j_0$. If $i_{1/2}\neq 1$,
\bea
\bar{\beta}(\mathsf{g},i_0-1,i_{1/2}+1,j_0,j_{1/2}\vert b)& = & \beta(\mathsf{g},i_0-1,i_{1/2}+1\vert b) + (j_0+j_{1/2})\tfrac{b}{2}\nonumber \\
& = & \beta(\mathsf{g},i_0,i_{1/2}\vert b)+\Delta + (j_0+j_{1/2})\tfrac{b}{2} \nonumber \\ 
& \geq &\beta(\mathsf{g},i_0,i_{1/2}\vert b) + (j_0+j_{1/2}-1)\tfrac{b}{2}\nonumber \\
&=& \bar{\beta}(\mathsf{g},i_0,i_{1/2},j_0-1,j_{1/2}\vert b). \nonumber
\eea
The equality still holds for $i_{1/2}=1$, more concretely
\beq
\bar{\beta}(\mathsf{g},i_0-1,2,j_0,j_{1/2}\vert b) =  \beta(\mathsf{g},i_0,1\vert b)-\tfrac{b}{2} + (j_0+j_{1/2})\tfrac{b}{2} = \bar{\beta}(\mathsf{g},i_0,1,j_0-1,j_{1/2}\vert b). \nonumber
\eeq

\smallskip

\noindent $\bullet$ Now, if $i_{1/2}\neq 1$, it is also better to increase $j_{1/2}$:
\bea
\bar{\beta}(\mathsf{g},i_0-1,i_{1/2}+1,j_0,j_{1/2}\vert b)& = & \beta(\mathsf{g},i_0,i_{1/2}\vert b)+\Delta + (j_0+j_{1/2})\tfrac{b}{2} \nonumber \\ 
& \geq &\beta(\mathsf{g},i_0,i_{1/2}) + (j_0+j_{1/2}+1\vert b)\tfrac{b}{2}\nonumber \\
&=& \bar{\beta}(\mathsf{g},i_0,i_{1/2},j_0,j_{1/2}+1\vert b). \nonumber
\eea
And if $i_{1/2}=1$, it is better to increase $j_{1/2}$ by $2$:
\bea
\bar{\beta}(\mathsf{g},k-2,2,j_0,j_{1/2}\vert b)& = & \beta(\mathsf{g},k,0\vert b)-\tfrac{b}{2}-\left(\mathfrak{d}\tfrac{b}{2}-1\right) + (j_0+j_{1/2})\tfrac{b}{2} \nonumber \\ 
& \geq &\beta(\mathsf{g},k,0) + b + (j_0+j_{1/2}\vert b)\tfrac{b}{2}\nonumber \\&=& \bar{\beta}(\mathsf{g},k,0,j_0,j_{1/2}+2\vert b). \nonumber
\eea
Observe that in the key case $i_{1/2}=1$, if we have $j_{1/2}=k_{1/2}-1$, i.e.~not the maximum but with no possibility of being increased by $2$, we will always have $j_0=1$ $(i_{1/2}+1=2)$ and if we decrease that before we will not lie in the case $i_{1/2}=1$ anymore. So this pathological case is not a real problem.

\smallskip

Therefore, the minimal exponent corresponds to the minimum $j_0$ and the maximum $j_{1/2}$, i.e.~$j_0=0$ and $j_{1/2}=k_{1/2}$, and $i_0=k$ and $i_{1/2}=0$:
\beq
\bar{\beta}(\mathsf{g},k,0,0,k_{1/2}\vert b)=\beta(\mathsf{g},k,0\vert b) + k_{1/2}\tfrac{b}{2}=(2\mathsf{g} - 2 + k)\left(\mathfrak{d}\tfrac{b}{2}-1\right)+ k_{1/2}\tfrac{b}{2}. \nonumber
\eeq
The final result follows from
$$
\mathbf{F}^{(\mathsf{g},k)}(x(v_1),\ldots,x(v_k)) \Big[\prod_{i  = 1}^k x'(v_i)\Big]= \mathbf{G}^{(\mathsf{g},k)}(v_1,\ldots,v_k),
$$
the critical behavior we just found for $\mathbf{G}^{(\mathsf{g},k)}(v_1,\ldots,v_k)$ and the asymptotic behavior for $x(v)$ in the two regimes $v = \tau\phi$ and $v = \tfrac{1}{2} + \tau\phi$ given in Appendix~\ref{App1}. The resulting power of $q$ is
\bea
&& (2\mathsf{g} - 2 + k)\left(\mathfrak{d}\tfrac{b}{2}-1\right)+k_{1/2}\tfrac{b}{2}-\tfrac{1}{2} k_0 \nonumber \\
& = & (2\mathsf{g} - 2 + k_0 + k_{1/2})\left(\mathfrak{d}\tfrac{b}{2}-1\right)-\tfrac{k_0 + k_{1/2}}{2} + k_{1/2}\,\tfrac{b + 1}{2}. \nonumber
\eea

For the $\bs{\mathcal{F}}$'s, the only differences compared to \eqref{bbar} are the factor $\frac{1}{4}$ instead of $\frac{b}{2}$ in the total exponent for fixed $j_0,j_{1/2}$ and $B=\frac{1}{2}$ instead of $b$ in $\beta$:
$$
\bar{\beta}(\mathsf{g},i_0,i_{1/2},j_0,j_{1/2}\vert \tfrac{1}{2})=\beta(\mathsf{g},i_0,i_{1/2}\vert\tfrac{1}{2}) + (j_0+j_{1/2})\tfrac{1}{4}.
$$
This is a particular case of the previous analysis, so the minimum of this exponent is again reached when $j_0 = 0$ and $j_{1/2} = k_{1/2}$, and $i_0=k$ and $i_{1/2}=0$, and this entails the claim. Since in this case, $\beta(\mathsf{g},k,0\vert b)=\beta(\mathsf{g},k,0\vert\frac{1}{2})$, in the end only the first difference matters.
\hfill $\Box$
\vspace{0.2cm}

\section{Critical behavior of nestings in the bending energy model}

\label{S6}

\subsection{Summary of strategy}

Our first goal here is to determine the behavior of the generating series $\pmb{\mathscr{F}}$ of maps realizing a given nesting graph $\Gamma$, without remembering  the arm lengths -- i.e.~setting $s(\mathsf{e}) = 1$ -- and in absence of marked points. For this purpose, we perform a saddle point analysis of the expression of Proposition~\ref{P212} using the previous results on the behavior of $\bs{\mathcal{F}}$, and of the generating series of cuffed cylinders $\hat{\mathbf{F}}_s^{(2)}$ and $\tilde{\mathbf{F}}_{s}^{(2)}$ in Section~\ref{armsS}. The final result is Theorem~\ref{CoscrF} below. The second goal is to extend these computations to the refined generating series $\pmb{\mathscr{F}}_{\Gamma,\star,\mathbf{s}}^{(\mathsf{g},k)}$ of maps realizing a given nesting graph. Here, we just need to repeat the computations of our first goal in presence of the variable $s$, which roughly amounts to replacing $b$ by $b(s)$ when necessary. The only important difference is that we wish to extract the leading contribution containing the dominant singularity in the variable $s$, and this sometimes brings some modification to the hierarchy of dominant terms. The result is described in Section~\ref{CoscrFs}.

In Section~\ref{Fixedas} we convert the critical behavior of all those generating series into asymptotics for fixed large volume $V$ and fixed boundary perimeters $(L_i)_i$ in the regime of small or large boundaries. In Section~\ref{Fixedasarms}, we also examine the critical behavior in this setting of the probability of having fixed arm lengths $P(\mathsf{e})$ tending to $\infty$ at rate $\ln V$ -- which naturally appears from the analysis. In particular, we compute the large deviation function for the arm lengths.

Finally, in Section~\ref{crititmarked}, we show that all these results continue to be valid in presence of marked points, provided one treats each marked point as a small boundary.

\subsection{Cylinders and cuffed cylinders}

\label{armsS}

In order to derive the critical behavior of $\pmb{\mathscr{F}}_{\Gamma,\star,\mathbf{s}}^{(\mathsf{g},k)}$, we need one more ingredient, namely the critical behaviors of $\tilde{\mathbf{F}}_{s}^{(2)}(x_1,x_2)$ and $\hat{\mathbf{F}}_{s}^{(2)}(x_1,x_2)$.

For this purpose, we first derive the critical behavior of $\mathbf{G}_{s}^{(2)}$ in the various regimes, which can be straightforwardly obtained using the expression in Proposition~\ref{p15} together with the asymptotic behavior of the special function $\Upsilon_b$ in Lemma~\ref{lemUp} in Appendix.

\begin{lemma}\label{behaviorG2}
Set $v_i = \varepsilon_i + \tau w_i$ for $\varepsilon_i \in \{0,\tfrac{1}{2}\}$. In the limit $q \rightarrow 0$, we have
\bea
\mathbf{G}_{s}^{(2)}(v_1,v_2) & = & \frac{(\frac{\pi}{T})^2}{4 - n^2s^2}\,\frac{q^{(\varepsilon_1 \oplus \varepsilon_2)b(s)}}{1 - q^{b(s)}} \nonumber \\
&& \times \left\{\begin{array}{lll} H_{b(s),0}(w_1,w_2) - q^{b(s)}H_{b(s) + 2,0}(w_1,w_2) + O(q^{2 - b(s)}) & & {\rm if}\,\,\varepsilon_1 = \varepsilon_2, \\ H_{b(s),\frac{1}{2}}(w_1,w_2) - q^{1 - b(s)}H_{b(s) - 2,\frac{1}{2}}(w_1,w_2) + O(q) && {\rm if}\,\,\varepsilon_1 \neq \varepsilon_2, \end{array}\right. \nonumber
\eea
where
\bea
H_{b,0}(w_1,w_2) & = & (b - 1)\Big(\frac{\sin \pi(b - 1)(w_1 + w_2)}{\sin \pi(w_1 + w_2)} - \frac{\sin \pi(b - 1)(w_1 - w_2)}{\sin \pi(w_1 - w_2)}\Big) \nonumber \\
&& + \frac{\cos \pi(w_1 + w_2)\cos\pi (b - 1)(w_1 + w_2)}{\sin^2 \pi(w_1 + w_2)} \nonumber \\
&& - \frac{\cos\pi (w_1 - w_2) \cos \pi(b - 1)(w_1 - w_2)}{\sin^2\pi(w_1 - w_2)}, \nonumber \\
H_{b,\frac{1}{2}}(w_1,w_2) & = & 8b\,\sin\pi b w_1\,\sin\pi b w_2. \nonumber
\eea
The errors are uniform for $w_1,w_2$ in any compact and stable under differentiation.
\end{lemma}

The first consequence of this Lemma is the critical behavior of the ``singular part'' of $\mathbf{F}_{s}^{(2)}(x_1,x_2)$ with respect to the variables $u$ and $(x_1,x_2)$, which will be used in Theorem~\ref{LAPB} to obtain the asymptotics of the cylinder generating series for fixed large volumes and fixed boundary perimeters. We warn the reader about two subtleties in this analysis regarding what we mean by this ``singular part''. $\mathbf{F}_{s}^{(2)}$ is directly expressed in terms of $\mathbf{G}_{s}^{(2)}$ in Proposition~\ref{p15} up to a shift term. This shift term can actually be dropped as far as fixing boundary perimeter is concerned, as it gives a zero contribution when performing contour integrations of the form $\oint \frac{\dd x_1\,x_1^{L_1}}{2{\rm i}\pi}\,\frac{\dd x_2\,x_2^{L_2}}{2{\rm i}\pi}\,\mathbf{F}_{s}^{(2)}(x_1,x_2)$. Powers $q^0$ should also be dropped from this ``singular term'' as they disappear in contour integrals $\oint \frac{\dd u}{2{\rm i}\pi u^{V + 1}}\mathbf{F}_{s}^{(2)}$ used to fix the volume; in such a case, the next-to-leading order will play the leading role in the computations for fixed volume. Taking these subtleties into account, the result for this ``singular part'' of $\mathbf{F}_{s}^{(2)}$ straightforwardly follows from Lemma~\ref{behaviorG2} and the behavior of $x(v)$ given in Lemma~\ref{LemB3} in Appendix:

\begin{corollary}
\label{dgfsgg} Set $x_i = x(\varepsilon_i + \tau\phi_i)$ for $\varepsilon_i \in \{0,\tfrac{1}{2}\}$. In the limit $q \rightarrow 0$, the singular parts (for this we use the sign $\equiv$) with respect to the variables $u$, $x_1,x_2$ of the cylinder generating series are
\bea 
\mathbf{F}^{(2)}_{s}(x_1,x_2) & \equiv & \frac{(\frac{\pi}{T})^{2}\,q^{\widetilde{\beta}^{(0,2)}(s,\varepsilon_1,\varepsilon_2)}}{1 - q^{b(s)}}\big\{\mathbf{F}^{(2)}_{s\,*}(\phi_1,\phi_2) + q^{b(s)}\mathbf{F}^{(2)}_{s\,**}(\phi_1,\phi_2) + O(q^{\min(2 - b(s),2b(s))})\big\}, \nonumber \\
\bs{\mathcal{F}}^{(2)}(x_1,x_2) & \equiv & \frac{(\frac{\pi}{T})^2\,q^{\widetilde{\beta}^{(0,2)}(1,\varepsilon_1,\varepsilon_2)}}{1 - q^{\frac{1}{2}}}\big\{\mathbf{F}^{(2)}_{s = 0\,*}(\phi_1,\phi_2) + q^{\frac{1}{2}}\,\mathbf{F}^{(2)}_{s = 0\,**}(\phi_1,\phi_2) + O(q)\big\}, \nonumber 
\eea
where 
\beq
\label{betti} \widetilde{\beta}^{(0,2)}(s,\varepsilon_1,\varepsilon_2) = \left\{\begin{array}{lll} -1 & & {\rm if}\,\,\varepsilon_1 =  \varepsilon_2 = 0, \\ \tfrac{b(s) - 1}{2} && {\rm if}\,\,\varepsilon_1 \neq \varepsilon_2, \\ b(s) & & {\rm if}\,\,\varepsilon_1 = \varepsilon_2 = \tfrac{1}{2},\end{array}\right.
\eeq
$$
\mathbf{F}_{s\,*}^{(2)} = \frac{1}{4 - n^2s^2}\,\frac{H_{b(s),\varepsilon_1 \oplus \varepsilon_2}(\phi_1,\phi_2)}{(x_{\varepsilon_1}^*)'(\phi_1)(x_{\varepsilon_2}^*)'(\phi_2)},
$$
and for $\varepsilon_1 = \varepsilon_2$:
$$
\mathbf{F}_{s\,**}^{(2)}(\phi_1,\phi_2) = \frac{1}{4 - n^2s^2}\,\frac{ - H_{b(s) + 2,0}(\phi_1,\phi_2)}{(x_{\varepsilon_1}^*)'(\phi_1)(x_{\varepsilon_2}^*)'(\phi_2)}.
$$
The value of $\mathbf{F}_{s\,**}^{(2)}$ for $\varepsilon_1 \neq \varepsilon_2$ will be irrelevant.
\end{corollary}

The second consequence of Lemma~\ref{behaviorG2} is the critical behavior of the generating series of cuffed cylinders $\hat{\mathbf{F}}_{s}^{(2)}$ and $\tilde{\mathbf{F}}_{s}^{(2)}$ which appear in the evaluation of $\pmb{\mathscr{F}}$ via Proposition~\ref{P212}.

\begin{lemma}\label{arms} Let $x_j=x(\varepsilon_j + \tau \phi_j)$ for $j=1,2$, and consider the critical regime $q \rightarrow 0$. Let $\bs{\mathcal{H}}(x)$ be a generating series which is holomorphic for $x \in \mathbb{C}\setminus[\gamma_-,\gamma_+]$ such that $\bs{\mathcal{H}}(x) \in O(1/x^2)$ when $x \rightarrow \infty$, and when $x = x(\varepsilon + \tau\phi)$ admits the critical behavior
$$
\bs{\mathcal{H}}(x) =  \Big(\frac{\pi}{T}\Big)^{C}\,q^{\frac{3}{2}\,\varepsilon}\big\{\bs{\mathcal{H}}_{\varepsilon,*}(\phi) + O(q^{b})\big\},
$$
where $C$ stands for an arbitrary real number. When computing the integral
\beq
\label{contH1} \oint_{\gamma} \frac{\dd x_1}{2{\rm i}\pi}\,\bs{\mathcal{H}}(x_1)\hat{\mathbf{F}}_{s}^{(2)}(x_1,x_2), 
\eeq
the relevant singular part (for this we use the sign $\equiv$) of $\hat{\mathbf{F}}_{s}^{(2)}\!\!$ is 
\beq
\label{FEF0} \hat{\mathbf{F}}_{s}^{(2)}(x_1,x_2) \equiv q^{\hat{\varkappa}(\varepsilon_2)}\big\{\hat{\mathbf{F}}_{s;\varepsilon_2,*}^{(2)}(\phi_1,\phi_2) + O(q^{b(s)})\big\},
\eeq
with $\varepsilon_1 = 0$ and the exponent
$$
\hat{\varkappa}(\varepsilon_2) = \left\{\begin{array}{lll} -\tfrac{1}{2} & & {\rm if}\,\,\varepsilon_2 = 0, \\ \tfrac{b(s)}{2} & & {\rm if}\,\,\varepsilon_2 = \tfrac{1}{2}. \end{array}\right.\,
$$
 
Likewise, let $\tilde{\bs{\mathcal{H}}}(x_1,x_2)$ be a generating series which is holomorphic for $(x_1,x_2) \in (\mathbb{C}\setminus[\gamma_-,\gamma_+])^2$ and such that $\tilde{\bs{\mathcal{H}}}(x_1,x_2) \in O(x_1^{-2}x_2^{-2})$ when $x_i \rightarrow \infty$, and admitting the following critical behavior when $x_j = x(\varepsilon_j + \tau\phi_j)$:
$$
\tilde{\bs{\mathcal{H}}}(x_1,x_2) =  \Big(\frac{\pi}{T}\Big)^{C} q^{\frac{3}{2}(\varepsilon_1 + \varepsilon_2)}\big\{\tilde{\bs{\mathcal{H}}}_{\varepsilon_1,\varepsilon_2,*}(\phi_1,\phi_2) + O(q^{b})\big\},
$$
where $C$ is an arbitrary number. When computing the contour integral
\beq 
\label{contH2} \oint_{\gamma} \frac{\dd x_1}{2{\rm i}\pi}\oint_{\gamma} \frac{\dd x_2}{2{\rm i}\pi}\,\tilde{\bs{\mathcal{H}}}(x_1,x_2)\,\tilde{\mathbf{F}}_{s}^{(2)}(x_1,x_2),
\eeq
the singular part of $\tilde{\mathbf{F}}_{s}^{(2)}$ is
\beq
\label{FEFE0}\tilde{\mathbf{F}}_{s}^{(2)}(x_1,x_2) \equiv \tilde{\mathbf{F}}_{s\,*}^{(2)}(\phi_1,\phi_2) + O(q^{b(s)}),
\eeq
with $\varepsilon_1 = \varepsilon_2 = 0$.
\end{lemma}

\vspace{0.2cm}

\noindent\textbf{Proof of Lemma~\ref{arms}.} We shall estimate the contour integrals \eqref{contH1} and \eqref{contH2} in the regime $q \rightarrow 0$ by the steepest descent method. In particular, we will have to determine which region of the complex plane gives the dominant contribution of the integral, and the proof will show that it is always the vicinity of $\gamma_+^*$. It is however convenient to first transform the expressions of $\hat{\mathbf{F}}_{s}^{(2)}$ and $\tilde{\mathbf{F}}_{s}^{(2)}$.

Using $\partial_{x} \mathbf{R}(x,y) = \mathbf{A}(x,y)$, the evaluation \eqref{contueq} of the contour integral of a function against $\mathbf{A}(x,y)$ and the definition of $\mathbf{G}_{s}^{(2)}(v_1,v_2)$ in Proposition~\ref{p15}, setting $x_i = x(v_i)$ and analytically continuing in $(v_1,v_2)$, we obtain
\bea
\hat{\mathbf{F}}_{s}^{(2)}(x_1,x_2) & = & s\oint_{\gamma}\frac{\dd y}{2{\rm i}\pi}  \mathbf{R}(x_1,y)\mathbf{F}_{s}^{(2)}(y,x_2) \nonumber \\
& = &  s\int^{x_1} \dd \tilde{x}_1 \oint_{\gamma} \frac{\dd y}{2{\rm i}\pi}\,\mathbf{A}(\tilde{x}_1,y)\mathbf{F}_{s}^{(2)}(y,x_2) + C(x_2) \nonumber\\
& = & -\int^{x_1} \dd \tilde{x}_1\,ns\, \varsigma'(\tilde{x}_1)\mathbf{F}_{s}^{(2)}(\varsigma(\tilde{x}_1),x_2) + C(x_2) \nonumber\\
& = & ns \int^{v(x_1)} \dd\tilde{v}_1\,\frac{\mathbf{G}_{s}^{(2)}(\tau - \tilde{v}_1, v_2)}{x'(v_2)}  \nonumber\\
\label{Fhat222} & + & \frac{ns}{4-n^2s^2}\left(\frac{2}{x_2-\varsigma(x_1)}+\frac{ns\,\varsigma'(x_2)}{\varsigma(x_2)-\varsigma(x_1)}\right) + C(x_2),
\eea
where we stress that $C(x_2)$ does not depend on $x_1$, and for this reason will disappear when performing contour integration against $\bs{\mathcal{H}}(x_1)$ as $\bs{\mathcal{H}}(x_1) \in O(x_1^{-2})$. We can then do a partial fraction expansion with respect to $x_1$:
\bea
\frac{1}{x_2-\varsigma(x_1)} & = & \frac{-\varsigma'(x_2)}{x_1 - \varsigma(x_2)} + \frac{1}{x_2 - \varsigma(\infty)}, \nonumber \\
\frac{\varsigma'(x_2)}{\varsigma(x_2) - \varsigma(x_1)} & = & -\frac{1}{x_1 - x_2} + \frac{\varsigma'(x_2)}{\varsigma(x_2) - \varsigma(\infty)}.\nonumber
\eea
Therefore:
\bea
\oint_{\gamma} \frac{\dd x_1}{2{\rm i}\pi}\,\bs{\mathcal{H}}(x_1)\,\hat{\mathbf{F}}_{s}^{(2)}(x_1,x_2) & = & ns \oint_{\gamma} \frac{\dd x_1}{2{\rm i}\pi}\, \bs{\mathcal{H}}(x_1) \int^{v(x_1)} \frac{\dd\tilde{v}_1\,\mathbf{G}^{(2)}_{s}(\tau - \tilde{v}_1,v_2)}{x'(v_2)} \nonumber \\
\label{contqq} && + \frac{ns}{4 - n^2s^2}\big(2\varsigma'(x_2)\bs{\mathcal{H}}(\varsigma(x_2)) + ns\bs{\mathcal{H}}(x_2)\big).
\eea

The second term is of order of magnitude $q^{\frac{3}{2}\,\varepsilon_2}$. To examine the behavior of the first term, we fix the value of $\varepsilon_2 \in \{0,\tfrac{1}{2}\}$. When the variable $x_1$ is in the regime $x_1 = x(\varepsilon_1 + \tau\phi_1)$, the integrand (including $\dd x_1$) is of order of magnitude
\beq
\label{qqpower} q^{\frac{3}{2}\,\varepsilon_1 + (\frac{1}{2} - \varepsilon_1) - (\frac{1}{2} - \varepsilon_2) + b(s) (\varepsilon_1 \oplus \varepsilon_2)}.
\eeq
If $\varepsilon_2 = 0$, this is for $\varepsilon_1 = 0$ equal to $q^{0}$, while for $\varepsilon_1 = \tfrac{1}{2}$ it is equal to $q^{\frac{1}{4} + \frac{b(s)}{2}}$ -- which is negligible compared to the former. If $\varepsilon_2 = \tfrac{1}{2}$, \eqref{qqpower} is equal for $\varepsilon_1 = 0$ to $q^{\frac{b(s) + 1}{2}}$, while for $\varepsilon_1 = \tfrac{1}{2}$ it is equal to $q^{\frac{3}{4}}$ -- which is negligible compared to the former. So, independently of the value of $\varepsilon_2$, we move the contour for $x_1$ to pass close to $\gamma_+^*$ and the integral will be dominated by the regime $x_1 = x(\varepsilon_1 + \tau\phi_1)$ with $\varepsilon_1 = 0$. And, the first term in \eqref{contqq} is of order $q^{0}$ when $\varepsilon_2 = 0$, and of order $q^{\frac{b(s) + 1}{2}}$ when $\varepsilon_2 = \tfrac{1}{2}$. Since $b(s) \in (0,\tfrac{1}{2})$, we deduce that \eqref{contqq} is of order $q^{0}$ if $\varepsilon_2 = 0$, and of order $q^{\frac{b(s) + 1}{2}}$ if $\varepsilon_2 = \tfrac{1}{2}$.

Combining everything, the effective part of $\hat{\mathbf{F}}_{s}^{(2)}$ which is relevant to extract the leading term in \eqref{contqq} is
$$
\hat{\mathbf{F}}_{s}^{(2)}(x_1,x_2) \equiv q^{\hat{\varkappa}(\varepsilon_2)}\big\{\hat{\mathbf{F}}_{s;\varepsilon_2,*}^{(2)}(\phi_1,\phi_2) + O(q^{b(s)})\big\},
$$
with $\varepsilon_1 = 0$, the exponent
$$
\hat{\varkappa}(\varepsilon_2) = \left\{\begin{array}{lll} -\tfrac{1}{2} & & {\rm if}\,\,\varepsilon_2 = 0, \\ \tfrac{b(s)}{2}  & & {\rm if}\,\,\varepsilon_2 = \tfrac{1}{2}, \end{array}\right.
$$
and the prefactors:
\bea
\label{FFFFFF0}  \hat{\mathbf{F}}_{s;0,*}^{(2)}(\phi_1,\phi_2) & = & \frac{ns}{4 - n^2s^2}\bigg\{ \int^{\phi_1} \dd\tilde{\phi}_1\,\frac{H_{b(s),0}(1 - \tilde{\phi}_1,\phi_2)}{(x^*_0)'(\phi_2)}  \\
&&  + \frac{(x_0^*)'(1 - \phi_2)}{(x_0^*)'(\phi_2)}\,\frac{2}{x_0^*(\phi_1) - x_0^*(1 - \phi_2)} - \frac{ns}{x_0^*(\phi_1) - x^*_0(\phi_2)}\bigg\}, \nonumber  \\
\label{FFFFFF} \hat{\mathbf{F}}_{s;\frac{1}{2},*}^{(2)}(\phi_1,\phi_2) & = & -\frac{8 b(s) ns\,\cos \pi b(s) \phi_1\,\sin \pi b(s) \phi_2}{4 - n^2s^2}.
\eea

Let us now turn to $\tilde{\mathbf{F}}^{(2)}_{s}$. We compute from the definition \eqref{Ftilded}:
\bea
&&  \tilde{\mathbf{F}}_{s}^{(2)}(x_1,x_2) \nonumber \\
& = & s\mathbf{R}(x_1,x_2) + s^2\oint_{\gamma^2}\frac{\dd y_1}{2{\rm i}\pi}\,\frac{\dd y_2}{2{\rm i}\pi} \mathbf{R}(x_1,y_1)\mathbf{R}(x_2,y_2)\mathbf{F}_{s}^{(2)}(y_1,y_2) \nonumber\\
& = & s\mathbf{R}(x_1,x_2) + s^2\int^{x_1} \dd\tilde{x}_1 \int^{x_2} \dd\tilde{x}_2 \oint_{\gamma} \frac{\dd y_1}{2{\rm i}\pi}\,\frac{\dd y_2}{2{\rm i}\pi}\,\mathbf{A}(\tilde{x}_1,y_1)\mathbf{A}(\tilde{x}_2,y_2)\mathbf{F}_{s}^{(2)}(y_1,y_2) \nonumber \\
&& + C_1(x_1) + C_2(x_2) \nonumber \\
& = & s\mathbf{R}(x_1,x_2) + n^2s^2\int^{x_1} \dd\tilde{x}_1 \int^{x_2}\dd\tilde{x}_2\, \varsigma'(\tilde{x}_1)\varsigma'(\tilde{x}_2) \mathbf{F}_{s}^{(2)}(\varsigma(\tilde{x}_1),\varsigma(\tilde{x}_2))
 \dd \tilde{x}_1 \dd \tilde{x}_2 \nonumber \\
 && + C_1(x_1) + C_2(x_2) \nonumber \\
 & = & s\mathbf{R}(x_1,x_2) + n^2s^2 \bigg(\int^{v(x_1)} \dd\tilde{v}_1\,\int^{v(x_2)}\dd\tilde{v}_2\,\mathbf{G}_{s}^{(2)}(\tau - \tilde{v}_1,\tau - \tilde{v}_2)  \nonumber \\
\label{Ftilde222} & &-  \frac{2\,\ln\big[\varsigma(x_1) - \varsigma(x_2)\big] + ns\,\ln\big[x_1 - \varsigma(x_2)\big]}{4 - n^2s^2} + \tilde{C}_1(x_1) + \tilde{C}_2(x_2)\bigg).
\eea
The functions $C_1(x_1)$, $\tilde{C}_1(x_1)$, $C_2(x_2)$ and $\tilde{C}_2(x_2)$ do not depend simultaneously on $x_1$ and $x_2$ and will thus disappear when we perform the contour integral against $\tilde{\bs{\mathcal{H}}}(x_1,x_2)$ as it behaves like $O(x_1^{-2}x_2^{-2})$ when $x_i \rightarrow \infty$. Given the expression \eqref{rin} for $\mathbf{R}$, the term $s\mathbf{R}(x_1,x_2)$ in the first line combines with the ratio in the second line, up to an extra term which only depends on $x_2$ and will also disappear:
\bea 
\tilde{\mathbf{F}}_{s}^{(2)}(x_1,x_2) & = & n^2s^2\int^{v(x_1)}\int^{v(x_2)}  \dd\tilde{v}_1\,\dd\tilde{v}_2\,\mathbf{G}_{s}^{(2)}(\tau - \tilde{v}_1,\tau - \tilde{v}_2) \nonumber \\
&& - \frac{2ns(ns\,\ln[x_1 - x_2] + 2\,\ln[x_1 - \varsigma(x_2)])}{4 - n^2s^2}+ \hat{C}_1(x_1) + \hat{C}_2(x_2),\nonumber 
\eea
where again $\hat{C}_i(x_i)$ does not depend simultaneously on both $x_1$ and $x_2$ so that they will disappear in the next step. Now, we can compute
\bea
&& \oint_{\gamma} \frac{\dd x_1}{2{\rm i}\pi}\,\oint_{\gamma} \frac{\dd x_2}{2{\rm i}\pi}\,\tilde{\bs{\mathcal{H}}}(x_1,x_2)\,\tilde{\mathbf{F}}_{s}^{(2)}(x_1,x_2)  \nonumber \\
&=& n^2s^2 \oint_{\gamma} \frac{\dd x_1}{2{\rm i}\pi} \oint_{\gamma} \frac{\dd x_2}{2{\rm i}\pi} \tilde{\bs{\mathcal{H}}}(x_1,x_2) \int^{v(x_1)} \bigg\{\dd\tilde{v}_1\,\int^{v(x_2)} \dd \tilde{v}_2\,\mathbf{G}_{s}^{(2)}(\tau - \tilde{v}_1,\tau - \tilde{v}_2) \nonumber \\&& \qquad \qquad - \frac{2 ns \, \dd \tilde{x}_1}{4 - n^2s^2}\Big(\frac{2}{\tilde{x}_1 - \varsigma(x_2)} + \frac{ns\varsigma'(\tilde{x}_1)}{\varsigma(\tilde{x}_1) - \varsigma(x_2)}\Big)\bigg\} \nonumber \\
&= &  n^2s^2 \bigg\{ \oint_{\gamma} \frac{\dd x_1}{2{\rm i}\pi} \oint_{\gamma} \frac{\dd x_2}{2{\rm i}\pi}\,\tilde{\bs{\mathcal{H}}}(x_1,x_2)  \int^{v(x_1)} \int^{v(x_2)} \dd\tilde{v}_1 \dd\tilde{v}_2\,\mathbf{G}_{s}^{(2)}(\tau - \tilde{v}_1,\tau - \tilde{v}_2)  \nonumber \\
\label{lassa}&& \qquad\qquad - \frac{2 ns}{4 - n^2s^2}\oint_{\gamma} \frac{\dd x_1}{2{\rm i}\pi}\bigg(2\int^{x_1} \dd\tilde{x}_1\,\tilde{\bs{\mathcal{H}}}(\tilde{x}_1,\varsigma(x_1))\,\varsigma'(x_1)  + ns \int^{x_1}  \dd\tilde{x}_1\,\tilde{\bs{\mathcal{H}}}(\tilde{x}_1,x_1)\bigg)\bigg\}.
\eea
The same arguments we used for $\hat{\mathbf{F}}_{s}^{(2)}$ show that the dominant contribution to the integrals always comes from the part of the integration where $x_j = x(\tau\phi_j)$ with $\phi_j$ of order $1$. So, the effective part of $\tilde{\mathbf{F}}_{s}^{(2)}$ which allows us to extract the dominant contribution of \eqref{lassa} is:
$$
\tilde{\mathbf{F}}_{s}^{(2)}(x_1,x_2) \equiv \tilde{\mathbf{F}}_{s\,*}^{(2)}(\phi_1,\phi_2) + O(q^{b(s)}),
$$
where $x_j = x(\varepsilon_j + \tau\phi_j)$ with $\varepsilon_1 = \varepsilon_2 = 0$ and:
\bea
\label{fnun0} \tilde{\mathbf{F}}_{s\,*}^{(2)} & = & \frac{n^2s^2}{4 - n^2s^2} \int^{\phi_1}\int^{\phi_2} \dd\tilde{\phi}_1\dd\tilde{\phi}_2\,H_{b(s),0}(1 - \tilde{\phi}_1,1 - \tilde{\phi}_2) \nonumber \\
&& - \frac{2ns}{4 - n^2s^2}\Big(2\ln[x_0^*(\phi_1) - x_0^*(1 - \phi_2)] + ns\ln[x_0^*(1 - \phi_1) - x_0^*(1 - \phi_2)]\Big). \nonumber
\eea
\hfill $\Box$
\vspace{0.2cm}

\subsection{Fixed nesting graph}

\label{CoscrFs}

Now we can deduce the critical behavior of the  generating series of maps with a fixed nesting graph $\Gamma$. Recall that we denoted $V_{0,2}(\Gamma)$ the set of univalent vertices of genus $0$ carrying exactly one boundary. Let us introduce the notations $V_{0,2}^{{\rm L}}(\Gamma)$ (resp. $V_{0,2}^{{\rm S}}(\Gamma)$) for the vertices for which we keep the boundary large (resp. small). Let $k^{(0,2)}$, $k_0^{(0,2)}$ and $k_{1/2}^{(0,2)}$ denote the cardinalities of $V_{0,2}(\Gamma)$, $V_{0,2}^{{\rm L}}(\Gamma)$ and $V_{0,2}^{{\rm S}}(\Gamma)$, respectively. We also denote $E_{0,2}^{{\rm S}}(\Gamma)$ the set of edges incident to a small boundary, and $E'(\Gamma)$ the set of edges which are not in $E_{0,2}^{{\rm S}}(\Gamma)$. 
\begin{theorem}
\label{CoscrF} Let $x_j=x(\varepsilon_j + \tau \phi_j)$ for $j=1, \ldots, k$, and $k_0$ and $k_{1/2}$ denote the number of $\varepsilon_j =0$ (large boundaries) and of $\varepsilon_j =1/2$ (small boundaries). When $q \rightarrow 0$, we have for the singular part with respect to $u$ and $x_i$'s:
$$ 
\pmb{\mathscr{F}}_{\Gamma,\star,\mathbf{s} = \mathbf{1}}^{(\mathsf{g},k)}(x_1,\ldots,x_k) = \Big(\frac{\pi}{T}\Big)^{k}\,q^{\varkappa(\mathsf{g},k,k_{1/2},k_{1/2}^{(0,2)})} \big\{[\pmb{\mathscr{F}}_{\Gamma,\star,\mathbf{s} = \mathbf{1}}^{(\mathsf{g},k)}]_{*}(\phi_1,\ldots, \phi_k) + O(q^{\frac{b}{2}})\big\},
$$
where
$$
\varkappa(\mathsf{g},k,k_{1/2},k_{1/2}^{(0,2)})=(2\mathsf{g} - 2 + k)(\mathfrak{d}\tfrac{b}{2} - 1) - \tfrac{k}{2} + \tfrac{3}{4}\,k_{1/2} + (\tfrac{b}{2}- \tfrac{1}{4})k_{1/2}^{(0,2)}.
$$
And, for the singular part with respect to $s,u$ and $x_i$'s:
\bea
\label{nuingfsg}\ \pmb{\mathscr{F}}_{\Gamma,\star,\mathbf{s}}^{(\mathsf{g},k)}(x_1,\ldots,x_k) & = & \Big(\frac{\pi}{T}\Big)^{k}\,\Big[\prod_{\mathsf{e} \in E(\Gamma)} \frac{1}{4 - n^2s^2}\Big]\,q^{\varkappa(\mathsf{g},k,k_{1/2},k_{1/2}^{(0,2)}) + \sum_{\mathsf{e} \in E_{0,2}^{{\rm S}}(\Gamma)} \frac{1}{2}(b[s(\mathsf{e})] - b)} \nonumber \\
&& \!\!\!\!\!\!\!\!\! \bigg\{\sum_{\tau\,:\,E'(\Gamma) \rightarrow \{0,1\}} \Big[\prod_{\mathsf{e} \in E'(\Gamma)} q^{\tau(e)\,b[s(\mathsf{e})]} \Big]\cdot \bigg([\pmb{\mathscr{F}}_{\Gamma,\star,\mathsf{s}}^{(\mathsf{g},k)}]_{*,\tau} + O\Big(\sum_{\mathsf{e} \in E(\Gamma)} q^{\frac{b[s(\mathsf{e})]}{2}}\Big)\bigg) \bigg\}.
\eea
\end{theorem} 
Remarkably, the result does not depend on the details of $\Gamma$, but only on its genus $\mathsf{g}$, and number of boundaries of different types. For a fixed topology $(\mathsf{g},k)$, the graphs minimizing the number of small boundaries have the biggest contribution and if we also fix a configuration $(k_0,k_{1/2})$, the graphs maximizing $k_{1/2}^{(0,2)}$ contribute the most.

\vspace{0.2cm}

\noindent\textbf{Proof.}
We want to estimate the expression of Proposition~\ref{P212} for $\pmb{\mathscr{F}}_{\Gamma,\star,\mathbf{s}}^{(\mathsf{g},k)}$ in the regime $q\rightarrow 0$. Given that the vertex weights are $\bs{\mathcal{F}}$'s whose leading term according to Theorem~\ref{ouqusf} has the property of receiving an extra factor $q^{\frac{3}{4}}$ whenever a boundary variable $x_i$ is not close to $\gamma_+^*$ at scale $q^{\frac{1}{2}}$, we are in the conditions of Lemma~\ref{arms}. We can apply the steepest descent method to approximate the integral, and we have argued in the proof of Lemma~\ref{arms} that the contour should be moved to pass close to $\gamma_+^*$ because the dominant contribution comes from the regime where each $y_{\mathsf{e}} - \gamma_+^* \in O(q^{\frac{1}{2}})$, i.e.~$y_{\mathsf{e}} = x(\tau\phi_{\mathsf{e}})$ for $\phi_{\mathsf{e}}$ of order $1$. Therefore, combining Theorem~\ref{ouqusf} for $\bs{\mathcal{F}}$'s and Lemma~\ref{arms} for $\hat{\mathbf{F}}_{s}^{(2)}$ and $\tilde{\mathbf{F}}_{s}^{(2)}$, we arrive to: 
\bea
&& \pmb{\mathscr{F}}^{(\mathsf{g},k)}_{\Gamma,\star,\mathbf{1}}(x_1,\ldots,x_k) \nonumber \\
& = & \oint_{\gamma^{E_{{\rm glue}}(\Gamma)}} \prod_{\mathsf{e} \in E_{{\rm glue}}(\Gamma)}  \frac{\dd y_{\mathsf{e}}}{2{\rm i}\pi} \prod_{\mathsf{v} \in \tilde{V}(\Gamma)} \frac{\bs{\mathcal{F}}^{(\mathsf{h}(\mathsf{v}),k(\mathsf{v}) + d(\mathsf{v}))}(x_{\partial(\mathsf{v})},y_{\mathsf{e}(\mathsf{v})})}{d(\mathsf{v})!} \nonumber \\
&& \qquad \qquad \times \prod_{\mathsf{e} \in \tilde{E}(\Gamma)} \tilde{\mathbf{F}}^{(2)}_{s = 1}(y_{\mathsf{e}_+},y_{\mathsf{e}_-}) \prod_{\mathsf{v} \in V_{0,2}(\Gamma)}\hat{\mathbf{F}}^{(2)}_{s = 1}(y_{\mathsf{e}_+(\mathsf{v})},x_{\partial(\mathsf{v})}) \nonumber\\
& = & \prod_{\mathsf{e} \in E_{{\rm glue}}(\Gamma)}q^{\frac{1}{2}}\prod_{\mathsf{v} \in \tilde{V}(\Gamma)}  q^{[2\mathsf{h}(\mathsf{v}) -2 +k(\mathsf{v})+d(\mathsf{v})](\mathfrak{d}\frac{b}{2} - 1)-\frac{k(\mathsf{v})+d(\mathsf{v})}{2}+\frac{3}{4}k_{1/2}(\mathsf{v})} \prod_{\mathsf{v} \in V_{0,2}^0}q^{-\frac{1}{2}} \prod_{\mathsf{v} \in V_{0,2}^{{\rm S}}} q^{\frac{b}{2}} \nonumber\\
\label{laisft}&&\times \big\{[\pmb{\mathscr{F}}_{\Gamma,\star,\mathbf{1}}^{(\mathsf{g},k)}]_{*}(\phi_1,\ldots, \phi_k) + O(q^{\frac{b}{2}})\big\},
\eea
with
\bea
[\pmb{\mathscr{F}}_{\Gamma,\star,\mathbf{1}}^{(\mathsf{g},k)}]_{*}(\phi_1,\ldots, \phi_k) & = & \oint_{\overline{\mathcal{C}}^{E_{{\rm glue}}(\Gamma)}} \prod_{\mathsf{e} \in E_{{\rm glue}}(\Gamma)} \frac{\dd x_0^*(\tilde{\phi}_{\mathsf{e}})}{2{\rm i}\pi}\,\prod_{\mathsf{v} \in \tilde{V}(\Gamma)} \frac{\bs{\mathcal{F}}_{*}^{(\mathsf{h}(\mathsf{v}),k(\mathsf{v}) + d(\mathsf{v}))}(\phi_{\partial(\mathsf{v})},\tilde{\phi}_{\mathsf{e}(\mathsf{v})})}{d(\mathsf{v})!} \nonumber \\
&& \times \prod_{\mathsf{e} \in \tilde{E}(\Gamma)} \tilde{\mathbf{F}}_{s = 1\,*}^{(2)}(\tilde{\phi}_{\mathsf{e}_+},\tilde{\phi}_{\mathsf{e}_{-}}) \prod_{\mathsf{v} \in V_{0,2}(\Gamma)} \hat{\mathbf{F}}^{(2)}_{s = 1\,*}(\tilde{\phi}_{\mathsf{e}_+(\mathsf{v})},\phi_{\partial(\mathsf{v})}). \nonumber
\eea

Since we refer all the time to a fixed  nesting graph $\Gamma$, we omit it in the notations for simplicity. Let us now simplify the total exponent. The first Betti number of the graph is 
$$
1-|V|+|E| = \mathsf{g}-\sum_{\mathsf{v}\in\tilde{V}}\mathsf{h}(\mathsf{v}),
$$
and we recall that $|V| = |\tilde{V}| + k^{(0,2)}$ and $|E| = |\tilde{E}| + k^{(0,2)}$. Then, we observe that
$$ 
\sum_{\mathsf{v} \in \tilde{V}} k(\mathsf{v}) = k - k^{(0,2)}\ \  \text{ and }\ \  \sum_{\mathsf{v} \in \tilde{V}} k_{1/2}(\mathsf{v}) = k_{1/2} - k_{1/2}^{(0,2)}.
$$
By counting inner half-edges we also find
$$
\sum_{\mathsf{v} \in \tilde{V}} d(\mathsf{v}) = 2|E| - |E_{{\rm un}}| = 2|\tilde{E}| + k^{(0,2)} = |E_{{\rm glue}}|.
$$
Moreover, we obviously have $k^{(0,2)} = k^{(0,2)}_{0} + k^{(0,2)}_{1/2}$. Substituting these relations in \eqref{laisft} gives a total exponent\bea
\varkappa & = & \tfrac{1}{2}\,|E_{{\rm glue}}| +\big(2(\mathsf{g}-|E|+|V|-1)-2|\tilde{V}| + k - k^{(0,2)} + 2|E|-k^{(0,2)}\big) \left(\mathfrak{d}\tfrac{b}{2}-1\right) \nonumber \\
&& -\tfrac{1}{2}(k-k^{(0,2)})-\tfrac{1}{2}\,|E_{{\rm glue}}|+\tfrac{3}{4}(k_{1/2}-k_{1/2}^{(0,2)})-\tfrac{1}{2}\,k_0^{(0,2)} + \tfrac{b}{2}\,k_{1/2}^{(0,2)} \nonumber \\
&= & (2\mathsf{g}-2+k)\left(\mathfrak{d}\tfrac{b}{2}-1\right)-\tfrac{k}{2}+\tfrac{3}{4}k_{1/2}+ (\tfrac{b}{2} - \tfrac{1}{4})k^{(0,2)}_{1/2}. \nonumber
\eea
A similar computation for general $\mathbf{s} = (s(\mathsf{e}))_{\mathsf{e} \in E(\Gamma)}$ using the behavior of the singular part of $\hat{\mathbf{F}}_{s}^{(2)}$ and $\tilde{\mathbf{F}}_{s}^{(2)}$ with respect to $s$, leads to the claim \eqref{nuingfsg} (we omit the expression of the prefactors).
\hfill $\Box$
\vspace{0.2cm}

\begin{remark}
Note that the analysis of the generating series of configurations with a fixed nesting graph performed in this section has shown that for the gluing annuli, which contain the inner boundaries of the arms, large lengths give effectively dominant contributions. This is the reason why the only sets of edges which play a role are: the set of edges $E_{0,2}^{\mathrm{S}}$ incident to one small boundary and one large boundary (either a boundary of the map or an inner annulus), and the set of edges $E_{0,2}^{\prime}$ incident to two large boundaries (either both boundaries of the map, if the map is a just cylinder; two inner annuli for internal edges; or a boundary of the map and an inner annulus).
\end{remark}

\section{Large volume asymptotics}
\label{FixedV}

Recall from Theorem~\ref{th38} the scaling of $q$ with respect to the variable $u$ coupled to the volume
$$
q \sim \Big(\frac{1 - u}{q_*}\Big)^{c},\qquad c = \frac{1}{1 - \frac{b}{2} - \mathfrak{d}\frac{b}{2}}, 
$$
with $\mathfrak{d} = 1$ in dense phase, $\mathfrak{d} = -1$ in dilute phase.

\subsection{Relative amplitude of nesting graphs}
\label{Fixedas}

We now extract from Theorem~\ref{CoscrF} the leading asymptotics of the generating series of maps of given volume $V$, given boundary perimeters, and given nesting graph $\Gamma$, not keeping track of the number of separating loops on each arm -- \textit{i.e.} for $s(\mathsf{e}) = 1$, in the regime $V \rightarrow \infty$, while we impose either small or large boundaries.

\begin{theorem}
\label{LAPA} Take $(g,h)$ on the non-generic critical line. Assume $2\mathsf{g} - 2 + k > 0$. The generating series of connected maps of volume $V$, of genus $\mathsf{g}$, with $k_{1/2}$ boundaries of finite perimeter $L_i=\ell_i$, among which $k_{1/2}^{(0,2)}$ are carried by a genus $0$ leaf as only mark, and $k_{0}$ boundaries of perimeters $L_i=\ell_i V^{c/2}$ -- for fixed positive $\bs{\ell} = (\ell_i)_{i = 1}^k$ -- and realizing the nesting graph $(\Gamma,\star)$, behaves when $V \rightarrow \infty$ as
\beq
\label{lneq} \Big[u^{V}\prod_{i = 1}^k x_i^{-(L_i+1)}\Big]\pmb{\mathscr{F}}_{\Gamma,\star,\mathbf{1}}^{(\mathsf{g},k)} \sim \pmb{\mathscr{A}}_{\Gamma,\star,\mathbf{1}}^{(\mathsf{g},k)}(\bs{\ell})\,V^{[-1 + c((2\mathsf{g} - 2 + k)(1 - \mathfrak{d}\frac{b}{2}) - \frac{1}{4}k_{1/2} + (\frac{1}{4} - \frac{b}{2})k_{1/2}^{(0,2)})]},
\eeq
where $k = k_0 + k_{1/2}$ is the total number of boundaries, and an expression for the non-zero prefactor is given in \eqref{Agks1}.
\end{theorem}

Several remarkable conclusions can be drawn from this result. Firstly, if we keep all boundaries large, we have
$$ 
\pmb{\mathscr{F}}_{\Gamma,\star,\mathbf{1}}^{(\mathsf{g},k)} \stackrel{\bigcdot}{\sim} V^{-1 + c(2\mathsf{g} - 2 + k)(1 - \mathfrak{d}\frac{b}{2})} 
$$
and the order of magnitude only depends on the global topology of $\Gamma$, i.e.~on the genus $\mathsf{g}$ and the number of boundaries $k$. In other words, for given $\mathsf{g}$ and $k$, all nesting graphs have comparable probabilities to be realized. 

\begin{center}
\begin{figure}[h!]
\includegraphics[width=0.8\textwidth]{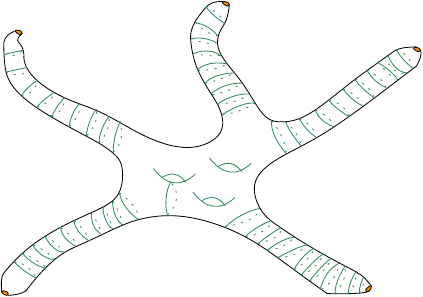}
\caption{\label{starfig} A typical map of the $O(n)$ model with small boundaries. These are most likely to be incident to distinct long arms (containing $O(\ln V)$ separating loops). We have only drawn in green the loops which are separating.}
\end{figure}
\end{center}

Secondly, if we keep a certain number $k_{1/2} > 0$ of small boundaries, the nesting graphs most likely to be realized when $V \rightarrow \infty$ at criticality are the ones with $k_{1/2}^{(0,2)} = k_{1/2}$, \textit{i.e.}, where each small boundary belongs as the only marked element to a connected component with the topology of a cylinder on the complement of all loops (see Figure~\ref{starfig}). And, all nesting graphs with this property have comparable probabilities.

For completeness, we also study the case of cylinders $(\mathsf{g},k) = (0,2)$ -- for which the computations already appeared in \cite{BBD}. There are only two possible nesting graphs:
\beq
\label{Gcyl} (\Gamma_1,\star) = \bullet^{1,2}\qquad\qquad (\Gamma_2,\star) = {}^{1}\!\bullet\!\!\!-\!\!\!-\!\!\bullet^{2}
\eeq
Before conditioning on the volume and the boundary perimeters, the generating series for $(\Gamma_1,\star)$ is $\bs{\mathcal{F}}^{(2)}(x_1,x_2)$, while the generating series for the $(\Gamma_2,\star)$ is $(\mathbf{F}_{s = 1}^{(2)} - \bs{\mathcal{F}}^{(2)})(x_1,x_2)$. We derive from Corollary~\ref{dgfsgg}:

\begin{theorem}
\label{LAPB} Take $(g,h)$ on the non-generic critical line. Fix $\ell_i$ positive independent of $V$, and $\varepsilon_i \in \{0,\tfrac{1}{2}\}$. If $\varepsilon_i = 0$, we choose $L_i = \ell_i V^{c/2}$, and if $\varepsilon_i = \tfrac{1}{2}$, we rather choose $L_i = \ell_i$. We have when $V \rightarrow \infty$:
\bea   
\label{qdoguns}\left[u^{V}x_1^{-(L_1+1)}x_2^{-(L_2+1)}\right]\,\pmb{\mathscr{F}}_{\Gamma_1,\star,s = 0}^{(0,2)}(x_1,x_2) & \sim & \pmb{\mathscr{F}}_{\Gamma_1,\star,s = 0}^{(0,2)}(\ell_1,\ell_2)\,V^{-1 - \frac{c}{2}(1-(\varepsilon_1 \oplus \varepsilon_2))}, \\
\label{qdoguns2} \left[u^{V}x_1^{-(L_1+1)}x_2^{-(L_2+1)}\right]\,\pmb{\mathscr{F}}_{\Gamma_2,\star,s = 1}^{(0,2)}(x_1,x_2) & \sim & \pmb{\mathscr{F}}_{\Gamma_2,\star,s = 1}^{(0,2)}(\ell_1,\ell_2)\,V^{-1 - cb(1-(\varepsilon_1 \oplus \varepsilon_2))},  
\eea
with a non-zero prefactor.
\end{theorem}

The constant prefactors $\pmb{\mathscr{A}}$ are computed in the course of the proofs.  Although their structure is combinatorially clear -- we essentially have to replace in the formula of Proposition~\ref{P212} all the factors by their effective leading asymptotics derived throughout the previous Section, and perform the extra contour integrations in $\tilde{u}$ and $\tilde{x}$ whose effect is simply displayed in \eqref{Agks1} -- it is however a formidable task to obtain explicit formulas (as functions of $\ell_i$) for a given nesting graph $\Gamma$. For us, the formula serves as showing that this prefactor is non trivial.

We remark that the formula for the exponent in Theorem \ref{LAPB} does not coincide with the one in Theorem \ref{LAPA} taking $(\mathsf{g},k) =(0,2)$.

\vspace{0.2cm}

\noindent \textbf{Proofs.} We briefly sketch the proof as the details of the saddle point analysis are essentially the same as in \cite[Section 6.5 and 6.6]{BBD}. Let $\partial_0(\Gamma)$ denote the set of boundaries for which we want to impose perimeter $L_i = \ell_i\,V^{c/2}$ (\textit{i.e.} we declare $\varepsilon_i = 0$), and $\partial_{1/2}(\Gamma)$ the set of boundaries for which we rather impose $L_i = \ell_i$ (\textit{i.e.} we declare $\varepsilon_i = \tfrac{1}{2}$). The analysis reveals that this scaling $V^{c/2}$ for large boundaries is the one for which a non-trivial behavior will be obtained.

\begin{center}
\textit{Conditioning on boundary perimeters}
\end{center}

\vspace{0.1cm}

\begin{figure}
\begin{center}
\includegraphics[width=0.6\textwidth]{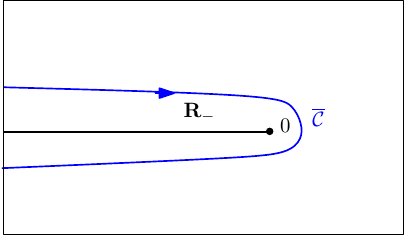}
\caption{\label{ContCbar} The contour $\overline{\mathcal{C}}$.}
\end{center}
\end{figure}

We first study integrals of the form
\beq
\label{Iform} \mathcal{I}(u) =   \prod_{i \in \partial_{0}(\Gamma)} \oint_{\gamma} \frac{x_i^{\ell_i V^{c/2}}\,\dd x_i}{2{\rm i}\pi}\ \prod_{i \in \partial_{1/2}(\Gamma)} \oint_{\gamma} \frac{x_i^{\ell_i}\,\dd x_i}{2{\rm i}\pi}\,\Phi\bigg[u;(x_i)_{i \in \partial_{1/2}(\Gamma)};\Big(\frac{x_i - \gamma_+}{q^{\frac{1}{2}}}\Big)_{i \in \partial_{0}(\Gamma)}\bigg],
\eeq
where $\Phi$ is a function which has a non-zero limit when $u \rightarrow 1$, and the convergence is uniform when its variables belong to any compact. We also take from Corollary~\ref{CoB7} in Appendix that
$$
\gamma_+^* - \gamma_+ = O(q).
$$
We use the change of variables
$$ 
x_i = \left\{\begin{array}{lll} \gamma_+^* + q^{\frac{1}{2}}\,x_0^*(\phi_i) && {\rm if}\,\,i \in \partial_{0}(\Gamma), \\ \gamma_+^* + x_{\frac{1}{2}}^*(\phi_i) && {\rm if}\,\,i \in \partial_{1/2}(\Gamma), \end{array}\right.
$$ 
and deform the contour in $\big(\tilde{x}_i = q^{-\frac{1}{2}}(x_i - \gamma_+^*)\big)_{i \in \partial_{0}(\Gamma)}$ so that it passes close to the cut (see Figure~\ref{ContCbar}). In the limit $u \rightarrow 1$, the properties of the integrand on those steepest descent contours ensure that we can use the monotone convergence theorem to find
\bea 
\mathcal{I}(u) & \sim & q^{\frac{1}{2}k_0} \prod_{i \in \partial_{0}(\Gamma)} \oint \frac{\dd x_0^*(\phi_i)\,e^{x_0^*(\phi_i)\ell_i/\gamma_+^*}}{2{\rm i}\pi}\prod_{i \in \partial_{1/2}(\Gamma)} \oint \frac{(x_{\frac{1}{2}}^*(\phi_i))^{\ell_i}\,\dd x_{\frac{1}{2}}^*(\phi_i)}{2{\rm i}\pi} \nonumber \\
 && \times\Phi\Big[u;\big(x_{\frac{1}{2}}^*(\phi_i)\big)_{i \in \partial_{1/2}(\Gamma)};\big(x_0^*(\phi_i)\big)_{i \in \partial_{0}(\Gamma)}\Big].  \nonumber
\eea

\vspace{0.2cm}

\begin{center}
\textit{Conditioning on volume $V$}
\end{center}

\vspace{0.1cm}

Next, we would like to estimate integrals of the form
$$
\overline{\mathcal{I}} = \oint \frac{\dd u}{2{\rm i}\pi\,u^{V + 1}}\,\mathcal{I}(u) q^{\eta}
$$
for some exponent $\eta$. We recall that $q$ is a function of $u$ for which Theorem~\ref{th38} gives
$$ 
q \sim \Big(\frac{1 - u}{q_*}\Big)^{c},\qquad u \rightarrow 1,
$$
and which is delta-analytic (at $1$) by Lemma~\ref{deltaanal}.

\begin{figure}[h!]
\begin{center}
\includegraphics[width=0.6\textwidth]{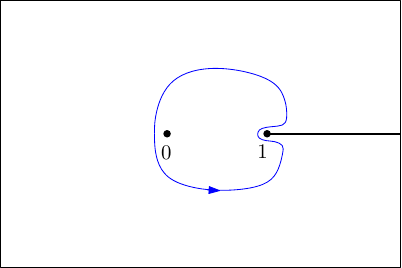}
\caption{\label{Cont1} The contour of integration for $\tilde{u}$.}
\end{center}
\end{figure}

We perform the change of variables
$$
u = 1 - \frac{\tilde{u}}{V}
$$
and deform the contour in $u$ to the one shown in Figure~\ref{Cont1}. Now assume
$$  
\eta + \tfrac{1}{2}k_0 \neq 0.
$$ 
In the limit $V \rightarrow \infty$, by the properties of the integrand on this steepest descent contour, we can complete the integral to a contour which is again $\overline{\mathcal{C}}$ shown in Figure~\ref{ContCbar} and find
\bea  
\overline{\mathcal{I}} & \sim & V^{-1 - c(\eta + \frac{1}{2}k_0)}\,\oint_{\overline{\mathcal{C}}} \frac{-\dd\tilde{u}\,e^{\tilde{u}}}{2{\rm i}\pi} \Big(\frac{\tilde{u}}{q_*}\Big)^{c(\eta + \frac{1}{2}k_0)} \prod_{i \in \partial_{0}(\Gamma)} \oint_{(x_0^*)^{-1}(\overline{\mathcal{C}})} \frac{\dd x_0^*(\phi_i)\,e^{x_0^*(\phi_i)\ell_i/\gamma_+^*}}{2{\rm i}\pi} \nonumber \\
&& \prod_{i \in \partial_{1/2}(\Gamma)} \oint_{(x_{\frac{1}{2}}^*)^{-1}(\gamma)} \frac{(x_{\frac{1}{2}}^*(\phi_i))^{\ell_i}\,\dd x_{\frac{1}{2}}^*(\phi_i)}{2{\rm i}\pi}\,\Phi\Big[1;(x_{\frac{1}{2}}^*(\phi_i))_{i \in \partial_{1/2}(\Gamma)};\big(x_0^*(\phi_i)\big)_{i \in \partial_{0}(\Gamma)}\Big].  \nonumber
\eea
The integral over $\tilde{u}$ factors out and yields a Gamma function
\bea 
\overline{\mathcal{I}} & = & \frac{V^{-1 - c(\eta + \frac{1}{2}k_0)}}{-\Gamma\big[-c\big(\eta + \frac{1}{2}k_0\big)\big]} \prod_{i \in \partial_{0}(\Gamma)} \oint_{(x_0^*)^{-1}(\overline{\mathcal{C}})} \frac{\dd x_0^*(\phi_i)\,e^{x_0^*(\phi_i)\ell_i/\gamma_+^*}}{2{\rm i}\pi}\nonumber \\
\label{Ibar} &&  \prod_{i \in \partial_{1/2}(\Gamma)} \oint_{(x_{\frac{1}{2}}^*)^{-1}(\gamma)} \frac{(x_{\frac{1}{2}}^*(\phi_i))^{\ell_i}\,\dd x_{\frac{1}{2}}^*(\phi_i)}{2{\rm i}\pi}\, \Phi\Big[1;(x_{\frac{1}{2}}^*(\phi_i))_{i \in \partial_{1/2}(\Gamma)};\big(x_0^*(\phi_i)\big)_{i \in \partial_{0}(\Gamma)}\Big] .
\eea

\vspace{0.2cm}

\begin{center}
\textit{Specialization to Theorem~\ref{LAPA}}
\end{center}
\vspace{0.1cm}
We obtain Theorem~\ref{LAPA} for $\pmb{\mathscr{F}}_{\Gamma,\star,\mathbf{1}}^{(\mathsf{g},k)}$ with $2\mathsf{g} - 2 + k > 0$ by taking from the proof of Theorem~\ref{CoscrF} the exponent
\beq
\label{etadenom} \eta \coloneqq (2\mathsf{g} - 2 + k)(\mathfrak{d}\tfrac{b}{2} - 1) - \tfrac{k}{2} + \tfrac{3}{4}\,k_{1/2} + (\tfrac{b}{2}- \tfrac{1}{4})k_{1/2}^{(0,2)}
\eeq
and
\beq
\Phi\Big[1;(x_{\frac{1}{2}}^*(\phi_i))_{i \in \partial_{1/2}(\Gamma)};\big(x_0^*(\phi_i)\big)_{i \in \partial_{0}(\Gamma)}\Big] = [\pmb{\mathscr{F}}_{\Gamma,\star,\mathbf{1}}^{(\mathsf{g},k)}]_{*}(\phi_1,\ldots,\phi_k).
\eeq
Since $k = k_0 + k_{1/2}$, we remark that
$$
\eta + \tfrac{1}{2}k_0 = (2\mathsf{g} - 2 + k)(\mathfrak{d}\tfrac{b}{2} - 1) + \tfrac{1}{4}\,k_{1/2} + (\tfrac{b}{2}- \tfrac{1}{4})k_{1/2}^{(0,2)}
$$
is non-zero. The constant prefactor is thus
\bea
\pmb{\mathscr{A}}^{(\mathsf{g},k)}_{\Gamma,\star,\mathbf{1}}(\bs{\ell}) & = & -\Gamma^{-1}\big[-c\big(\eta + \frac{1}{2}k_0\big)\big] \prod_{i \in \partial_{0}} \oint_{(x_0^*)^{-1}(\overline{\mathcal{C}})} \frac{\dd x_0^*(\phi_i)\,e^{x_0^*(\phi_i)\ell_i/\gamma_+^*}}{2{\rm i}\pi}\nonumber \\
\label{Agks1} &&  \prod_{i \in \partial_{1/2}} \oint_{(x_{\frac{1}{2}}^*)^{-1}(\gamma)} \frac{(x_{\frac{1}{2}}^*(\phi_i))^{\ell_i}\,\dd x_{\frac{1}{2}}^*(\phi_i)}{2{\rm i}\pi}\, [\pmb{\mathscr{F}}_{\Gamma,\star,\mathbf{1}}^{(\mathsf{g},k)}]_{*}(\phi_1,\ldots,\phi_k).
\eea

\vspace{0.2cm}

\begin{center}
\textit{Specialization to Theorem~\ref{LAPB}}
\end{center}

\vspace{0.1cm}

We first consider $\pmb{\mathscr{F}}^{(0,2)}_{\Gamma,\star,s = 1}$. From Corollary~\ref{dgfsgg}, the first term leads us to the previous setting with
\beq
\label{exonde}\eta + \tfrac{1}{2}k_0 = \widetilde{\beta}^{(0,2)}(s,\varepsilon_1,\varepsilon_2) + \tfrac{1}{2}k_0 = \left\{\begin{array}{lll} 0 & & {\rm if}\,\,\varepsilon_1 =  \varepsilon_2 = 0, \\ \tfrac{b(s)}{2} && {\rm if}\,\,\varepsilon_1 \neq \varepsilon_2, \\ b(s) & & {\rm if}\,\,\varepsilon_1 = \varepsilon_2 = \tfrac{1}{2}, \end{array}\right.
\eeq
with $s = 0$ for $(\Gamma_1,\star)$ and $s = 1$ for $(\Gamma_2,\star)$. However, in the case of two large boundaries ($\varepsilon_1 = \varepsilon_2 = 0$), we see that this first term contains no power of $q$, so is regular in $u$. The leading contribution in this case comes from the second term, hence corresponds to an exponent
\beq
\label{exonde2} \eta + \tfrac{1}{2}k_0 = b(s), \quad {\rm if}\,\,\varepsilon_1 = \varepsilon_2 = 0.
\eeq
So, we obtain the desired result by specializing \eqref{Ibar} to the exponent \eqref{exonde} corrected by \eqref{exonde2} and
\beq
\label{Phoih} \Phi\Big[1;(x_{\frac{1}{2}}^*(\phi_i))_{i \in \partial_{1/2}(\Gamma)};\big(x_0^*(\phi_i)\big)_{i \in \partial_{0}(\Gamma)}\Big] = \left\{\begin{array}{lll} \mathbf{F}_{s\,**}^{(2)}(\phi_1,\phi_2) & & {\rm if}\,\,\varepsilon_1 = \varepsilon_2 = 0, \\ \mathbf{F}_{s\,*}^{(2)}(\phi_1,\phi_2) && {\rm otherwise}, \end{array}\right. 
\eeq
with again $s = 0$ for $(\Gamma_1,\star)$ and $s = 1$ for $(\Gamma_2,\star)$. If we define $[\pmb{\mathscr{F}}^{(0,2)}_{\Gamma,\star,\mathbf{1}}]_*(\phi_1,\phi_2)$ to be the right-hand side of \eqref{Phoih}, the prefactors in \eqref{qdoguns}-\eqref{qdoguns2} are then also given by \eqref{Agks1} with the exponents $\eta$ we just saw.

\hfill $\Box$

\subsection{Large deviation for arm lengths in a fixed nesting graph}
\label{Fixedasarms}

\begin{center}
\begin{figure}
\includegraphics[width=0.8\textwidth]{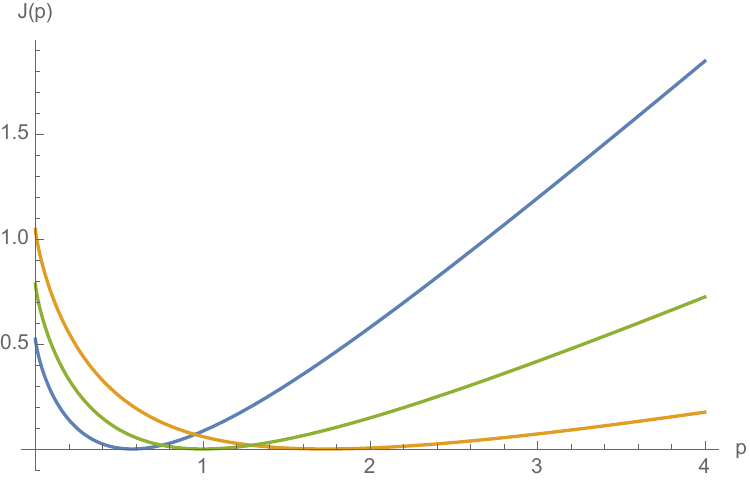}
\caption{\label{Jplot} The function $J(p)$: blue for $n = 1$, green for $n = \sqrt{2}$ (Ising), and orange for $n = \sqrt{3}$ ($3$-Potts).}
\end{figure}
\end{center}

Next, we also determine the asymptotics of the probability
\beq
\label{Pgk} \mathbb{P}^{(\mathsf{g},k)}\big[\mathbf{P}|\Gamma,\star,V,\mathbf{L}\big] \coloneqq \frac{\Big[u^{V} \prod_{\mathsf{e} \in E(\Gamma)} s(\mathsf{e})^{P(\mathsf{e})} \prod_{i = 1}^k x_i^{-(L_i+1)}\Big]\,\pmb{\mathscr{F}}^{(\mathsf{g},k)}_{\Gamma,\star,\mathbf{s}}(x_1,\ldots,x_k)}{\Big[ u^{V} \prod_{i = 1}^k x_i^{-(L_i+1)}\Big]\,\pmb{\mathscr{F}}^{(\mathsf{g},k)}_{\Gamma,\star,\mathbf{1}}(x_1,\ldots,x_k)}
\eeq
that a connected map of genus $\mathsf{g}$, of fixed volume $V$, with $k$ boundaries of fixed perimeters $\mathbf{L} = (L_i)_{i = 1}^k$, fixed nesting graph $(\Gamma,\star)$, has a number $P(\mathsf{e})$ of separating loops on every arm $\mathsf{e} \in E(\Gamma)$. We assume that $\Gamma$ has at least one arm for this to make sense. We introduce
\bea
J(p) & = & \sup_{s \in [0,2/n]} \big\{p\ln(s) + {\arccos}(ns/2) - {\rm arccos}(n/2)\big\} \nonumber \\
\label{Jdef} & = & p\ln\Big(\frac{2}{n}\,\frac{p}{\sqrt{1 + p^2}}\Big) + {\rm arccot}(p) - {\rm arccos}(n/2).
\eea
This function is plotted in Figure~\ref{Jplot}.

\begin{theorem}
\label{igfsgb}Take $(g,h)$ on the non-generic critical line. Assume $2\mathsf{g} - 2 + k > 0$, fix positive variables $\bs{\ell} = (\ell_i)_{i = 1}^k$ independent of $V$, and positive $\mathbf{p} = \big(p(\mathsf{e})\big)_{\mathsf{e} \in E(\Gamma)}$ such that $p(\mathsf{e}) \ll \ln V$. We consider the regime where $k_0$ boundaries have perimeter $L_i = \ell_i V^{c/2}$, $k_{1/2}$ boundaries have perimeter $L_i = \ell_i$, and
\beq
\label{factor2} P(\mathsf{e}) = \frac{c\ln V\,p(\mathsf{e})}{\pi}\cdot \left\{\begin{array}{lll} \frac{1}{2}, & & {\rm if}\,\,\mathsf{e} \in E_{0,2}^{{\rm S}}(\Gamma) \\ 1 & & {\rm if}\,\,\mathsf{e} \in E'(\Gamma). \end{array}\right. 
\eeq
In the limit $V \rightarrow \infty$, we have
\beq
\label{Omun} \mathbb{P}^{(\mathsf{g},k)}\big[\mathbf{P}|\Gamma,\star,V,\mathbf{L}\big] \sim \mathscr{P}^{(\mathsf{g},k)}\big[\mathbf{p}|\Gamma,\star,\bs{\ell}\big]\,\prod_{\mathsf{e} \in E'(\Gamma)} V^{-\frac{c}{\pi}\,p(\mathsf{e})\ln \frac{2}{n}}\,\prod_{\mathsf{e} \in E_{0,2}^{{\rm S}}(\Gamma)} \frac{V^{-\frac{c}{2\pi}\,J[p(\mathsf{e})]}}{\sqrt{\ln V}},
\eeq
We omit the expression of the prefactor.
\end{theorem}

For completeness we recall the result for $(\mathsf{g},k) = (0,2)$ from \cite[Theorem 7.1]{BBD}. Let $\mathbb{P}^{(0,2)}[P|V,L_1,L_2]$ be the probability that a cylinder with boundaries of perimeters $L_1$ and $L_2$ has exactly $P$ separating loops.

\begin{theorem}
\label{CylGG}Take $(g,h)$ on the non-generic critical line. Fix positive variables $(\ell_1,\ell_2)$ independent of $V$, and $p$ positive such that $p \ll \ln V$. We have when $V \rightarrow \infty$
\bea
\label{O92}\mathbb{P}^{(0,2)}\Big[P = \frac{c\ln V}{\pi}\,p\,\Big|\,V\,,\,L_1 = \ell_1\,,\,L_2 = \ell_2\Big] & \sim & \mathscr{P}^{(0,2)}_{{\rm 1}}(p,\ell_1,\ell_2)\,\frac{V^{-\frac{c}{\pi} J(p)}}{\sqrt{\ln V}},  \\
\mathbb{P}^{(0,2)}\Big[P = \frac{c\ln V}{2\pi}\,p\,\Big|\,V\,,\, L_1 = \ell_1\,,\,L_2 = \ell_2 V^{c/2}\Big] & \sim & \mathscr{P}^{(0,2)}_{2}(p,\ell_1,\ell_2)\,\frac{V^{-\frac{c}{2\pi}\,J(p)}}{\sqrt{\ln V}},  \nonumber \\
\mathbb{P}^{(0,2)}\Big[P = \frac{c\ln V}{\pi}\,p\,\Big|\,V\,,\,L_1 = \ell_1 V^{c/2}\,,\,L_2 = \ell_2 V^{c/2} \Big]  & \sim & \mathscr{P}^{(0,2)}_{3}(p,\ell_1,\ell_2)\,V^{-\frac{c}{\pi}\,p\,\ln\tfrac{2}{n}}. \nonumber  
\eea
\end{theorem}

From Theorem~\ref{igfsgb} one concludes that, for a given nesting graph $\Gamma$, the arm lengths for configurations of higher topology also behave typically like independent random variables. Recall that the analysis of the generating series of configurations with a fixed nesting graph showed that for the gluing annuli, which contain the inner boundaries of the arms, large lengths give effectively dominant contributions. The arms that correspond to an edge $\mathsf{e}$ incident to a small boundary have a depth of order typically $\ln V$, with large deviation function proportional to $J(p)$. This is exactly the same behavior as for the depth of a cylinder with at least one small boundary (see Theorem~\ref{CylGG}, first and second case). 

All the other arms \,--\, for higher topologies, arms with both boundaries large (either both interior, or one interior and one exterior) \,--\, have typically a finite depth, and we only see the exponential tail $(n/2)^{P}$ for the distribution of their depth. This is exactly the same behavior as for the depth of a cylinder with two large boundaries (see Theorem~\ref{CylGG}, third case).

Moreover, when we consider a configuration of higher topology and force the volume $V\rightarrow \infty$, the infinite volume will typically concentrate in the components corresponding to vertices in $\tilde{V}(\Gamma)$ (\textit{i.e.} vertices which are not univalent of genus $0$) and the arms that correspond to an edge incident to a vertex in $V_{0,2}^{{\rm S}}(\Gamma)$. In other words, the arms that correspond to an edge not incident to a vertex in $V_{0,2}^{{\rm S}}(\Gamma)$ will typically have finite volume and will contain finitely many separating loops.

The function $J(p)$ is universal as large deviation function for depths of cylinders of large volume, up to some factors of $2$ that are prescribed by the geometry. Focusing around the point
$$
p_{\rm opt} = \frac{n}{\sqrt{4 - n^2}},
$$
where $J$ reaches its minimum value $0$, we obtain:
\begin{corollary}
\label{Gaussflu} Consider the ensemble of connected maps of genus $\mathsf{g}$ with $k$ boundaries of perimeters $\mathbf{L}$, with volume $V$, realizing a fixed nesting graph $(\Gamma,\star)$. Under the assumptions and the regime described in Theorem~\ref{igfsgb}, the vector of random variables
$$
\left(\frac{P(\mathsf{e}) - \frac{c\,p_{{\rm opt}}\ln V}{2\pi}}{\sqrt{\ln V}}\right)_{\mathsf{e}\in E_{0,2}^{{\rm S}}(\Gamma)}
$$
converges in law when $V \rightarrow \infty$ to the random Gaussian vector $\big(\mathscr{N}(0,\sigma^2)\big)_{\mathsf{e}\in E_{0,2}^{{\rm S}}(\Gamma)}$ with variances
$$
\sigma^2 = \frac{2 \, n \, c}{\pi(4 - n^2)^{\frac{3}{2}}}.$$
\end{corollary}

\noindent \textbf{Proof of Theorem~\ref{igfsgb}.} Our starting point is \eqref{nuingfsg}:
\bea
\pmb{\mathscr{F}}_{\Gamma,\star,\mathbf{s}}^{(\mathsf{g},k)}(x_1,\ldots,x_k)\! & = &\! \Big[\prod_{\mathsf{e} \in E} \frac{1}{4 - n^2s^2}\Big]\,q^{\varkappa(\mathsf{g},k,k_{1/2},k_{1/2}^{(0,2)}) + \sum_{\mathsf{e} \in E_{0,2}^{{\rm S}}} \frac{1}{2}(b[s(\mathsf{e})] - b)} \nonumber \\
\!&&\! \bigg\{\sum_{\tau\,:\,E'(\Gamma) \rightarrow \{0,1\}} \Big[\prod_{\mathsf{e} \in E'(\Gamma)} q^{\tau(e)\,b[s(\mathsf{e})]} \Big]\cdot \bigg([\pmb{\mathscr{F}}_{\Gamma,\star,\mathsf{s}}^{(\mathsf{g},k)}]_{*,\tau} + O\Big(\sum_{\mathsf{e} \in E(\Gamma)} q^{\frac{b[s(\mathsf{e})]}{2}}\Big)\bigg) \bigg\}. \nonumber \\
\!&&\! \label{718e} 
\eea 
We denote $\overline{\mathcal{I}}^{(\tau)}_{\mathbf{s}}$ the contribution attached to $\tau\,:\,E'(\Gamma) \rightarrow \{0,1\}$ in the above sum and get an expression of the form
\bea
\mathcal{J}^{(\tau)} & = & \oint \prod_{\mathsf{e} \in E} \frac{\dd s(\mathsf{e})}{2\pi {\rm i}\,s(\mathsf{e})^{P(\mathsf{e}) + 1}}\,\prod_{i = 1}^{k} \frac{\dd x_i}{2\pi {\rm i}}\,x_i^{L_i}\,\frac{\dd u}{2 \pi {\rm i}\,u^{V + 1}}\,\overline{\mathcal{I}}^{(\tau)}_{\mathsf{s}} \nonumber \\
& \sim & V^{-1 - c(\varkappa + \tfrac{1}{2}k_0)} \oint \prod_{\mathsf{e} \in E} \frac{1}{4 - \mathsf{n}^2s^2} \prod_{\mathsf{e} \in E} \frac{\dd s(\mathsf{e})}{2\pi {\rm i}\,s(\mathsf{e})^{P(\mathsf{e}) + 1}} \nonumber \\
& & \cdot\, \frac{V^{-c\eta_{\tau}(\mathbf{s})}}{-\Gamma\big[-c(\varkappa + \frac{1}{2}k_0 + \eta_{\tau}(\mathbf{s}))\big]} \nonumber \\
&& \cdot\prod_{i \in \partial_{0}(\Gamma)} \oint_{(x_0^*)^{-1}(\overline{\mathcal{C}})} \frac{\dd x_0^*(\phi_i)\,e^{x_0^*(\phi_i)\ell_i/\gamma_+^*}}{2\pi {\rm i}} \prod_{i \in \partial_{1/2}(\Gamma)} \oint_{(x_{1/2}^*)^{-1}(\gamma)} \frac{(x_{1/2}^*(\phi_i))^{\ell_i}\,\dd x_{1/2}^*(\phi_i)}{2\pi {\rm i}} \nonumber \\
&& \cdot\, \Phi_{\mathbf{s}}^{(\tau)}\Big[1\,;\,\big(x_{1/2}(\phi_i)\big)_{i \in \partial_{1/2}(\Gamma)}\,;\,\big(x_0^*(\phi_i)\big)_{i \in \partial_0(\Gamma)}\Big], \nonumber 
\eea 
with
$$
\eta_{\tau}(\mathbf{s}) = \sum_{\mathsf{e} \in E_{0,2}^{{\rm S}}(\Gamma)} \tfrac{1}{2}(b[s(\mathsf{e})] - b) + \sum_{\mathsf{e} \in E'(\Gamma)} \tau(e)\,b[s(\mathsf{e})].
$$

It is natural to study the regime
\beq
P(\mathsf{e}) = \frac{c\, p(\mathsf{e}) \ln V}{\jmath(\mathsf{e})\pi}, \; \text{ with } \jmath(\mathsf{e}) = \left\{\begin{array}{lll} 2, & &  \text{ if }\,\mathsf{e}\in E_{0,2}^{{\rm S}}(\Gamma), \\ 1, & & \text{ otherwise}, \end{array}\right. \nonumber
\eeq 
for $p(\mathsf{e}) > 0$ independent of $V$. If we extend the map $\tau\,:\,E'(\Gamma) \rightarrow \{0,1\}$ to a map $\tau\,:\,E(\Gamma) \rightarrow \{0,1\}$ by declaring $\tau(\mathsf{e}) = 1$ for $\mathsf{e} \in E_{0,2}^{{\rm S}}(\Gamma)$, the singular part of the integrand is of the form
\bea
\prod_{\mathsf{e}\in E} s(\mathsf{e})^{-P(\mathsf{e})} \prod_{\mathsf{e}\in\tau^{-1}(1)} V^{-c \frac{b(s(\mathsf{e}))}{\jmath(\mathsf{e})}} \prod_{\mathsf{e}\in\tau^{-1}(0)}\frac{1}{4-n^2s(\mathsf{e})^2}= \nonumber\\ \prod_{\mathsf{e} \in \tau^{-1}(1)} \exp\Big(\jmath(\mathsf{e})^{-1}\,\ln V\,\mathcal{S}_{p(\mathsf{e})}[s(\mathsf{e})]\Big) \prod_{\mathsf{e} \in \tau^{-1}(0)} \frac{s(\mathsf{e})^{-\frac{cp}{\pi}\,\ln V}}{4 - n^2s(\mathsf{e})^2},
\eea
with
$$
\mathcal{S}_{p}(s) = -\frac{cp\ln s}{\pi} - cb(s).
$$
We first compute the saddle point $\mathfrak{s}(p)$ of $\mathcal{S}_{p}$, i.e., the point such that $\mathcal{S}'_{p}(\mathfrak{s}(p)) = 0$. We find
$$
\mathfrak{s}(p) = \frac{2}{n}\,\frac{p}{\sqrt{1 + p^2}}.
$$
We also compute in terms of the function $J$ introduced in \eqref{Jdef}
$$
\mathcal{S}_{p}(\mathfrak{s}(p)) = \frac{c (\pi b - J(p))}{\pi}.
$$
For $\mathsf{e} \in \tau^{-1}(1)$ we perform the change of variables
$$
s(\mathsf{e}) = \mathfrak{s}[p(\mathsf{e})] + \frac{\tilde{s}(\mathsf{e})}{\sqrt{\ln V}}
$$
and find by Taylor expansion of $\mathcal{S}_{p}$ at order $2$ around $s = \mathfrak{s}(p)$:
$$ 
\frac{\dd s(\mathsf{e})}{2\pi {\rm i}\,s(\mathsf{e})^{P + 1}}\,V^{-cb(s)/\jmath(\mathsf{e})} \sim \frac{\dd \tilde{s}(\mathsf{e})}{2\pi {\rm i}\,\mathfrak{s}(p(\mathsf{e}))}\, (\ln V)^{-\frac{1}{2}}\,V^{\frac{\mathcal{S}_{p(\mathsf{e})}(\mathfrak{s}(p(\mathsf{e})))}{\jmath(\mathsf{e})}}\,\exp\Big(\frac{cn^2(p(\mathsf{e})^2 + 1)^2}{8\jmath(\mathsf{e})\pi p(\mathsf{e})}\,\tilde{s}(\mathsf{e})^2\Big),
$$ 
which remains valid when $p(\mathsf{e})$ is allowed to depend on $V$ such that $p(\mathsf{e}) \ll \ln V$ and $P(\mathsf{e}) \gg 1$. We then deform the contour in $\tilde{s}(\mathsf{e})$ to a steepest descent contour ${\rm i}\mathbb{R}$, and the properties of the integrand imply we can apply the monotone convergence theorem and computation of the Gaussian integral in $\tilde{s}$ yields when $V \rightarrow \infty$:
\bea 
\mathcal{J}^{(\tau)} & \dot{\sim} & \frac{V^{-1 - c(\varkappa + \frac{1}{2}k_0)}}{-\Gamma\big[-c\big(\varkappa + \frac{1}{2}k_0 +\eta_{\tau}(\underline{\mathfrak{s}}_{\tau}(\mathbf{p}))\big)\big]}\,\prod_{\mathsf{e} \in \tau^{-1}(1)} \frac{V^{cb/\jmath(\mathsf{e}) -cJ(p(\mathsf{e}))/\jmath(\mathsf{e})\pi}}{\sqrt{2j(\mathsf{e})^{-1}cp(\mathsf{e})(p^2(\mathsf{e}) + 1)\ln V}} \prod_{\mathsf{e} \in E_{0,2}^{{\rm S}}} V^{-\frac{b}{2}} \nonumber \\
\label{Seconffu}&& \cdot \prod_{\mathsf{e} \in \tau^{-1}(0)} V^{-\frac{c}{\pi}\,p(\mathsf{e})\,\ln(2/n)}
\eea
where
$$
\underline{\mathfrak{s}}_{\tau}(\mathbf{p}) = \Big(\big(\mathfrak{s}[p(\mathsf{e})]\big)_{\mathsf{e} \in \tau^{-1}(1)}\,,\,\big(\tfrac{2}{n}\big)_{\mathsf{e} \in \tau^{-1}(0)}\Big)\,.
$$
The contour integral over $s(\mathsf{e})$ for $\tau(\mathsf{e}) = 0$ was easy to calculate and just produces $(n/2)^{P(\mathsf{e})}$ and appears in an equivalent form in the second line of \eqref{Seconffu}. Here we had to separate cases for $P(\mathsf{e})$ even or odd, and check they both give the same contribution, taking into account the prefactors $\hat{\mathcal{W}}^{[0,2]}_{s;0,*}$ \eqref{FFFFFF0} and $\tilde{\mathcal{W}}^{[0,2]}_{s;*}$ \eqref{fnun0}.

As we need to sum over $\tau$ as in \eqref{718e}, we have to compare for $\mathsf{e} \in E'(\Gamma)$ the factor coming from $\tau(\mathsf{e}) = 0$
$$
V^{-\frac{c}{\pi}\,p(\mathsf{e})\,\ln(2/n)}
$$ 
to the factor coming from $\tau(\mathsf{e}) = 1$
$$
V^{-c(\frac{J(p(\mathsf{e})}{\pi} + b)}.
$$
Since
$$
c\left(\frac{J(p)}{\pi}+b\right) > \ln\left(\frac{2}{n}\right)\frac{cp}{\pi}, \;\; \text{ for all } p \geq 0,
$$
the term with $\tau(\mathsf{e}) = 0$ for all $\mathsf{e} \in E'(\Gamma)$ dominates. We conclude by dividing by the asymptotic exponent of the numerator which has been previously obtained.
\hfill $\Box$

\subsection{Addendum: generating series of maps with marked points}

\label{crititmarked}

We now generalize Theorem~\ref{ouqusf} to allow marked points. 
\begin{lemma}
Let $k = k_0 + k_{1/2} \geq 1$ and $\mathsf{g} \geq 0$ such that $(\mathsf{g},k) \neq (0,1)$. Let $x_j = x(\tfrac{1}{2} + \tau \phi_j)$ for $j \in \{1,\ldots,k_{1/2}\}$, i.e.~$x_j$ remains finite and away from $[\gamma_-^*,\gamma_+^*]$. Let $y_j = x(\tau \psi_j)$
 for $j \in \{1,\ldots,k_0\}$, i.e.~$y_j$ scales with $q \rightarrow 0$ such that $y_j - \gamma_+ \in O(q^{\frac{1}{2}})$. We have in the critical regime $q \rightarrow 0$
\bea
&& \bs{\mathcal{F}}^{(\mathsf{g},k,\bullet k')}(x_1,\ldots,x_{k_{1/2}},y_{1},\ldots,y_{k_0}) \nonumber \\
& = & \Big(\frac{\pi}{T}\Big)^{k}\,q^{\widetilde{\beta}(\mathsf{g},k + k',k_{1/2} + k')}\big\{\bs{\mathcal{F}}^{(\mathsf{g},k,\bullet k')}_*(\phi_1,\ldots,\phi_{k_{1/2}},\psi_{1},\ldots,\psi_{k_0}) + O(q^{\frac{b}{2}})\big\}. \nonumber
\eea
This is also true for $(\mathsf{g},k) = (0,1)$.
\end{lemma} 
\vspace{0.2cm}
The outcome is that marked points behave as small boundaries. Subsequently, the asymptotics of the generating series $\pmb{\mathscr{F}}_{\Gamma,\star,\mathbf{s}}^{(\mathsf{g},k)}$ given by Proposition~\ref{P212} in presence of $k'$ marked points are the same as obtained in Theorem~\ref{CoscrF}, provided one replaces $k_{1/2}$ with $k_{1/2} + k'$, and likewise for Theorem~\ref{LAPA} concerning fixed volume asymptotics, and Theorem~\ref{igfsgb} concerning fixed volume and fixed arm lengths asymptotics.

\vspace{0.2cm}

\noindent \textbf{Proof.} First assume $(\mathsf{g},k) \neq (0,1)$. We proceed by recursion, starting from the base case $k' = 0$ obtained in Theorem~\ref{ouqusf}:
$$
\bs{\mathcal{F}}^{(\mathsf{g},k)}(\mathbf{x},\mathbf{y}) = \Big(\frac{\pi}{T}\Big)^{k}\,q^{\widetilde{\beta}(\mathsf{g},k,k_{1/2})}\,\Phi\Big[u;(x_i)_{i = 1}^{k_{1/2}};\Big(\frac{y_i - \gamma_+^*}{q^{\frac{1}{2}}}\Big)_{i = 1}^{k_0}\Big],
$$
where $\Phi$ is a function which has a uniform limit when $u \rightarrow 1$ and its other variables remain in a compact, and
\beq
\widetilde{\beta}(\mathsf{g},k,k_{1/2}) = (2\mathsf{g} - 2 + k)(\mathfrak{d}\tfrac{b}{2} - 1) - \tfrac{k}{2} + \tfrac{3}{4}k_{1/2}.
\eeq
We shall use \eqref{renFmarked} to decrease the value of $k'$. Assume the claim holds for $k''$ marked points with $k'' < k'$. Equation \eqref{renFmarked} gives us
\bea
\bs{\mathcal{F}}^{(\mathsf{g},k,\bullet k')}(\mathbf{x},\mathbf{y}) & = & \Big(2 - 2\mathsf{g} - k - \sum_{i = 1}^{k_{1/2}} \tfrac{1}{2}\,\partial_{x_i} x_i - \sum_{i = 1}^{k_{0}} \tfrac{1}{2}\,\partial_{y_i}y_i\Big)\bs{\mathcal{F}}^{(\mathsf{g},k,\bullet (k' - 1))}(\mathbf{x},\mathbf{y}) \nonumber \\
\label{tehrer} & & - \oint_{\gamma} \frac{\dd z}{2{\rm i}\pi}\,\big(\tfrac{z}{2}\tilde{\mathbf{V}}'(z) - \tilde{\mathbf{V}}(z)\big)\bs{\mathcal{F}}^{(\mathsf{g},k + 1,\bullet (k' - 1))}(z,\mathbf{x},\mathbf{y}), 
\eea
with
\bea 
\tilde{\mathbf{V}}'(x) & = & \mathbf{V}'(x) - \oint_{\gamma} \mathbf{A}(x,z)\mathbf{F}(z) \nonumber \\
& = & \mathbf{V}'(x) + n\varsigma'(x)\mathbf{F}(\varsigma(x)) - \frac{nu\varsigma''(x)}{2\varsigma'(x)}. \nonumber
\eea
We can substitute in this expression the function $\mathbf{G}$ introduced in \eqref{EquationG}:
\beq
\tilde{\mathbf{V}}'(x) = \mathbf{V}'(x) -n\frac{\mathbf{G}(\tau - v)}{x'(v)} + \frac{n\big(2\varsigma'(x)\mathbf{V}'(\varsigma(x)) + n\mathbf{V}'(x)\big)}{4 - n^2} - \frac{nu\varsigma''(x)}{\varsigma'(x)}. \nonumber
\eeq
The critical behavior of $\mathbf{G}(v)$ when $v = \varepsilon + \tau w$ with $\varepsilon \in \{0,\tfrac{1}{2}\}$, and $q = e^{{\rm i}\pi \tau} \rightarrow 0$ is obtained from substituting its expression from Proposition~\ref{theimdisk}, using the asymptotics of the function $\Upsilon_b$ in \ref{lemUp}, and the identities \eqref{D1}-\eqref{D2}. The result takes the form
$$
\mathbf{G}(\tau - v) = q^{(1 - 2\varepsilon)(1 - \mathfrak{d}\frac{b}{2})}\big\{\tilde{\mathbf{G}}^*_{\varepsilon}(\phi) + O(q^{\frac{b}{2}})\big\}.
$$ 
Besides, the induction hypothesis tells us that the order of magnitude of$$
\bs{\mathcal{F}}^{(\mathsf{g},k + 1,\bullet (k' - 1))}(x(v),\mathbf{x},\mathbf{y})
$$
receives an extra factor of $q^{\frac{3}{4}}$ when $v = \frac{1}{2} + \tau\phi$ with $\phi$ in a compact. As $b \in (0,\tfrac{1}{2})$, in any case we have $\tfrac{3}{4} < 1 - \mathfrak{d}\tfrac{b}{2}$ and therefore the contribution of the vicinity (at scale $q^{\frac{1}{2}}$) of $\gamma_+^*$ in the contour integral over $\gamma$ in the second line of  \eqref{tehrer} remains negligible compared to the contribution of the bulk of the contour (given by the regime $\varepsilon = \tfrac{1}{2}$). And, by induction hypothesis, this contribution is of order $q^{\widetilde{\beta}(\mathsf{g},(k + k' - 1) + 1,(k_{1/2} + k' - 1) + 1)}$, where the $+1$ come from the variable $z \in \gamma$. On the other hand, the first line in \eqref{tehrer} has a contribution of order $q^{\widetilde{\beta}(\mathsf{g},k + k' - 1,k_{1/2} + k' - 1)}$. As 
$$
\widetilde{\beta}(\mathsf{g},k + k' - 1,k_{1/2} + k' - 1) - \widetilde{\beta}(\mathsf{g},k + k',k_{1/2} + k') = \tfrac{1}{2} - \mathfrak{d}\tfrac{b}{2}  > 0
$$
the first line is always negligible compared to the second line, and this gives the claim for $k'$ marked points. We conclude for all $(\mathsf{g},k) \neq (0,1)$ by induction.
 
 \vspace{0.2cm}
 
Now consider $(\mathsf{g},k) = (0,1)$. For $k' = 1$, we have from \eqref{diskpointed}:
$$
\bs{\mathcal{F}}^{\bullet}(x) = \frac{1}{\sqrt{(x - \gamma_+)(x - \gamma_-)}},
$$
Therefore with $x = x(\tau \phi) = \gamma_+^* + q^{\frac{1}{2}}x_0^*(\phi)$ in the critical regime
$$
\bs{\mathcal{F}}^{\bullet}(x) \sim q^{-\frac{1}{4}}\,\bs{\mathcal{F}}^{\bullet}_{*}(\phi),
$$
whose exponent agrees with  $\widetilde{\beta}(\mathsf{g},k + k' = 2,k_{1/2} + k' = 1)$. On the other hand, for $x = x(\tfrac{1}{2} + \tau\phi)$ in the critical regime, we have
$$
\bs{\mathcal{F}}^{\bullet}(x) = \frac{1}{\sqrt{(x - \gamma_+^*)(x - \gamma_-^*)}} + q^{\frac{1}{2}}\,\tilde{\bs{\mathcal{F}}}^{\bullet}_{*}(\phi) + O(q)
$$
coming from the behavior of $\gamma_+$ when $q \rightarrow 0$ as given by Corollary~\ref{CoB5}. This exponent $\frac{1}{2}$ agrees with $\widetilde{\beta}(\mathsf{g} = 0,k + k' = 2,k_{1/2} + k' = 2)$. With these two cases as initial conditions and the previous results, we can repeat the previous steps to show from \eqref{tehrer} that the claim holds for $(\mathsf{g},k) = (0,1)$ for any $k' > 0$. \hfill $\Box$

\appendix
\newpage

\section{The special function $\Upsilon_b$}

\label{AppA}
\label{AppUp}
Let $\tau$ be a complex number in the upper-half plane. The Jacobi theta function is the entire function of $v \in \mathbb{C}$ defined by
\beq
\vartheta_{1}(v|\tau) = -\sum_{m \in \mathbb{Z}} e^{{\rm i}\pi \tau (m + \frac{1}{2})^2 + {\rm i}\pi (w  + \frac{1}{2})(2m + 1)}.
\eeq
Its main properties are
\beq
\vartheta_{1}(-v|\tau) = \vartheta_{1}(v + 1|\tau) = -\vartheta_{1}(v|\tau),\qquad \vartheta_{1}(v + \tau|\tau) = -e^{-2{\rm i}\pi (v  + \frac{\tau}{2})}\,\vartheta_{1}(v|\tau)
\eeq
and the effect of the modular transformation:
\beq
\label{modularS}\vartheta_{1}(v|\tau) = \frac{e^{-\frac{{\rm i}\pi v^2}{\tau}}}{\sqrt{-{\rm i}\tau}}\,\vartheta_{1}(\tfrac{v}{\tau}|-\tfrac{1}{\tau}).
\eeq

\begin{definition}
We define $\Upsilon_{b}(v)$ as the unique meromorphic function with a simple pole at $v = 0$ with residue $1$, and the pseudo-periodicity properties:
$$
\Upsilon_{b}(v + 1) = \Upsilon_{b}(v),\qquad \Upsilon_{b}(v + \tau) = e^{{\rm i}\pi b}\Upsilon_{b}(v).
$$
We have several expressions:
\bea
\Upsilon_{b}(v) & = & \sum_{m \in \mathbb{Z}} e^{-{\rm i}\pi b m}\,\mathrm{cotan}\,\pi(v + m\tau) \nonumber \\
& = & \frac{\vartheta_{1}'(0|\tau)}{\vartheta_{1}(-\frac{b}{2}|\tau)}\,\frac{\vartheta_{1}(v - \frac{b}{2}|\tau)}{\vartheta_{1}(v|\tau)} \nonumber \\
\label{210eq}& = &  \frac{e^{\frac{{\rm i}\pi b v}{\tau}}}{{\rm i}T}\,\frac{\vartheta_1'\big(0|-\frac{1}{\tau}\big)}{\vartheta_1\big(-\frac{b}{2\tau}\big|-\frac{1}{\tau}\big)}\,\frac{\vartheta_{1}\big(\frac{v - \frac{b}{2}}{\tau}\big|-\frac{1}{\tau}\big)}{\vartheta_{1}\big(\frac{v}{\tau}|-\frac{1}{\tau}\big)}.
\eea
\end{definition}

We have the expansion:
\beq
\label{expord1Upsilon}\Upsilon_{b}(w) = \frac{1}{w} + \sum_{j\geq 0}  \upsilon_{b,j} w,\qquad w \rightarrow 0,
\eeq
with
\bea
\label{upb1def} \upsilon_{b,1} & = &  \frac{1}{2}\,\frac{\vartheta_{1}''(\frac{b}{2}|\tau)}{\vartheta_{1}(\frac{b}{2}|\tau)} - \frac{1}{6}\,\frac{\vartheta_{1}'''(0|\tau)}{\vartheta_{1}'(0|\tau)}  \\
& = & \frac{1}{({\rm i}T)^2}\Big(\frac{1}{2}\,\frac{\vartheta_{1}''(\frac{b\tilde{\tau}}{2}|\tilde{\tau})}{\vartheta_{1}(\frac{b\tilde{\tau}}{2}|\tilde{\tau})} -\frac{1}{6}\,\frac{\vartheta_{1}'''(0|\tilde{\tau})}{\vartheta_{1}'(0|\tilde{\tau})} + {\rm i}\pi b \frac{\vartheta_{1}'(\frac{b\tilde{\tau}}{2}|\tilde{\tau})}{\vartheta_{1}(\frac{b\tilde{\tau}}{2}|\tilde{\tau})} - \frac{\pi^2b^2}{2}\Big), \nonumber
\eea 
where $\tilde{\tau} = -1/\tau$ and $\tau = {\rm i}T$. The value of the constant term in \eqref{expord1Upsilon} is irrelevant for our purposes. The expressions involving $\tilde{\tau}$ or
$$
q = e^{{\rm i}\pi\tilde{\tau}} = e^{-\frac{\pi}{T}}
$$
are convenient to study the regime $T \rightarrow 0$, i.e.~$q \rightarrow 0$.
\begin{lemma}
\label{lemUp}Let $v = \varepsilon + \tau w$. We have, for $b \in (0,1)$:
{\small $$
\Upsilon_{b}(v) = \frac{2\pi q^{\varepsilon b}}{T(1 - q^b)} \cdot \left\{\begin{array}{lll} \Upsilon_{b,0}^*(w)  - q^b\Upsilon_{b + 2,0}^*(w) + O(q^{2 - b}) & & {\rm if}\,\,\varepsilon = 0, \\ \Upsilon_{b,\frac{1}{2}}^*(w) - (q^{1 - b} - q)\Upsilon_{b - 2,\frac{1}{2}}^*(w) + q\Upsilon_{b + 2,\frac{1}{2}}^*(w) + O(q^{1 + b}) & & {\rm if}\,\,\varepsilon = \tfrac{1}{2}.\end{array}\right.
$$}
The errors are uniform for $w$ in any compact independent of $\tau \rightarrow 0$, stable under differentiation, and the expressions for the limit functions are
\bea
\label{theone}\Upsilon_{b,0}^*(w) & = & \frac{e^{{\rm i}\pi (b -1)w}}{2{\rm i}\sin(\pi w)}, \\
\label{thehalf}\Upsilon_{b,\frac{1}{2}}^*(w) & = & - e^{{\rm i}\pi b w}.\eea
We also have
\beq
\upsilon_{b,1} = \Big(\frac{\pi}{T}\Big)^2\Big\{\frac{1}{3} + b + \frac{b^2}{2} + O(q^{b})\Big\}.
\eeq
\hfill $\Box$
\end{lemma}

\section{The parametrization $x \leftrightarrow v$}
\label{App1}
\label{xbeh}

Consider given values of $\gamma_{\pm}$ and $\varsigma(\gamma_{\pm})$ such that
\beq
\label{orderg}\gamma_- < \gamma_+ < \varsigma(\gamma_+) < \varsigma(\gamma_-).
\eeq
We set
\beq
\label{io} v = {\rm i}C\,\int^{x}_{\varsigma(\gamma_+)} \frac{\dd y}{\sqrt{(y - \varsigma(\gamma_-))(y - \varsigma(\gamma_+))(y - \gamma_+)(y - \gamma_-)}}.
\eeq
The normalizing constant is chosen such that, for $x$ moving from the origin $\varsigma(\gamma_+)$ to $\varsigma(\gamma_-)$ with a small negative imaginary part, $v$ is moving from $0$ to $\tfrac{1}{2}$. When $x$ moves on the real axis from $\varsigma(\gamma_+)$ to $\gamma_+$, $v$ moves from $0$ to a purely imaginary value denoted $\tau = {\rm i}T$. Then, the function $v \mapsto x(v)$ has the properties:
$$
x(v + 2\tau) = x(v + 1) = x(-v) = x(v),\qquad \varsigma(x(v)) = x(v - \tau),  
$$
and is depicted in Figure~\ref{ParamF}. $x'(v)$ has zeroes when $v \in \tfrac{1}{2}\mathbb{Z} + \tau\mathbb{Z}$, and double poles at $v = v_{\infty} + \mathbb{Z} + 2\tau\mathbb{Z}$. From \eqref{io}, paying attention to the determination of the squareroot at infinity obtained by analytic continuation, we can read in particular:
\beq
x'(v) \sim \frac{{\rm i}C}{(v - v_{\infty})^2},\qquad v \rightarrow v_{\infty}.
\eeq
From \eqref{orderg}, we know that $v_{\infty} = \frac{1}{2} + \tau w_{\infty}$, where $w_{\infty} \in (0,1)$ is determined as a function of $\gamma_{\pm}$ and $\varsigma(\gamma_{\pm})$. 

There is an alternative expression for \eqref{io} in terms of Jacobi functions:
$$
v = \frac{2{\rm i}C\,{\rm arcsn}^{-1}\Big[\sqrt{\frac{\varsigma(\gamma_+) - \gamma_-}{\varsigma(\gamma_-) - \gamma_-}\,\frac{x - \varsigma(\gamma_+)}{x - \varsigma(\gamma_-)}}\,;\,k\Big]}{\sqrt{(\varsigma(\gamma_+) - \gamma_-)(\varsigma(\gamma_-) - \gamma_+)}},
$$
with
$$
k = \sqrt{\frac{(\varsigma(\gamma_-) - \gamma_-)(\varsigma(\gamma_+) - \gamma_+)}{(\varsigma(\gamma_-) - \gamma_+)(\varsigma(\gamma_+) - \gamma_-)}}.
$$
By specialization at $x = \gamma_-$ and $x = \varsigma(\gamma_-)$, we deduce the expressions:
\bea
\label{XCeq} C & = & \frac{\sqrt{(\varsigma(\gamma_+) - \gamma_-)(\varsigma(\gamma_-) - \gamma_+)}}{4K'(k)}, \\
\label{TCeq} T & = & \frac{K(k)}{2K'(k)},
\eea
in terms of the complete elliptic integrals. By matching poles and zeroes, we can infer an expression for $x(v) - \gamma_+$ in terms of Jacobi theta functions:
\beq
\label{Xtheta} x(v) - \gamma_+ = -{\rm i}C\,\frac{\vartheta_{1}'(0|2\tau)\vartheta_{1}(2v_{\infty}|2\tau)}{\vartheta_{1}(v_{\infty} - \tau|2\tau)\vartheta_{1}(v_{\infty} + \tau|2\tau)}\,\frac{\vartheta_{1}(v - \tau|2\tau)\vartheta_{1}(v + \tau|2\tau)}{\vartheta_{1}(v - v_{\infty}|2\tau)\vartheta_{1}(v + v_{\infty}|2\tau)}.
\eeq
From \eqref{io}, one can derive the expansion of $x(v)$ when $v \rightarrow v_{\infty}$. 
\begin{lemma}
\label{Xinfexp} When $v \rightarrow v_{\infty}$, we have the expansion
$$
x(v) = \frac{-{\rm i}C}{v - v_{\infty}} + \frac{E_{1}}{4} + \frac{{\rm i}}{C}\,\frac{3E_{1}^2 - 8E_{2}}{48}\,(v - v_{\infty}) + \frac{-E_1^3 + 4E_1E_2 - 8E_3}{64C^2}\,(v - v_{\infty})^2 + O(v - v_{\infty})^3,
$$
where we introduced the symmetric polynomials in the endpoints:
\bea
 E_{1} & = & \gamma_- + \gamma_+ + \varsigma(\gamma_+) + \varsigma(\gamma_-), \label{DefE} \\
 E_{2} & = & \gamma_-\big\{\gamma_+ + \varsigma(\gamma_+) + \varsigma(\gamma_-)\big\} + \gamma_+\big\{\varsigma(\gamma_+) + \varsigma(\gamma_-)\big\} + \varsigma(\gamma_+)\varsigma(\gamma_-),  \label{DefE2} \\
 E_{3} & = & \gamma_-\gamma_+\varsigma(\gamma_+) + \gamma_-\gamma_+\varsigma(\gamma_-)  + \gamma_-\varsigma(\gamma_-)\varsigma(\gamma_+) + \gamma_+\varsigma(\gamma_+)\varsigma(\gamma_-). \label{DefE3}
\eea
More generally, the coefficient of $(v - v_{\infty})^k$ in this expansion is a homogeneous symmetric polynomial of degree $(k + 1)$ with respect to the endpoints, with rational coefficients up to an overall factor $({\rm i}C)^{-k}$.\hfill $\Box$
\end{lemma}

In the study of non-generic critical points, we want to take the limit where $\gamma_+$ and $\varsigma(\gamma_+)$ collide to the fixed point of the involution:
$$
\gamma_+^* = \frac{1}{(\alpha + 1)h},
$$
while $\gamma_- \rightarrow \gamma_-^*$ remains distinct from $\varsigma(\gamma_-^*)$. This implies $T\rightarrow 0$, or equivalently $k \rightarrow 0$. This limit is easily studied using the modular transformation \eqref{modularS} in \eqref{Xtheta}, or the properties of the elliptic integrals. If we set
$$
q = e^{-\frac{\pi}{T}},
$$
we arrive to:
\begin{lemma}
\bea
q & = & \Big(\frac{k}{4}\Big)^{4}\Big\{1 + O(k^2)\Big\}, \nonumber \\
w_{\infty} & = & w_{\infty}^*\big\{1 + O(q^{\frac{1}{2}})\big\}. \nonumber 
\eea
\hfill $\Box$
\end{lemma}

We can then derive the critical behavior of the parametrization $x(v)$ in the two regimes of interest:
\begin{lemma}
\label{LemB3}Let $v = \varepsilon + \tau w$ for $\varepsilon \in \{0,\frac{1}{2}\}$. We have
$$
x(v) - \gamma_+ = q^{\frac{1}{2} - \varepsilon}\,\big\{x_{\varepsilon}^*(w) + O(q^{\frac{1}{2}})\big\}.
$$
The error is uniform for $w$ in any compact independent of $\tau \rightarrow 0$, and this is stable under differentiation with respect to $v$. It is actually an asymptotic series in $q^{\frac{1}{2}}$. The limit functions are
\bea
x_{0}^*(w) & = & 8\sqrt{(\varsigma(\gamma_-^*) - \gamma_+^*)(\gamma_+^* - \gamma_-^*)}\,\sin(\pi w_{\infty}^*)\,\cos^2\Big(\frac{\pi w}{2}\Big), \nonumber \\
x_{\frac{1}{2}}^*(w) & = & \sqrt{(\varsigma(\gamma_-^*) - \gamma_+^*)(\gamma_+^* - \gamma_-^*)}\,\frac{\sin(\pi w_{\infty}^*)}{\cos(\pi w) - \cos (\pi w_{\infty}^*)}. \nonumber
\eea
\hfill $\Box$
\end{lemma}
If we specialize the second equation to $v = \frac{1}{2} + \tau$, use the expression \eqref{varsigma} of $\varsigma(x)$ and perform elementary trigonometric manipulations, we find:
\begin{corollary}
\label{CoB4}
$$
{\rm cos}(\pi w_{\infty}^*) = \frac{1 - \alpha}{1 + \alpha}\cdot \frac{1 - h(1 + \alpha)\gamma_-^*}{1 + h(1 - \alpha)\gamma_-^*}.
$$ 
\hfill $\Box$ 
\end{corollary}
We may consider $w_{\infty}^*$ as a parameter for the non-generic critical line.  Specializing again Lemma~\ref{LemB3} to $v = \varepsilon + \tau$ and using Corollary~\ref{CoB4} yields:
\begin{corollary}
\label{CoB5} There exists a constant $\rho_1$ such that:
\bea
2h(\gamma_+^* - \gamma_+) & = & \frac{16\cos(\pi w_{\infty}^*)}{(1 - \alpha^2)}\,q^{\frac{1}{2}} + O(q), \nonumber \\
2h(\gamma_-^* - \gamma_-) & = & \rho_{1} q^{\frac{1}{2}} + O(q),  \nonumber
\eea
and
\bea
E_1 & = & \frac{1 - \alpha\sin^2(\pi w_{\infty}^*)}{(1 - \alpha^2)h \sin^2(\pi w_{\infty}^*)}+ \frac{2\rho_1\cos(\pi w_{\infty}^*)}{h(1 - \cos(\pi w_{\infty}^*))^2}\,q^{\frac{1}{2}} + O(q), \nonumber \\
E_2 & = & \frac{2\big((3\alpha^2 - 1)\sin^2(\pi w_{\infty}^*) - 2(3\alpha - 2)\big)}{(\alpha^2 - 1)^2h^2\sin^2(\pi w_{\infty}^*)} + \frac{2\rho_1(3\alpha - 2)}{h^2(1 - \alpha^2)(1 - \cos(\pi w_{\infty}^*))^2}\,q^{\frac{1}{2}} + O(q), \nonumber \\
E_3 & = & \frac{4\big(\alpha^2\sin^2(\pi w_{\infty}^*) - \alpha(2 + \cos^2(\pi w_{\infty}^*) + 1)\big)}{(1 - \alpha)^2(1 + \alpha)^3\sin^2(\pi w_{\infty}^*)h^3} + O(q^{\frac{1}{2}}). \nonumber
\eea
\end{corollary}
The first four lines are used in \cite{BBD} to describe the phase diagram (reviewed here in Section~\ref{Secphase}) and the critical exponents of the model. Straightforward computations with \eqref{XCeq}-\eqref{TCeq} yield:
\begin{corollary}
\label{CBCB6} \bea
\frac{\pi C}{T} & = & \sqrt{(\varsigma(\gamma_-) - \gamma_+^*)(\gamma_+^* - \gamma_-)} + O(q), \nonumber \\
& = & \frac{2\,{\rm cot}(\pi w_{\infty}^*)}{(1 - \alpha^2)h} + \frac{(1 + \cos(\pi w_{\infty}^*))\rho_1}{2(1 - \cos(\pi w_{\infty}^*))\sin(\pi w_{\infty}^*)}\,q^{\frac{1}{2}} + O(q). \nonumber
\eea
\hfill $\Box$
\end{corollary}

There are some simplifications in absence of bending energy, \textit{i.e.}, $\alpha = 1$. We then have $w_{\infty}^* = \frac{1}{2}$ which is in agreement with Corollary~\ref{CoB4}. The non-generic critical line is then parametrized by $\rho = 1 - 2h\gamma_-^*$, which is related to the former parametrization by letting $\alpha \rightarrow 1$ and $w_{\infty}^* \rightarrow \frac{1}{2}$ in such a way that
\beq
\label{conflua}\Big(\frac{1}{2} - w_{\infty}^*\Big) \sim \frac{(1 - \alpha)}{2\pi}\,\rho.
\eeq
Corollary~\ref{CoB5} specializes to:
\begin{corollary}
\label{CoB7} For $\alpha  = 1$,  we have:
\bea
2h(\gamma_+^* - \gamma_+) & = & O(q), \nonumber \\
E_1 & = & \frac{2}{h} + O(q), \nonumber \\
E_2 & = & \frac{6 - \rho^2}{4h^2} - \frac{\rho\rho_1}{2h^2}\,q^{\frac{1}{2}} + O(q), \nonumber \\
E_3 & = & \frac{2 - \rho^2}{4h^3} + O(q^{\frac{1}{2}}), \nonumber \\ 
\frac{\pi C}{T} & = & \frac{\rho}{2h} + \frac{\rho_1}{2h}\,q^{\frac{1}{2}} + O(q). \nonumber
\eea
\hfill $\Box$
\end{corollary}
The fact that $\varsigma(x) = \frac{1}{h} - x$ and $\gamma_+^* = \frac{1}{2h}$ gives the exact relation $E_1 = \frac{2}{h}$, in agreement with the second line.

\section{The coefficients $\tilde{g}_k$}
\label{Appgdeter}
In the loop model with bending energy where all faces are triangles, the parameters are: $g$ (resp. $h$) the weight per face not visited (resp. visited) by a loop, $\alpha$ the bending energy, and $n$ the weight per loop. We can compute $\tilde{g}_{k}$ from their definition \eqref{deftildeg} if we insert the expansion of Lemma~\ref{Xinfexp}. We recall that $C$ is the constant in \eqref{io}, and $E$'s are symmetric polynomials in the endpoints defined in Lemma~\eqref{Xinfexp}. If we introduce
$$
\tilde{g}_k = ({\rm i}C)^k\,\widehat{g}_{k},
$$
we find
\bea
\widehat{g}_{3} & = & \frac{2g}{4 - n^2}, \nonumber \\
\widehat{g}_{2} & = & \frac{2 - gE_1}{4 - n^2}, \nonumber \\
\widehat{g}_{1} & = & \frac{g(3E_{1}^2 - 4E_2) - 6E_1}{12(4 - n^2)}, \nonumber \\
\widehat{g}_{0} & = & -\frac{2u}{2 + n}. \nonumber
\eea
We remark that $\widehat{g}_{3}$ and $\widehat{g}_{0}$ depend on the parameters of the model in a very simple way, whereas $\widehat{g}_{1}$ and $\widehat{g}_{2}$ have a non-trivial behavior in the non-generic critical regime, which can be deduced up to $O(q)$ from Corollary~\ref{CoB5}, either in terms of the parameter $w_{\infty}^*$, or the parameter $\rho$ if $\alpha = 1$.

\begin{corollary}
We have:
\bea 
\widehat{g}_{2} & = & \frac{1}{4 - n^2}\Big[1 + \frac{2g}{h}\Big(\alpha - \frac{1}{\sin^2(\pi w_{\infty}^*)}\Big)\Big]  \nonumber \\
&& - \frac{g}{h}\,\frac{\rho_1\cos(\pi w_{\infty}^*)}{(1 - \cos(\pi w_{\infty}^*))^2(4 - n^2)}\,q^{\frac{1}{2}} + O(q), \nonumber \\
\widehat{g}_{1} & = & \frac{2g\big[(3\alpha^2 + 1)\sin^4(\pi w_{\infty}^*) + 2(3\alpha - 2)\sin^2(\pi w_{\infty}^*) + 6\big]}{3(1 - \alpha^2)^2h^2(4 - n^2)\sin^4(\pi w_{\infty}^*)} \nonumber \\  
&& + \frac{3h\sin^2(\pi w_{\infty}^*)(1 - \alpha^2)(\alpha\sin^2(\pi w_{\infty}^*) + 1)}{3(1 - \alpha^2)^2h^2(4 - n^2)\sin^4(\pi w_{\infty}^*)} \nonumber \\
&& + \frac{\rho_1\cos(w)\big\{2g\big[4 -3\alpha\sin^2(\pi w_{\infty}^*) + 2\cos^2(\pi w_{\infty}^*)\big] -  3h\sin^2(\pi w_{\infty}^*)(1 - \alpha^2)\big\}}{(1 - \cos(\pi w_{\infty}^*))^2\sin^2(\pi w_{\infty}^*)(1 - \alpha^2)h^2(4 - n^2)}\,q^{\frac{1}{2}} \nonumber \\
&& + O(q). \nonumber
\eea 

\hfill $\Box$
\end{corollary}

There are some simplifications for $\alpha = 1$. Owing to the exact relation $E_1 = \frac{2}{h}$, only $\widehat{g}_{1}$ has a non-trivial dependence in the non-critical regime:
\begin{corollary}
For $\alpha = 1$, we have:
\bea
\widehat{g}_{2} & = & \frac{2}{4 - n^2}\Big(1 - \frac{g}{h}\Big), \nonumber \\
\widehat{g}_{1} & = & \frac{1}{h(4 - n^2)}\Big(- 1 + \frac{g}{h}(\rho^2 + 6)\Big) + \frac{g\rho\rho_1}{h^2(4 - n^2)}\,q^{\frac{1}{2}} + O(q). \nonumber 
\eea
\hfill $\Box$
\end{corollary}

\section{Determination of the endpoints and phase diagram}
\label{proofbeh}
\label{AppD}
In this section, we recall the elements leading to the proofs of the theorems of Section~\ref{Secphase}, see \cite{BBD} for more details. The equations $\Delta_{\varepsilon}\mathbf{G}^{\bullet}(0) = 0$ for $\varepsilon \in \{0,\frac{1}{2}\}$ determine $\gamma_{\pm}$ in terms of the weights of the model. We compute from Proposition~\ref{theimdisk} and the behavior of $\Upsilon_{b}(\tau\phi + 1/2)$ given in Lemma~\ref{lemUp}:
\bea
\label{D1} \mathcal{D}Y_{b,0}(\pi w_{\infty}) - q^{1 - b}\mathcal{D}Y_{b - 2,0}(\pi w_{\infty}) + O(q) & = & 0, \label{eq1} \\
\label{D2} \mathcal{D}Y_{b,\frac{1}{2}}(\pi w_{\infty}) - q^{b} Y_{b + 2,\frac{1}{2}}^{(k)}(\pi w_{\infty}) + O(q) & = & 0, \label{eq2}
\eea
where
\beq
\label{DYB} Y_{b,0}(w) = \cos(b w),\qquad Y_{b,\frac{1}{2}}(w) = \frac{\sin[(1 - b)w]}{\sin w},\qquad \mathcal{D} = \sum_{l = 0}^3 \frac{(-1)^l \widehat{g}_{l}}{l!}\,\Big(\frac{\pi C}{T}\Big)^{l} \partial_{\pi w_{\infty}}^l.
\eeq

Exactly at criticality, we must have $u = 1$ and $q = 0$, thus using Corollary~\ref{CBCB6}:
$$
-\frac{2}{2 + n} + \sum_{k = 1}^{3} \frac{(-1)^k \widehat{g}_k^*}{k!}\,\Big(\frac{2{\rm cot}(\pi w_{\infty}^*)}{(1 - \alpha^2)h}\Big)^k\,\frac{Y_{b,\varepsilon}^{(k)}(\pi w_{\infty}^*)}{Y_{b,\varepsilon}(\pi w_{\infty}^*)} = 0,\qquad \varepsilon \in \{0,\tfrac{1}{2}\}.
$$
We note that the critical values $\widehat{g}_k^*$ obtained in Section~\ref{Appgdeter} are such that \eqref{eq1}-\eqref{eq2} give a system of two linear equations determining $\frac{g}{h}$ and $h^2$ in terms of the parameter $w_{\infty}^*$. For $\alpha = 1$, we rather use $\rho$ as parameter, and the solution is
\bea
\label{gsurh} \frac{g}{h} & = & \frac{4(\rho b\sqrt{2 + n} - \sqrt{2 - n})}{\rho^2(b^2 - 1)\sqrt{2 - n} + 4\rho b\sqrt{2 + n} - 2\sqrt{2 - n}}, \\
\label{h2eq} h^2 & = & \frac{\rho^2 b}{24\sqrt{4 - n^2}}\,\frac{\rho^2\,b(1 - b^2)\sqrt{2 + n}  - 4\rho\sqrt{2 - n} + 6b\sqrt{2 + n}}{-\rho^2(1 - b^2)\sqrt{2 - n} + 4\rho b\sqrt{2 + n} - 2\sqrt{2 - n}}.
\eea
Since $\tfrac{g}{h}$ and $h^2$ must be nonnegative, we must have $\rho \in [\rho_{\min}',\rho_{\max}]$ with
\bea
\label{E55} \rho'_{\min}  & = & \frac{2\sqrt{1 - b^2}\sqrt{2 - n} - \sqrt{2}\sqrt{(10 + n)b^2 - 4 + 2n}}{b\sqrt{1 - b^2}\sqrt{2 - n}},  \\
\label{E56} \rho_{\max}  & = & \frac{1}{b}\,\sqrt{\frac{2 - n}{2 + n}}.\eea
However, we will see later that the non-generic critical line only exists until some value $\rho_{\min} > \rho'_{\min}$, so \eqref{E55} will become irrelevant.
For $\alpha \neq 1$, see \cite[Appendix D]{BBD}.

Now, let us examine the approach of criticality. We fix $(g,h)$ on the non-generic critical line for $u = 1$, and we now study the behavior when $u \neq 1$ but $u \rightarrow 1$ of the endpoints $\gamma_{\pm}$. In particular, since the behavior of the elliptic functions is conveniently expressed in this regime in terms of $q = e^{-\frac{\pi}{T}}$, our first task is to relate $(1 - u)$ to $q \rightarrow 0$. For this purpose, we look at \eqref{eq1}, and note that $u$ only appears in $\widehat{g}_{0}$. There could be a term of order $q^{\frac{1}{2}}$ stemming from near-criticality corrections to $w_{\infty}$, $\widehat{g}_{k}$ and $\frac{\pi C}{T}$, but computation reveals that it is absent. Therefore, we obtain:
$$
1 - u = \frac{n + 2}{2}\bigg(\sum_{l = 0}^3 \frac{(-1)^{l} \widehat{g}_{l}^*}{l!}\Big(\frac{2\,{\rm cot}(\pi w_{\infty}^*)}{(1 - \alpha^2)h}\Big)^l\,\frac{Y_{b - 2,0}^{(l)}(\pi w_{\infty}^*)}{Y_{b,0}(\pi w_{\infty}^*)}\bigg) q^{1 - b} + O(q),
$$
where $\widehat{g}^*_{0} = -\frac{2}{2 + n}$ and $(\widehat{g}^*_k)_{k \geq 1}$ should be replaced by their values in terms of $(g,h,w_{\infty}^*)$ from Section~\ref{Appgdeter}, and $(g,h)$ by their parametrization \eqref{gsurh}-\eqref{h2eq} on the critical line.

We examine the case $\alpha = 1$. Using the parametrization \eqref{gsurh}-\eqref{h2eq}, the resulting formula is:
\beq
\label{Deltastar}1 - u = q_*\,q^{1 - b} + (q_{*,1} + c'\rho_1)q + o(q).
\eeq
with:
\bea
q_* & = & \frac{12}{b}\,\frac{\rho^2(1 - b)^2\sqrt{2 + n} + 2\rho(1 - b)\sqrt{2 - n} - 2\sqrt{2 + n}}{-\rho^2 b(1 - b^2)\sqrt{2 + n} + 4\rho(1 - b^2)\sqrt{2 - n} - 6b\sqrt{2 + n}}, \nonumber \\
q_{*,1} & = & \frac{24}{b}\,\frac{-\rho^2(b^2 + 1)\sqrt{2 + n} + 2\rho b\sqrt{2 - n} + 2\sqrt{2 + n}}{-\rho^2b(1 - b^2)\sqrt{2 + n} + 4\rho(1 - b^2)\sqrt{2 - n} - 6b\sqrt{2 + n}}. \nonumber
\eea
The value of $c'$ is irrelevant because we will soon show that $\rho_1 = 0$. As $(1 - u)$ should be nonnegative for $q > 0$, we must have $q_* \geq 0$, which demands $\rho \in [\rho_{\min},\rho_{\max}]$ with:
\beq
\label{rhominf}\rho_{\min} = \frac{\sqrt{6 + n} - \sqrt{2 - n}}{(1 - b)\sqrt{2 + n}}.
\eeq
We observe that this lower bound is larger than $\rho_{\min}'$ given by \eqref{E55} for any $n \in [0,2]$, therefore the non-generic critical line can only exist in the range $\rho \in [\rho_{\min},\rho_{\max}]$ provided by \eqref{rhominf}-\eqref{E56}. These necessary conditions were also obtained in \cite{BBG12b} -- where the lower bound arose from the constraint of positivity of the spectral density associated with the generating series of disks $\mathbf{F}(x)$ -- and it was checked that these conditions are sufficient.

We now turn to the second equation \eqref{eq2}. We have checked that the term in $q^{b}$ vanishes, as we expect by consistency with \eqref{Deltastar}. Then, the term of order $q^{\frac{1}{2}}$ is proportional to $\rho_{1}$, therefore we must have, in both dense and dilute phase:
$$
\rho_{1} = 0,
$$
which means that $\gamma_{-} - \gamma_{-}^* \in O(q)$.

\vspace{0.2cm}

We see that for $\rho \in (\rho_{\min},\rho_{\max}]$:
$$
q \sim q_*\,(1 - u)^{\frac{1}{1 - b}}.
$$
for some constant $q_* > 0$. This corresponds, by definition, to the dense phase. For $\rho = \rho_{\min}$ (\textit{i.e.} the dilute phase) we have $q_* = 0$, and \eqref{Deltastar} specializes to:
$$
1 - u = \frac{24}{b(1 - b)(2 - b)}\,q + o(q).
$$

For general $\alpha$ not too large (see the statement of Theorem~\ref{alphanotlarge}), the result is qualitatively the same, only the non-zero constant prefactor differs -- see \cite[Appendix D]{BBD}.

\section{Proof of Lemma~\ref{pieces}}
\label{AppBB}

The goal in this appendix is to obtain the critical behavior of the building blocks. We give expressions valid for both universality classes using
$$
\mathfrak{d} = \left\{\begin{array}{lll} 1 & & {\rm dense}, \\ -1 & & {\rm dilute}. \end{array} \right.
$$
 From the expression \eqref{BB0}-\eqref{BB1} and the behavior of the special function $\Upsilon_{b}$ from Lemma~\ref{lemUp} we find:
\label{aabas}
\begin{lemma}
\label{lmH5} We have in the regime $T \rightarrow 0$:
\bea
\mathbf{B}_{\varepsilon,l}(\varepsilon' + \tau\phi) & = & \frac{2(-1)^{l + 1}}{\sqrt{4 - n^2}}\,\Big(\frac{\pi}{T}\Big)^{2l + 2}\,q^{b(\varepsilon \oplus \varepsilon')}\big\{B_{\varepsilon\oplus \varepsilon',b}^{*,(2l + 1)}(\pi\phi) + O(q^{b})\big\}, \nonumber \\
\bs{\mathcal{B}}_{\varepsilon,l}(\varepsilon' + \tau\phi) & = & (-1)^{l + 1} \Big(\frac{\pi}{T}\Big)^{2l + 2} q^{\frac{1}{2}(\varepsilon \oplus \varepsilon')}\big\{B_{\varepsilon \oplus \varepsilon',\frac{1}{2}}^{*,(2l + 1)}(\pi\phi) + O(q^{\frac{1}{2}})\big\} ,\nonumber
\eea
where
$$
B_{0,b}^*(\phi) = \frac{\sin (1 - b)\phi}{\sin \phi},\qquad B_{\frac{1}{2},b}^*(\phi) = 2\cos b\phi.
$$
The error is uniform for $\phi$ in any compact, and stable by differentiation.
\end{lemma}

We next focus on the denominator of the recursion kernel.

\begin{lemma}
\label{lmH6} We have in the regime $T \rightarrow 0$
\beq
(\Delta_{\varepsilon}\mathbf{G})(\tau\phi) = \Big(\frac{\pi}{T}\Big)\,q^{(1 - \mathfrak{d}\frac{b}{2})(1 - 2\varepsilon)}\,\big\{G^*_{\varepsilon}(\phi) + O(q^b)\big\}. \nonumber
\eeq
We have
$$
G^*_{0}(\phi) = \left\{\begin{array}{lll} -D^*_{b - 2}\,\sin \pi\phi\,\sin \pi(1 - b)\phi & & {\rm in}\,\,{\rm dense}\,\,{\rm phase}, \\ D^*_{b + 2}\,\sin\pi\phi\,\sin\pi(1 + b)\phi & & {\rm in}\,\,{\rm dilute}\,\,{\rm phase}, \end{array}\right.
$$
with $D^*_b$ given in \eqref{Dstarb} below. In each phase, $G_0^*(\phi) \neq 0$. We have
$$
G_{1/2}^*(\phi) = {\rm i}\sqrt{4 - n^2}\,\mathcal{D}^*\bigg(2\frac{\sin\pi(1 - b)w_{\infty}^*}{\sin\pi w_{\infty}^*} - \frac{\sin\pi(1 - b)(w_{\infty}^* - \phi)}{\sin\pi(w_{\infty}^* - \phi)} - \frac{\sin\pi(1 - b)(w_{\infty}^* + \phi)}{\sin\pi(w_{\infty}^* + \phi)}\bigg),
$$
where $\mathcal{D}^*$ is a differential operator given in \eqref{Diffopsa} below.
\end{lemma}
\noindent\textbf{Proof.} From Proposition~\ref{theimdisk} and the behavior of $\Upsilon_{b}(\tau\phi + \tfrac{1}{2})$ given in Lemma~\ref{lemUp}, we repeat in a finer way the computation of the beginning of Section~\ref{Secphase}.:
\bea
\label{D0G}  && (\Delta_{0}\mathbf{G})(\tau\phi) \\
& = & \sqrt{4 - n^2}\,\frac{8{\rm i}\pi}{T}\,\frac{q^{b/2}}{1 - q^b}\Big\{ -\cos \pi b \phi\,\mathcal{D}Y_{b,0}(\pi w_{\infty}) + q^{1 - b}\,\cos \pi(b - 2)\phi\,\mathcal{D}Y_{b - 2,0}(\pi w_{\infty}) \nonumber \\
&& - q\big(\cos\pi(b - 2)\phi\,\mathcal{D}Y_{b - 2,0}(\pi w_{\infty}) + \cos\pi(b + 2)\phi\,\mathcal{D} Y_{b + 2,0}(\pi w_{\infty})\big) + O(q^{1 + b})\Big\}, \nonumber
\eea
where $Y_{b,0}$ are $\mathcal{D}$ were introduced in \eqref{DYB}. One of the exact condition determining the endpoint was $\Delta_0\mathbf{G}(0) = 0$, \textit{i.e.}
$$
\mathcal{D}Y_{b,0}(\pi w_{\infty}) = q^{1 - b}\mathcal{D} Y_{b - 2,0}(\pi w_{\infty}) - q \mathcal{D}(Y_{b - 2,0} + Y_{b + 2,0})(\pi w_{\infty}) + O(q^{1 + b}) = 0,
$$
which we can substitute in \eqref{D0G} to obtain
\bea
&& (\Delta_{0}\mathbf{G})(\tau\phi) \nonumber \\
& = & \sqrt{4 - n^2}\,\frac{16{\rm i}\pi}{T}\,\frac{q^{b/2}}{1 - q^b}\bigg\{q^{1 - b}\sin\pi\phi \sin\pi(1 - b)\phi \mathcal{D}Y_{b - 2,0}(\pi w_{\infty}) \nonumber \\
&& + q \sin\pi\phi \Big(\sin\pi(1 - b)\phi\,\mathcal{D} Y_{b - 2,0}(\pi w_{\infty}) + \sin\pi(1 + b)\phi\,\mathcal{D}Y_{b + 2,0}(\pi w_{\infty})\Big) + O(q^{1 + b})\bigg\}. \nonumber
\eea
The dense phase was characterized by
\beq
\label{Diffopsa} \mathcal{D}^*Y_{b - 2,0}(\pi w_{\infty}^*) \neq 0,\qquad \mathcal{D}^* = \sum_{l \geq 0} \frac{(-1)^l\widehat{g}_{l}^*}{l!}\,\partial_{\pi w_{\infty}^*}^{l}.
\eeq
Therefore, the first term, of order $q^{1 - b}$, is indeed the leading term. The dilute phase is characterized by $\mathcal{D}^*Y_{b - 2}(\pi w_{\infty}^*) = 0$ and then one can check that $\mathcal{D}^*Y_{b + 2}(\pi w_{\infty}^*) \neq 0$. So, in the dilute phase the leading term is of order $O(q)$. This gives the announced results with
\beq
\label{Dstarb} D^*_{b} = 16{\rm i}\sqrt{4 - n^2}\,\mathcal{D}^*Y_{b}(\pi w_{\infty}^*).
\eeq
For $(\Delta_{\frac{1}{2}}\mathbf{G})(\tau\phi)$, we easily arrive to the result using the beh\-avior of $\Upsilon_{b}(\tau\phi)$ from Lem\-ma~\ref{lemUp}, and exploiting the freedom to subtract $\Delta_{\frac{1}{2}}\mathbf{G}(0) = 0$. \hfill $\Box$

\begin{corollary}
\label{coE3} We have when $T \rightarrow 0$, for $r = 1,2$ and $\varepsilon \in \{0,\tfrac{1}{2}\}$:
$$
y_{\varepsilon,r} = \Big(\frac{\pi}{T}\Big)^{2r + 1}q^{(1 - 2\varepsilon)(1 - \mathfrak{d}b/2)}\big\{y_{\varepsilon,r}^* + O(q^{b})\big\}, 
$$
with $y_{\varepsilon,r}^* \neq 0$ computable from Lemma~\ref{lmH6}, and $y_{0,2}^*/y_{0,1}^* = - 2 + 2b - b^2$.
\end{corollary}

Inserting the previous results into the expressions \eqref{Coini}-\eqref{C1ini} for the initial conditions, we find:
\begin{corollary}
When $T \rightarrow 0$, we have
\bea
\mathsf{C}^{(0,3)}\bigl[{}^{0}_{\varepsilon}\,{}^{0}_{\varepsilon}\,{}^{0}_{\varepsilon}\bigr] & = & \Big(\frac{\pi}{T}\Big)^{-3}\,q^{(1 - 2\varepsilon)(\mathfrak{d}b/2 - 1)}\Big\{\mathsf{C}^{(0,3)}_{*}\bigl[{}^{0}_{\varepsilon}\,{}^{0}_{\varepsilon}\,{}^{0}_{\varepsilon}\bigr] + O(q^{b})\Big\}, \nonumber \\
\mathsf{C}^{(1,1)}\bigl[{}^{0}_{\varepsilon}\bigr] & = & \Big(\frac{\pi}{T}\Big)^{-1}\,q^{(1 - 2\varepsilon)(\mathfrak{d}b/2 - 1)}\Big\{\mathsf{C}^{(1,1)}_{*}\bigl[{}^{0}_{\varepsilon}\bigr] + O(q^{b})\Big\}, \nonumber \\
\mathsf{C}^{(1,1)}\bigl[{}^{1}_{\varepsilon}\bigr] & = & \Big(\frac{\pi}{T}\Big)^{-3}\,q^{(1 - 2\varepsilon)(\mathfrak{d}b/2 - 1)}\Big\{\mathsf{C}^{(1,1)}_{*}\bigl[{}^{1}_{\varepsilon}\bigr] + O(q^{b})\Big\}, \nonumber \\
\eea
and likewise for the $\mathcal{C}$'s, with: 
$$
\begin{aligned}
\mathsf{C}^{(0,3)}_{*}\bigl[{}^{0}_{\varepsilon}\,{}^{0}_{\varepsilon}\,{}^{0}_{\varepsilon}\bigr] \ & = \  -\frac{2}{y_{\varepsilon,1}^{*}}, & \ \ \ \  \mathcal{C}^{(0,3)}_{*}\bigl[{}^{0}_{\varepsilon}\,{}^{0}_{\varepsilon}\,{}^{0}_{\varepsilon}\bigr] \ & = \  -\frac{2}{y_{\varepsilon,1}^{*}}, \\
\mathsf{C}^{(1,1)}_{*}\bigl[{}^{0}_{0}\bigr] \ & = \  \frac{6 + 26b + 11b^2}{24y_{0,1}^*}, & \ \ \ \  \mathcal{C}^{(1,1)}_{*}\bigl[{}^{0}_{0}\bigr] \ & = \  \frac{29 + 26b + 11b^2}{24y_{0,1}^*},   \\
\mathsf{C}^{(1,1)}_{*}\bigl[{}^{\,0}_{\frac{1}{2}}\bigr] \ & = \ \frac{y_{\frac{1}{2},2}^*}{24(y_{\frac{1}{2},1}^*)^2} + \frac{2 + 6b + 3b^2}{6y_{\frac{1}{2},1}^*}, & \ \ \ \ \mathcal{C}^{(1,1)}_{*}\bigl[{}^{\,0}_{\frac{1}{2}}\bigr] \ & = \  \frac{y_{\frac{1}{2},2}^*}{24(y_{\frac{1}{2},1}^*)^2} + \frac{23}{24y_{\frac{1}{2},1}^*}, \\
\mathsf{C}^{(1,1)}_*\bigl[{}^{1}_{\varepsilon}\bigr] \ & = \  -\frac{1}{24y_{\varepsilon,1}^{*}}, & \ \ \ \  \mathcal{C}^{(1,1)}_*\bigl[{}^{1}_{\varepsilon}\bigr] \ & = \   -\frac{1}{24y_{\varepsilon,1}^{*}}.
\end{aligned}
$$
\end{corollary}

From Corollary~\ref{coE3} we can then deduce the critical behavior of $K$'s and $\tilde{K}$'s.

\begin{corollary}
For $\varepsilon,\sigma,\sigma' \in \{0,\frac{1}{2}\}$, we denote:
$$
f(\varepsilon,\sigma,\sigma'\vert B) \coloneqq B\big[(\varepsilon \oplus \sigma) + (\varepsilon \oplus \sigma')\big] + (\mathfrak{d}\tfrac{b}{2} - 1)(1 - 2\varepsilon).
$$
When $T \rightarrow 0$, we have
\bea
K\bigl[{}^{l}_{\varepsilon}\,{}^{m}_{\sigma}\,{}^{m'}_{\sigma'}\bigr] & = & \Big(\frac{\pi}{T}\Big)^{2(m + m' - l) + 1}\,q^{f(\varepsilon,\sigma,\sigma'\vert b)}\Big\{K^*\bigl[{}^{l}_{\varepsilon}\,{}^{m}_{\sigma}\,{}^{m'}_{\sigma'}\bigr] + O(q^{b})\Big\}, \nonumber \\
\tilde{K}\bigl[{}^{l}_{\varepsilon}\,{}^{l'}_{\varepsilon}\,{}^{m}_{\sigma}\bigr] & = & \Big(\frac{\pi}{T}\Big)^{2(m - l - l') - 1}\,q^{f(\varepsilon,\varepsilon,\sigma\vert b)}\Big\{\tilde{K}^*\bigl[{}^{l}_{\varepsilon}\,{}^{l'}_{\varepsilon}\,{}^{m}_{\sigma}\bigr] + O(q^{b})\Big\}, \nonumber
\eea
and 
\bea
\mathcal{K}\bigl[{}^{l}_{\varepsilon}\,{}^{m}_{\sigma}\,{}^{m'}_{\sigma'}\bigr] & = & \Big(\frac{\pi}{T}\Big)^{2(m + m' - l) + 1}\,q^{f\left(\varepsilon,\sigma,\sigma'\vert \frac{1}{2}\right)}\Big\{K^*\bigl[{}^{l}_{\varepsilon}\,{}^{m}_{\sigma}\,{}^{m'}_{\sigma'}\bigr] + O(q^{b})\Big\}, \nonumber \\
\tilde{\mathcal{K}}\bigl[{}^{l}_{\varepsilon}\,{}^{l'}_{\varepsilon}\,{}^{m}_{\sigma}\bigr] & = & \Big(\frac{\pi}{T}\Big)^{2(m - l - l') - 1}\,q^{f\left(\varepsilon,\varepsilon,\sigma\vert \frac{1}{2}\right)}\Big\{\tilde{K}^*\bigl[{}^{l}_{\varepsilon}\,{}^{l'}_{\varepsilon}\,{}^{m}_{\sigma}\bigr] + O(q^{b})\Big\}, \nonumber
\eea
with
\bea
K^*\bigl[{}^{l}_{\varepsilon}\,{}^{m}_{\sigma}\,{}^{m'}_{\sigma'}\bigr] & = & \frac{4(-1)^{l + m + m'}}{4 - n^2} \Res_{\phi \rightarrow 0} \frac{\dd \phi\,\phi^{2l + 1}}{(2l + 1)!\,G^*_{\varepsilon}(\phi)}\,B_{\varepsilon \oplus \sigma}^{*,(2m + 1)}[\pi(\phi + 1)]\,B_{\varepsilon \oplus \sigma'}^{*,(2m' + 1)}[\pi(\phi - 1)], \nonumber \\
\tilde{K}^*\bigl[{}^{l}_{\varepsilon}\,{}^{l'}_{\varepsilon}\,{}^{m}_{\sigma}\bigr] & = & \frac{2(-1)^{m + l + l' + 1}}{\sqrt{4 - n^2}}\,\Res_{\phi \rightarrow 0} \frac{\dd \phi\,\phi^{2(l + l') + 1}}{(2l + 1)!(2l')!\,G^*_{\varepsilon}(\phi)}\,B_{\varepsilon \oplus \sigma}^{*,(2m + 1)}[\pi(\phi + 1)], \nonumber
\eea
and likewise for the $\mathcal{K}^*$ and $\tilde{\mathcal{K}}^*$.
\end{corollary}
\noindent \textbf{Proof.} This is a direct computation from Lemma~\ref{lmH5}-\ref{lmH6}. We note that for $\tilde{K}$ (and resp. $\tilde{\mathcal{K}}$), we find an exponent $q^{\tilde{f}(\varepsilon,\sigma\vert B)}$, with $\tilde{f}(\varepsilon,\sigma\vert B) = B(\varepsilon \oplus \sigma) + (\mathfrak{d}\tfrac{b}{2} - 1)(1 - 2\varepsilon)$, with $B=b$ (and resp. $B=\frac{1}{2}$). But since $\varepsilon \oplus \varepsilon = 0$, this is also equal to $f(\varepsilon,\varepsilon,\sigma\vert B)$. \hfill $\Box$

We also remark that the order of magnitude of $\mathsf{C}^{(0,3)}\bigl[{}^{0}_{\varepsilon}\,{}^{0}_{\varepsilon}\,{}^{0}_{\varepsilon}\bigr]$ and $\mathsf{C}^{(1,1)}\bigl[{}^{l}_{\varepsilon}]$, and of $\mathcal{C}^{(0,3)}\bigl[{}^{0}_{\varepsilon}\,{}^{0}_{\varepsilon}\,{}^{0}_{\varepsilon}\bigr]$ and $\mathcal{C}^{(1,1)}\bigl[{}^{l}_{\varepsilon}]$, is $q^{f(\varepsilon,\varepsilon,\varepsilon\vert b)}=q^{f\left(\varepsilon,\varepsilon,\varepsilon\vert \frac{1}{2}\right)}$. Therefore, for a given graph $\mathcal{G}$ and coloring $\bs{\sigma}$ of its edges appearing in the sum of Proposition~\ref{cosums}, and any vertex $\mathsf{v} \in V(\mathcal{G})$, the factor associated to $\mathsf{v}$ -- either $K$, $\tilde{K}$, $\mathsf{C}^{(0,3)}$ or $\mathsf{C}^{(1,1)}$ -- is of order of magnitude $q^{f_{\mathsf{v}}(b)}$ with
$$
f_{\mathsf{v}}(b) = f\big(\sigma(\mathsf{e}_{\mathsf{v}}^{0}),\sigma(\mathsf{e}_{\mathsf{v}}^{1}),\sigma(\mathsf{e}_{\mathsf{v}}^{2})\vert b\big).
$$
Similarly, any factor $\mathcal{K}$, $\tilde{\mathcal{K}}$, $\mathsf{\mathcal{C}}^{(0,3)}$ or $\mathsf{\mathcal{C}}^{(1,1)}$ associated to a vertex $\mathsf{v}\in V(\mathcal{G})$ scales like $q^{f_{\mathsf{v}}\left(\frac{1}{2}\right)}$.

\providecommand{\bysame}{\leavevmode\hbox to3em{\hrulefill}\thinspace}

\end{document}